\documentclass[letterpaper,12pt]{article}
\pdfoutput=1

\usepackage{amsthm,amsfonts,amssymb,amsmath,dsfont}
\usepackage[pdftex]{graphicx,graphics,color}
\usepackage{color,colortbl}
\usepackage{lmodern}
\usepackage[english]{babel}
\definecolor{citecl}{rgb}{0.55,0.55,0.55}
\usepackage[colorlinks=true,linkcolor=black,citecolor=citecl,anchorcolor=blue,filecolor=blue,urlcolor=blue]{hyperref}
\usepackage[super,sort&compress]{natbib}
\usepackage[affil-it]{authblk}

\DeclareMathOperator{\tr}{tr}
\DeclareMathOperator{\re}{re}
\DeclareMathOperator{\im}{im}

\begin{document}


\newcommand{\nn}{\nonumber}
\newcommand{\bra}{\langle}
\newcommand{\ket}{\rangle}
\newcommand{\del}{\partial}
\newcommand{\vt}{\vec}
\newcommand{\dg}{^{\dag}}
\newcommand{\cg}{^{*}}
\newcommand{\T}{^{T}}
\newcommand{\vep}{\varepsilon}
\newcommand{\vth}{\vartheta}
\newcommand{\suml}{\sum\limits}
\newcommand{\prodl}{\prod\limits}
\newcommand{\intl}{\int\limits}
\newcommand{\til}{\tilde}
\newcommand{\wb}{\overline}
\newcommand{\mcl}{\mathcal}
\newcommand{\mfk}{\mathfrak}
\newcommand{\mds}{\mathds}
\newcommand{\mbb}{\mathbb}
\newcommand{\mrm}{\mathrm}
\newcommand{\ds}{\displaystyle}
\newcommand{\rmi}{\mathrm{i}}
\newcommand{\rme}{\mathrm{e}}
\newcommand{\rmd}{\mathrm{d}}
\newcommand{\rmD}{\mathrm{D}}
\newcommand{\vphi}{\varphi}
\newcommand{\stm}{\text{\textsf{s}}}
\newcommand{\bth}{\text{\textsf{b}}}
\newcommand{\itr}{\text{\textsf{i}}}
\renewcommand{\mod}{\mathrm{\,mod\,}}
\renewcommand{\b}{\bar}
\renewcommand{\dim}{\mbox{dim}}
\renewcommand{\-}{-}


\interfootnotelinepenalty=10000

\renewcommand{\topfraction}{0.85}
\renewcommand{\textfraction}{0.1}

\allowdisplaybreaks[1]

\setlength{\jot}{1ex}

\renewcommand\arraystretch{1}

\renewcommand{\theequation}{\thesection.\arabic{equation}}
\numberwithin{equation}{section}
\renewcommand{\thefigure}{\thesection.\arabic{figure}}
\numberwithin{figure}{section}
\renewcommand{\thefootnote}{\roman{footnote}}


\title{Semiclassical Propagator in the Generalized Coherent-State Representation}

\author[1]{Thiago F. Viscondi}
\affil[1]{Instituto de Física, Universidade de S\~ao Paulo, S\~ao Paulo, SP, Brazil}
\author[2]{Adriano Grigolo}
\author[2]{Marcus A. M. de Aguiar}
\affil[2]{Instituto de Física Gleb Wataghin, Universidade Estadual de Campinas, Campinas, SP, Brazil}

\date{}
\maketitle


\begin{abstract}
A detailed derivation of the semiclassical propagator in the generalized coherent-state representation 
is performed by applying the saddle-point method to a path integral over the classical phase space. 
With the purpose of providing greater accessibility and applicability to the developed formalism,
a brief review of the generalized concept of coherent states is presented, in which three examples 
of coherent-state sets are examined, namely, the canonical, spin, and $\mrm{SU}(n)$ bosonic coherent 
states.
\end{abstract}



\section{Introduction}
\label{sec:1}

Semiclassical methods have their origin in the early stages of quantum mechanics, when 
attempts to understand quantum phenomena in terms of classical mechanics were a natural approach.
Currently, as a consequence of the progress made in quantum physics, semiclassical formulas 
are interpreted as approximate solutions to the Schr\"odinger equation, which are based on the 
hypothesis that typical values of the classical action for the observed system are substantially 
larger than the Planck constant.

The development of semiclassical approximations has two main motivations. First, in the investigation of physical 
systems with many degrees of freedom, it is believed that the proper application of semiclassical methods leads 
to significant computational advantages over exact quantum calculations. In other words, within their range of 
validity, the semiclassical formulas stand as a viable alternative in the analysis of complex quantum systems. 
For this reason, semiclassical approaches have great relevance in theoretical chemistry, especially in the study 
of atomic and molecular dynamics.\cite{Miller01,Thoss04,VanVoorhis04,Kay05,Miller06,Saha07,Harabati07,Moix09,Goletz09a} 
Second, the semiclassical formalism provides important tools for the deep understanding of the connections between 
classical and quantum theories. According to the correspondence principle, classical physics constitutes a limiting 
case of quantum mechanics. Although this statement is based on solid foundations, there are many unanswered questions 
concerning the transition region between the two theoretical fields. In particular, the use of semiclassical 
treatments enables a careful examination of fundamental topics such as quantum chaos\cite{Tomsovic91,Kay94c}, 
entanglement\cite{Ribeiro12} and open quantum systems\cite{Grossmann06,Koch08,Moix08,Goletz09b,Goletz10}.

In the present paper, we are primarily interested in the semiclassical dynamics of isolated quantum systems. 
Consequently, the central element of our work corresponds to the \textit{quantum propagator}, which is defined 
as the probability amplitude for the transition between two quantum states.

The semiclassical formulation of the quantum propagator, in the position representation, was originally obtained by 
Van Vleck in the first half of the last century.\cite{VanVleck28} As an attempt to overcome the technical difficulties 
in the practical application of Van Vleck's formula, several studies were directed to the derivation and discussion 
of the semiclassical propagator in the representation of canonical coherent states, also known as harmonic-oscillator 
or Heisenberg-Weyl coherent states.\cite{Klauder79,Weissman82,Weissman83,Solari86,Huber87,Huber88,Baranger01,Braun07a} 
Even though the complications of the Van Vleck's propagator have persisted, the coherent-state approach 
to the semiclassical dynamics exhibits some substantial advantages over the position representation, 
such as an immediate phase-space visualization of the system, thus enabling a proper comparison with 
classical mechanics.

With the purpose of describing the dynamics of systems with spin degrees of freedom, several independent 
derivations of the semiclassical propagator in the representation of $\mrm{SU}(2)$ coherent states were 
performed.\cite{Solari87,Vieira95,Kochetov95a,Stone00,Ribeiro06,Braun07b} As a natural consequence of these 
theoretical investigations, Kochetov extended his results to arbitrary dynamical groups.\cite{Kochetov95b,Kochetov98} 
Therefore, with the aid of a broad definition of coherent states,\cite{Perelomov72,Gilmore72} the application 
of semiclassical methods became possible for a wide variety of physical systems, including cases whose classical 
analogues were not evident.

In a few words, the derivation of the semiclassical propagator amounts to explicitly evaluating the path integral 
of a quantum system by quadratically approximating the action functional. This task may prove to be fairly complicated, 
particularly in systems with multiple degrees of freedom or non-Euclidean metric. For this reason, we present here an 
independent derivation of the generalized coherent-state semiclassical propagator, which focuses on its accessibility 
and applicability. Although our final result has the same form of the Kochetov's propagator,\cite{Kochetov98} our 
line of reasoning closely follows the derivation procedure proposed by Braun and Garg,\cite{Braun07a,Braun07b} who 
obtained specific formulas for the semiclassical propagator in the representations of canonical and spin coherent 
states with an arbitrary number of degrees of freedom.

The paper is organized as follows. In section~\ref{sec:2}, we briefly review the basic formalism of the generalized 
coherent states. In this way, we introduce important definitions for the construction of our subsequent results, such 
as the analytic parametrization of a Lie group's quotient space and the analytic continuations of the K\"ahler potential 
and metric. Subsection~\ref{ssc:2.2} provides three examples of coherent-state sets, which we designate as canonical, spin, 
and $\mrm{SU}(n)$ bosonic coherent states. Section~\ref{sec:3} comprises the main result of this paper, a detailed derivation 
of the semiclassical propagator in the coherent-state representation. First, in subsection~\ref{ssc:3.1}, we promptly present 
the final expression for the semiclassical propagator along with its constituent elements, such as the action functional and 
the effective classical Hamiltonian. With this initial presentation, we intend to facilitate reference to a comprehensive set 
of semiclassical formulas from within the present paper and in future work. Then, in subsection~\ref{sss:3.2.1}, we begin the 
actual derivation of the semiclassical propagator by obtaining the coherent-state path integral. In subsection~\ref{sss:3.2.1.add}, 
we properly determine the classical equations of motion, which are subject to two-time boundary conditions as a natural imposition 
from the coherent-state propagator. In this context, a very subtle concept is introduced, the phase-space duplication. 
Subsection~\ref{sss:3.2.2} shows the evaluation of the first variation of the action functional. As a result of this 
auxiliary calculation, we obtain some important formulas, such as the first and second derivatives of the classical 
action. In subsection~\ref{sss:3.2.3}, we perform the semiclassical approximation of the coherent-state path integral, 
which amounts to the second-order expansion of the action functional around a classical trajectory. As a consequence 
of this perturbative procedure, the semiclassical propagator is formulated as the product between a zero-order factor 
and a Gaussian path integral. This remaining integral, to which we refer as the reduced propagator, is evaluated in 
subsection~\ref{sss:3.2.4}. The value of the reduced propagator is given in terms of the solution of a second-order 
homogeneous differential equation, which is then related to the Jacobi equation in subsection~\ref{sss:3.2.5}. 
In subsection~\ref{sss:3.2.6}, after a final change of dynamical variables, we conclude the derivation of 
the semiclassical propagator. Finally, in section~\ref{sec:4}, we summarize our main findings and discuss 
the applicability of the developed semiclassical formalism. In appendix~\ref{app:A}, we present the definitions 
of three matrix quantities, which are particularly relevant to semiclassical dynamics. For completeness, some 
auxiliary results employed in subsections \ref{sss:3.2.2} and \ref{sss:3.2.3} are shown in appendix~\ref{app:B}.



\section{Coherent States}
\label{sec:2}

The coherent states were initially envisioned by Schrödinger\cite{Schrodinger26} as minimum-uncertainty Gaussian states, 
whose dynamics demonstrate maximum similarity to the classical harmonic oscillator. The interest in these specific states, 
which are related to the Heisenberg-Weyl group, was again stimulated by the works of Glauber\cite{Glauber63a,Glauber63b,Glauber63c}, 
Klauder\cite{Klauder60,Klauder63a,Klauder63b} and Sudarshan\cite{Sudarshan63}, with first applications emerging in the areas 
of quantum optics and semiclassical approximation.

The generalization of the coherent states to arbitrary Lie groups was originally examined by Klauder, but a complete and detailed 
discussion of this subject was developed only later by Perelomov\cite{Perelomov72} and Gilmore\cite{Gilmore72}. In this way, the 
interesting properties of the Heisenberg-Weyl coherent states were extended to other groups of physical interest, resulting in an 
ideal setting for the study of classical-quantum correspondence.

In this section, we review the generalized concept of coherent state. In addition, we present 
the main assumptions used in the subsequent development of the semiclassical propagator.

\subsection{Definitions}
\label{ssc:2.1}

\subsubsection{Dynamical Group}
\label{sss:2.1.1}

Consider a connected Lie group~$\mrm{G}$ and its corresponding algebra~$\mrm{g}$. These two sets are related by the exponential 
map:\cite{Hall03}
\begin{equation}
 a=\rme^{\rmi A},
 \label{eq:2.1}
\end{equation}
\noindent with $a\in\mrm{G}$ and $A\in\mrm{g}$. If the Hamiltonian operator~$H$ of a given physical system can be expressed 
as a function of the generators of the group~$\mrm{G}$, that is,\footnote{In general, we shall not use the hat symbol to indicate 
quantum operators, except in cases where its absence could generate ambiguities.}
\begin{equation}
 H=H(A_{1},A_{2},\ldots,A_{d_{\mrm{g}}}),
 \label{eq:2.2}
\end{equation}
\noindent in which the set of operators $\{A_{1},A_{2},\ldots,A_{d_{\mrm{g}}}\}$ constitutes a basis for the algebra~$\mrm{g}$, 
we say that the physical system possesses \textit{dynamical group}~$\mrm{G}$.

The Hamiltonian operator determines the state space dynamically accessible to the physical system, which we denote 
by $\mds{H}_{l}$. As the first fundamental hypothesis of this section, we assume that the Hilbert space~$\mds{H}_{l}$ 
supports an \textit{irreducible unitary representation} of the group~$\mrm{G}$, denoted by $\mrm{U}_{l}$ and specified 
by the set of indices~$l$. According to the map~\eqref{eq:2.1}, the unitary representations of the group~$\mrm{G}$ are 
associated with Hermitian representations of the algebra~$\mrm{g}$. Then, as a second hypothesis, we assume that the 
basis of $\mrm{g}$ can be decomposed as follows:
\begin{equation}
 \left\{A_{j}\right\}=\left\{Q_{k},E\dg_{m}+E_{m},\rmi(E\dg_{m}-E_{m})\right\},
 \label{eq:2.3}
\end{equation}
\noindent with $j=1,2,\ldots,d_{\mrm{g}}$, $k=1,2,\ldots,r_{\mrm{g}}$ and $m=1,2,\ldots,(d_{\mrm{g}}-r_{\mrm{g}})/2$. 
The generators~$Q_{k}$ constitute a set of mutually commuting operators among the basis elements of the algebra~$\mrm{g}$. 
Also in equation~\eqref{eq:2.3}, we introduce the non-Hermitian quantities $E_{m}$ and $E_{m}\dg$, which are known 
respectively as the \textit{lowering and raising operators} on the space $\mds{H}_{l}$.

As a consequence of the commutativity of the generators~$Q_{k}$, the matrix representations of these Hermitian operators can 
be simultaneously diagonalized by a suitable choice of basis in the space~$\mds{H}_{l}$. Among the elements of this particular 
basis, there is a single normalized state~$|\psi_{0}^{l}\ket$, known as \textit{minimum weight state}, such that
\begin{equation}
 E_{m}|\psi_{0}^{l}\ket=0,
 \label{eq:2.4}
\end{equation}
\noindent for all possible values of $m$. Furthermore, there can also be a subset of values of the index~$m$, which we 
indicate by $m'$, so that
\begin{equation}
 E\dg_{m'}|\psi_{0}^{l}\ket=0.
 \label{eq:2.5}
\end{equation}

In order to explicitly present all the defining relations for the state~$|\psi_{0}^{l}\ket$, we write its eigenvalue 
equations associated with generators~$Q_{k}$:
\begin{equation}
 Q_{k}|\psi_{0}^{l}\ket=\lambda_{k}(l)|\psi_{0}^{l}\ket,
 \label{eq:2.6}
\end{equation}
\noindent in which $\lambda_{k}(l)$ stands for the eigenvalue of the operator~$Q_{k}$ in the space~$\mds{H}_{l}$.

Finally, considering a particular choice for the decomposition~\eqref{eq:2.3}, we observe that the dynamical group~$\mrm{G}$ 
and the Hilbert space~$\mds{H}_{l}$ unambiguously determine the state~$|\psi_{0}^{l}\ket$, except for a global phase without 
physical meaning.

\subsubsection{Quotient Space and Coherent States}
\label{sss:2.1.2}

The \textit{stability subgroup}~$\mrm{S}$ is defined as the subset of unitary operators~$s$, belonging to the group~$\mrm{G}$, 
which preserve the \textit{reference state}~$|\psi_{0}^{l}\ket$ invariant within a global phase factor, that is,
\begin{equation}
 s|\psi_{0}^{l}\ket=\rme^{\rmi\phi(s)}|\psi_{0}^{l}\ket,
 \label{eq:2.7}
\end{equation}
\noindent for $s\in\mrm{S}$ and $\phi(s)\in\mds{R}$. With the assistance of the definition for the stability subgroup, we are 
able to discriminate the elements of the \textit{quotient space}~$\mrm{G}/\mrm{S}$, which comprises the operations in $\mrm{G}$ 
that produce mutually non-equivalent physical states when applied individually on $|\psi_{0}^{l}\ket$. It follows directly from 
the fundamental axioms of group theory\cite{Hamermesh62} that, for each operation~$a\in\mrm{G}$, there is a unique decomposition
\begin{equation}
 a=qs,
 \label{eq:2.8}
\end{equation}
\noindent so that $s\in\mrm{S}$ and, by definition, $q\in\mrm{G}/\mrm{S}$. Therefore, the application of any element $a\in\mrm{G}$ 
on the reference state results in an identity of the following form:
\begin{equation}
 a|\psi_{0}^{l}\ket=qs|\psi_{0}^{l}\ket=\rme^{\rmi\phi(s)}q|\psi_{0}^{l}\ket.
 \label{eq:2.9}
\end{equation}

As evidenced in the above equation, two different operations of the group~$\mrm{G}$ on the state~$|\psi_{0}^{l}\ket$ can produce 
results with equivalent physical content, arising from the action of the same operator~$q\in\mrm{G}/\mrm{S}$, but differing by a
global phase factor resulting from the application of distinct elements in the subgroup~$\mrm{S}$. In order to avoid this ambiguity 
of physical origin, we define the \textit{coherent states}\cite{Perelomov72,Gilmore72} as the action of the quotient space on the 
reference state:
\begin{equation}
 |\psi^{l}(q)\ket=q|\psi_{0}^{l}\ket.
 \label{eq:2.10}
\end{equation}

Observe that the coherent states are in bijective correspondence with the elements of the quotient space and, as a consequence,
they inherit from $\mrm{G}/\mrm{S}$ their geometric and algebraic properties. Also note that, for the purposes of this paper, 
we deliberately choose the reference state as a minimum weight state. This choice is not unique, but it gives rise to numerous 
interesting physical and mathematical properties, which will be essential to our future developments.

\subsubsection{Parametrization}
\label{sss:2.1.3}

Considering the decomposition~\eqref{eq:2.3} and identities~(\ref{eq:2.4}\-\ref{eq:2.6}), we conclude that the stability subgroup 
results from the exponentiation of linear combinations of the generators $Q_{k}$, $(E\dg_{m'}+E_{m'})$ and $\rmi(E\dg_{m'}-E_{m'})$, 
for $k=1,2,\ldots,r_{\mrm{g}}$ and values of $m'$ defined by equation~\eqref{eq:2.5}. That is, according to the exponential 
map~\eqref{eq:2.1}, the elements of $\mrm{S}$ are given by
\begin{equation}
 s=\exp\left[\rmi\suml_{k}\sigma_{k}Q_{k}+\suml_{m'}\left(\gamma_{m'}E\dg_{m'}-\gamma\cg_{m'}E_{m'}\right)\right],
 \label{eq:2.11}
\end{equation}
\noindent for $\sigma_{k}\in\mds{R}$ e $\gamma_{m'}\in\mds{C}$. Then, once again employing expressions~(\ref{eq:2.4}\-\ref{eq:2.6}), 
we can easily calculate the action of the subgroup $\mrm{S}$ on the reference state:
\begin{equation}
 s|\psi_{0}^{l}\ket=\exp\left[\rmi\suml_{k}\sigma_{k}\lambda_{k}(l)\right]|\psi_{0}^{l}\ket.
 \label{eq:2.12}
\end{equation}

On the other hand, the quotient space is generated by the basis elements of $\mrm{g}$ which are not present 
in equation~\eqref{eq:2.11}. Thus, an unitary operation belonging to $\mrm{G}/\mrm{S}$ has the following 
exponential parametrization:
\begin{equation}
 q=\exp\left[\suml_{m\neq m'}\left(\tau_{m}E\dg_{m}-\tau\cg_{m}E_{m}\right)\right],
 \label{eq:2.13}
\end{equation}
\noindent for $\tau_{m}\in\mds{C}$. Note that the summation in the exponent of the previous equation only covers 
the values of $m$ absent in identity~\eqref{eq:2.5}. As established earlier, the action of an element~$q$ on the 
reference state results in a coherent state associated with the group~$\mrm{G}$ in the space~$\mds{H}_{l}$:
\begin{equation}
 |\psi^{l}(\tau)\ket=\exp\left[\suml_{m\neq m'}\left(\tau_{m}E\dg_{m}-\tau\cg_{m}E_{m}\right)\right]|\psi_{0}^{l}\ket,
 \label{eq:2.14}
\end{equation}
\noindent where $\tau$ is the $d$-dimensional complex vector\footnote{Throughout this paper, vectors are symbolized in 
the same way as scalar variables, without the use of boldface letters or a superscript arrow. By convention, vector variables 
are represented, when necessary, by a column matrix. However, left multiplication by a vector indicates the product by a row 
matrix. In this case, the transposition symbol is omitted without any loss to the comprehension of the pertinent equations.} 
that naturally parametrizes the coherent states, while $d\leq(d_{\mrm{g}}-r_{\mrm{g}})/2$ is the number possible values of 
the index~$m$ excluding the elements present in the subset indicated by $m'$. Henceforth, aiming at a more convenient and 
compact notation, but preserving the generality of the discussion, we deliberately choose a specific order for the values 
of index~$m$, so that $\tau=\left(\tau_{1}\;\tau_{2}\,\ldots\,\tau_{d}\right)\T$. This procedure is equivalent to reordering 
the raising operators, by placing those that do not satisfy the relation~\eqref{eq:2.5} in the first $d$ positions.

In general, by using the pertinent commutation relations, the exponential found in equation~\eqref{eq:2.14} can be rewritten as 
a product of new exponentials of the same operators. In this way, we are able to factor the dependence of the original exponential 
on the lowering operators, whose action on the reference state can be trivially calculated, as implied by identity~\eqref{eq:2.4}. 
So, after performing the indicated manipulations, we obtain a reparametrization for the set of coherent states:\cite{Zhang90a}
\begin{equation}
 |\psi^{l}(z)\ket=\mcl{N}(z\cg,z)\exp\!\left(\suml_{m=1}^{d} z_{m}E\dg_{m}\right)|\psi_{0}^{l}\ket,
 \label{eq:2.15}
\end{equation}
\noindent with $z=\left(z_{1}\;z_{2}\,\ldots\,z_{d}\right)\T$ and $z_{m}\in\mds{C}$. The quantity~$\mcl{N}(z\cg,z)$ represents a 
scalar normalization factor, which is responsible for all the dependence of coherent state on the complex conjugate variables~$z\cg$. 
In this paper, we do not address the algebraic relations between the parameters $\tau$ and $z$, since only the properties of the latter
parametrization are interesting for our subsequent applications. For the same reason, from now on, we adopt equation~\eqref{eq:2.15} 
as the working definition of coherent states.

Notice that the parameters~$z$ are presented as complex variables without any restriction on their possible values. This property 
implicitly corresponds to the third fundamental hypothesis of the present section, with which we demand that the variables~$z_{m}$ 
have the entire complex plane as domain. The parametrization introduced by equation~\eqref{eq:2.15} constitutes a very particular 
way of describing the coherent states, since this expression is analytic in $z$, except for the normalization factor. However, 
for certain choices of dynamical group, it is not possible to satisfy the requirement of an unrestricted complex domain along 
with the option for an analytic parametrization.\cite{Perelomov86} Therefore, the results elaborated in the later stages of 
this paper possess some limitations in their applicability, as a consequence of the conditions imposed on the group~$\mrm{G}$.

\subsubsection{K\"ahler Potential and Metric}
\label{sss:2.1.4}

According to the previous subsections, a set of coherent states is determined unequivocally by the choices of dynamical group~$\mrm{G}$ 
and Hilbert space~$\mds{H}_{l}$. Once this underlying geometric structure is well established, in order to simplify the notation, we 
can start to label the coherent states solely by their complex parameters:
\begin{equation}
 |z\ket=|z_{1},z_{2},\ldots,z_{d}\ket.
 \label{eq:2.16}
\end{equation}

As discussed earlier, the coherent states described by the equation~\eqref{eq:2.15} are analytic functions of their complex 
parameters~$z$, except for the normalization factor. For this reason, it becomes relevant to define the \textit{non-normalized 
coherent states}:
\begin{equation}
 |z\}=\mcl{N}(z\cg,z)^{-1}|z\ket=\exp\!\left(\suml_{m=1}^{d} z_{m}E\dg_{m}\right)|\psi_{0}^{l}\ket,
 \label{eq:2.17}
\end{equation}
\noindent which clearly manifest the analytic properties of the chosen parametrization. From the above definition, it follows 
immediately the construction of an analytic continuation for the \textit{K\"ahler potential}:\footnote{The K\"ahler potential is
usually defined for ${z'}=z$. For this reason, we emphasize that the formulation described by equation~\eqref{eq:2.18} is 
an analytic continuation.}
\begin{equation}
 f({z'}\cg,z)=\ln\{z'|z\},
 \label{eq:2.18}
\end{equation}
\noindent in which is encoded all geometric information of the non-linear subspace of $\mds{H}_{l}$ composed only of coherent 
states, whose topology is inherited from the quotient space~$\mrm{G}/\mrm{S}$. Notice that the quantity~$f({z'}\cg,z)$ is a 
well-defined analytic function of the variables ${z'}\cg$ and $z$. Also note that the normalization factor can be rewritten 
in terms of the K\"ahler potential:
\begin{equation}
 \mcl{N}(z\cg,z)=\exp\left[-\frac{1}{2}f(z\cg,z)\right].
 \label{eq:2.19}
\end{equation}

At this point, due to its recurrent use in later developments, it is convenient to define the analytic continuation 
of the \textit{quotient-space metric}:
\begin{equation}
 g({z'}\cg,z)=\frac{\del^{2}f({z'}\cg,z)}{\del z\del{z'}\cg},
 \label{eq:2.20}
\end{equation}
\noindent which is simply given by the matrix of second-order cross derivatives of the K\"ahler potential.\footnote{The 
second derivative of a scalar function~$h(u, v)$ with respect to two vectors $u$ and $v$, with respective dimensions $d_{1}$ 
and $d_{2}$, denotes the matrix with entries determined by the relation $\left[\frac{\del^{2}h(u,v)}{\del u\del v}\right]_{jk}
=\frac{\del^{2}h(u,v)}{\del u_{j}\del v_{k}}$, for $j=1,2,\ldots,d_{1}$ and $k=1,2,\ldots,d_{2}$.} Once the metric 
of the parameter space is defined, we employ the overcompleteness property of the set of coherent states to formulate 
a \textit{resolution of the identity} in the space~$\mds{H}_{l}$:
\begin{equation}
 \intl_{z\in\mds{C}^{d}}\rmd\mu(z\cg,z)|z\ket\bra z|=\mds{1}.
 \label{eq:2.21}
\end{equation}

In the above equation, observe that the unrestricted character of the variables~$z$ is evidenced in the domain of integration.
The \textit{invariant measure}~$\rmd\mu(z\cg,z)$ is essentially constituted by the determinant of the metric and, consequently, 
can exhibit explicit dependence on the quotient-space coordinates:
\begin{equation}
 \rmd\mu(z\cg,z)=\kappa(l)\det\!\left[g(z\cg,z)\right]\frac{\rmd^{2}z}{\pi^{d}}.
 \label{eq:2.22}
\end{equation}

The normalization constant~$\kappa(l)$ depends only on the choice of the Hilbert space~$\mds{H}_{l}$ and, therefore, is also 
exclusively specified by the set of indices~$l$. We recall that, by assumption, the space~$\mds{H}_{l}$ supports an unitary 
irreducible representation~$\mrm{U}_{l}$ of the group~$\mrm{G}$. Thus, $\kappa(l)$ is another quantity determined directly 
by the representation of the dynamical group on the accessible Hilbert space. In equation~\eqref{eq:2.22}, we also introduce 
the following notation:\footnote{The element of integration for a $d$-dimensional vector variable~$u$ is denoted by 
$\rmd u=\prodl_{j=1}^{d}\rmd u_{j}$.}
\begin{equation}
 \begin{aligned}
 \rmd^{2}z
 &=\prodl_{j=1}^{d}\rmd^{2}z_{j}
 =\prodl_{j=1}^{d}\rmd x_{j}\rmd y_{j}\\
 &=\prodl_{j=1}^{d}\frac{\rmd z\cg_{j}\rmd z_{j}}{2\rmi}
 =\frac{\rmd z\cg\rmd z}{(2\rmi)^{d}},
 \end{aligned}
 \label{eq:2.23}
\end{equation}
\noindent in which $x_{j}=\re\left(z_{j}\right)$ and $y_{j}=\im\left(z_{j}\right)$.

\subsubsection{Symplectic Structure}
\label{sss:2.1.5}

The choice of the reference state as a minimum weight state confers a natural symplectic structure upon 
the set of coherent states determined by the equation~\eqref{eq:2.15}.\cite{Onofri75,Kramer81} In other words, 
for a reference state satisfying the condition~\eqref{eq:2.4}, the quotient space~$\mrm{G}/\mrm{S}$ constitutes 
a \textit{symplectic space}. Consequently, it is possible to equip the parameter space with a very particular 
bilinear operation, widely known as \textit{Poisson bracket}:\footnote{The derivative of a scalar function~$h(v)$ 
with respect to a $d$-dimensional vector quantity~$v$ represents a new vector with entries defined by 
$\left[\frac{\del h(v)}{\del v}\right]_{j}=\frac{\del h(v)}{\del v_{j}}$, for $j=1,2,\ldots,d$.}
\begin{equation}
 \{\mcl{A}_{1}(z\cg,z),\mcl{A}_{2}(z\cg,z)\}=
 -\rmi\left[\frac{\del\mcl{A}_{1}}{\del z}\xi\T(z\cg,z)\frac{\del\mcl{A}_{2}}{\del z\cg}
 -\frac{\del\mcl{A}_{1}}{\del z\cg}\xi(z\cg,z)\frac{\del\mcl{A}_{2}}{\del z}\right],
 \label{eq:2.24}
\end{equation}
\noindent in which $\mcl{A}_{1}(z\cg,z)$ and $\mcl{A}_{2}(z\cg,z)$ denote two arbitrary differentiable functions 
on the quotient space and $\xi({z'}\cg,z)=g^{-1}({z'}\cg,z)$ symbolizes the inverse of the metric. As expected, 
the expression~\eqref{eq:2.24} satisfies the defining properties of a Poisson bracket:
\begin{subequations}
 \label{eq:2.25}
 \begin{align}
 &\{\mcl{A}_{1},\mcl{A}_{2}\}=-\{\mcl{A}_{2},\mcl{A}_{1}\},
 \label{eq:2.25a}\\[1.2ex]
 &\{\alpha\mcl{A}_{1}+\beta\mcl{A}_{2},\mcl{A}_{3}\}=
 \alpha\{\mcl{A}_{1},\mcl{A}_{3}\}
 +\beta\{\mcl{A}_{2},\mcl{A}_{3}\},
 \label{eq:2.25b}\\[1.2ex]
 &\{\mcl{A}_{1},\mcl{A}_{2}\mcl{A}_{3}\}=
 \mcl{A}_{2}\{\mcl{A}_{1},\mcl{A}_{3}\}
 +\{\mcl{A}_{1},\mcl{A}_{2}\}\mcl{A}_{3},
 \label{eq:2.25c}\\[1.2ex]
 &\left\{\{\mcl{A}_{1},\mcl{A}_{2}\},\mcl{A}_{3}\right\}
 +\left\{\{\mcl{A}_{2},\mcl{A}_{3}\},\mcl{A}_{1}\right\}
 +\left\{\{\mcl{A}_{3},\mcl{A}_{1}\},\mcl{A}_{2}\right\}=0,
 \label{eq:2.25d}
 \end{align}
\end{subequations}
\noindent where $\alpha$ and $\beta$ are constants.

As a consequence of the definition~\eqref{eq:2.24}, we observe that the quotient space intrinsically incorporates a Hamiltonian 
formalism, that is, the complex variables~$z$ can be regarded as coordinates of a \textit{Hamiltonian phase space}. For 
this reason, the coherent states provide an ideal framework for studying the correspondence between classical and quantum 
mechanics.

Finally, in order to conclude our general discussion on coherent states, we summarize the three main assumptions 
made in the present section:
\begin{enumerate}
 \item The effective Hilbert space of the physical system of interest supports an irreducible unitary representation 
 of the chosen dynamical group.
 \item The basis of the algebra associated with the dynamical group admits a decomposition of the form indicated 
 by equation~\eqref{eq:2.3}.
 \item The variables~$z$, corresponding to an analytic complex parametrization of the coherent states, do not exhibit 
 any restriction on their domains in the complex plane.
\end{enumerate}

In general, the construction of coherent states for groups not satisfying the above requirements is also feasible. 
However, many of the results elaborated in this paper are restricted to dynamical groups specified by the three 
aforementioned conditions. In other words, these assumptions determine the sets of coherent states used in our 
future developments.

\subsection{Examples}
\label{ssc:2.2}

\subsubsection{Canonical Coherent States}
\label{sss:2.2.1}

Consider a system of identical bosonic particles\cite{Negele98} with $d$ modes\footnote{That is, the single-particle Hilbert 
space has dimension $d$.}. In this case, it is convenient to define the \textit{creation operator}~$a\dg_{j}$, whose action 
on an arbitrary state of the Hilbert space results in the addition of a particle to the $j$-th mode. Oppositely, the 
\textit{annihilation operator}~$a_{j}$ is responsible for the subtraction of a boson from the $j$-th mode. The operators 
$a\dg_{j}$ and $a_{j}$, for $j=1,2,\ldots,d$, are subject to the \textit{canonical bosonic commutation relations}:
\begin{subequations}
 \label{eq:2.26}
 \begin{align}
 &[a_{j},a\dg_{k}]=\delta_{jk},
 \label{eq:2.26a}\\[1.2ex]
 &[a_{j},a_{k}]=[a\dg_{j},a\dg_{k}]=0.
 \label{eq:2.26b}
 \end{align}
\end{subequations}

By inspecting the previous identities, we note that the set $\{\mds{1},a\dg_{j}+a_{j},\rmi(a\dg_{j}-a_{j})\}$, 
for a fixed value of $j$, constitutes a basis for a Lie algebra, since its elements generate a vector space closed 
under the bilinear operation of commutation. The algebra formed by the bosonic creation and annihilation operators 
is known as \textit{Heisenberg-Weyl algebra}, which we denote by $\mrm{h}$.

The Hilbert space accessible to an arbitrary system with $d$ bosonic modes, also called the bosonic Fock space, is symbolized 
by $\mds{B}^{d}$. An usual basis for $\mds{B}^{d}$ is formed by the set of normalized states $\{|m_{1},m_{2},\ldots,m_{d}\ket\}$, 
known as \textit{occupation-number states}, in which the non-negative integer $m_{j}$ stands for the number of bosons in the 
$j$-th mode. As a result of the equations~\eqref{eq:2.26}, the following identities hold:
\begin{subequations}
 \label{eq:2.27}
 \begin{align}
 &a_{j}|m_{1},\ldots,m_{j},\ldots,m_{d}\ket=\sqrt{m_{j}}|m_{1},\ldots,m_{j}-1,\ldots,m_{d}\ket,
 \label{eq:2.27a}\\[1.2ex]
 &a\dg_{j}|m_{1},\ldots,m_{j},\ldots,m_{d}\ket=\sqrt{m_{j}+1}|m_{1},\ldots,m_{j}+1,\ldots,m_{d}\ket.
 \label{eq:2.27b}
 \end{align}
\end{subequations}

Observe that, as a consequence of their construction, the occupation-number states are eigenstates of the operators~$a\dg_{j}a_{j}$, 
that is, $a\dg_{j}a_{j}|m_{1},m_{2},\ldots,m_{d}\ket=m_{j}|m_{1},m_{2},\ldots,m_{d}\ket$.

The Fock space~$\mds{B}^{1}$, resulting from the restriction of equations~\eqref{eq:2.27} to a fixed value of $j$, supports 
the only unitary irreducible representation of the Heisenberg-Weyl group~$\mrm{H}$, associated with the algebra~$\mrm{h}$ by 
the exponential map~\eqref{eq:2.1}. Consequently, $\mds{B}^{d}$ constitutes the support space for the irreducible representation 
of the direct product $\mrm{H}^{d}=\mrm{H}^{(1)}\times\mrm{H}^{(2)}\times\ldots\times\mrm{H}^{(d)}$, in which $\mrm{H}^{(j)}$ 
is the dynamical group of the $j$-th mode taken alone. Comparing the definitions~\eqref{eq:2.4} and \eqref{eq:2.5} to 
the identities~\eqref{eq:2.27}, we conclude that the \textit{vacuum state}~$|0,0,\ldots,0\ket$, characterized by 
zero occupation number in all $d$ modes, corresponds to the minimum weight state of this single unitary representation 
of $\mrm{H}^{d}$.

Once we know the lowering and raising operators associated with the dynamical group~$\mrm{H}^{d}$ and the reference 
state of its irreducible representation, we can perform a direct application of the definition~\eqref{eq:2.17}. In 
this way, we obtain the \textit{canonical coherent states} in their non-normalized form:\footnote{The juxtaposition 
of two vectors $u$ and $v$, both with dimension $d$, denotes the matrix product $uv=u_{1}v_{1}+u_{2}v_{2}+\ldots+u_{d}v_{d}$.}
\begin{equation}
 \begin{aligned}
 |z\}&=\exp\!\left(z a\dg\right)|0,0,\ldots,0\ket\\
 &=\suml_{m_{1},m_{2},\ldots,m_{d}=0}^{\infty}
 \left[\prodl_{j=1}^{d}\frac{z_{j}^{m_{j}}}{\sqrt{m_{j}!}}\right]
 |m_{1},m_{2},\ldots,m_{d}\ket,
 \end{aligned}
 \label{eq:2.28}
\end{equation}
\noindent where, in order to simplify the notation, we introduce the vector~$a\dg=(a\dg_{1}\;a\dg_{2}\,\ldots\,a\dg_{d})\T$. 
From the above expression, we can immediately calculate the overlap between coherent states:
\begin{equation}
 \{z'|z\}=\exp\left({z'}\cg z\right),
 \label{eq:2.29}
\end{equation}
\noindent from which follows, according to equation~\eqref{eq:2.18}, the K\"ahler potential:
\begin{equation}
 f({z'}\cg,z)={z'}\cg z.
 \label{eq:2.30}
\end{equation}

As expected, $f({z'}\cg,z)$ is an analytic function of its two arguments. Then, with the aid of identity~\eqref{eq:2.19}, 
we can readily calculate the factor $\mcl{N}(z\cg,z)$, which allows us to write the normalized expression for the canonical 
coherent states: 
\begin{equation}
 |z\ket=\rme^{-\frac{z\cg z}{2}}
 \suml_{m_{1},m_{2},\ldots,m_{d}=0}^{\infty}
 \left[\prodl_{j=1}^{d}\frac{z_{j}^{m_{j}}}{\sqrt{m_{j}!}}\right]
 |m_{1},m_{2},\ldots,m_{d}\ket.
 \label{eq:2.31}
\end{equation}

Substituting the result~\eqref{eq:2.30} into the definition~\eqref{eq:2.20}, we obtain the metric of the parameter space:
\begin{equation}
 g({z'}\cg,z)=\mds{1}.
 \label{eq:2.32}
\end{equation}

Notice that, according to the above identity, the canonical coherent states are associated with a quotient space endowed 
with an Euclidean metric. Another relevant quantity, which composes many results in later sections, is the determinant 
of the metric:\footnote{Due to the extremely simple form of the metric~\eqref{eq:2.32}, the determinant~\eqref{eq:2.33} 
is trivially calculated. We explicitly display this result here in order to provide a direct comparison with the examples
shown in the following subsections.}
\begin{equation}
 \det[g({z'}\cg,z)]=1.
 \label{eq:2.33}
\end{equation}

Next, by using the identity~\eqref{eq:2.22}, we can readily find the the expression for the volume element in the quotient space:
\begin{equation}
 \rmd\mu(z\cg,z)=\kappa\frac{d^{2}z}{\pi^{d}},
 \label{eq:2.34}
\end{equation}
\noindent whose normalization factor is determined as
\begin{equation}
 \kappa=1.
 \label{eq:2.35}
\end{equation}

Note that the factor~$\kappa$ does not depend on a possible set of indices, which would specify the irreducible representation
of the dynamical group employed in the construction of the canonical coherent states. Evidently, this result was expected, since 
$\mrm{H}^{d}$ has only a single unitary irreducible representation.

Upon substitution of the results~\eqref{eq:2.34} and \eqref{eq:2.35} into the expression~\eqref{eq:2.21}, we obtain the resolution 
of the identity in the space~$\mds{B}^{d}$:
\begin{equation}
 \int\frac{d^{2}z}{\pi^{d}}|z\ket\bra z|=\mds{1}.
 \label{eq:2.36}
\end{equation}

The canonical coherent states establish a theoretical framework applicable to a wide variety of physical systems, 
due to the high recurrence of the dynamical group~$\mrm{H}^{d}$. In particular, we mention the Hamiltonians of 
the form~$H\!=\!H(\hat{q},\hat{p})$, in which $\hat{q}\!=\!\left(\hat{q}_{1}\;\hat{q}_{2}\,\ldots\,\hat{q}_{d}\right)\T$ 
and $\hat{p}\!=\!\left(\hat{p}_{1}\;\hat{p}_{2}\,\ldots\,\hat{p}_{d}\right)\T$ respectively represent the position and 
momentum vector operators. This possibility of application arises from the following homomorphism:
\begin{subequations}
 \label{eq:2.37}
 \begin{align}
 &\hat{q}_{j}=\frac{\zeta_{j}}{\sqrt{2}}(a\dg_{j}+a_{j}),
 \label{eq:2.37a}\\[1.2ex]
 &\hat{p}_{j}=\frac{\rmi\hbar}{\sqrt{2}\zeta_{j}}(a\dg_{j}-a_{j}),
 \label{eq:2.37b}
 \end{align}
\end{subequations}
\noindent where $\zeta_{j}$ symbolizes a real parameter with dimension of length and arbitrary value.

\subsubsection{Spin Coherent States}
\label{sss:2.2.2}

Consider a system consisting of $d$ localized particles with spin. In general, we can describe the Hamiltonian of this 
system as a function of the set of operators~$\{\hat{J}_{x,k},\hat{J}_{y,k},\hat{J}_{z,k}\}$, for $k=1,2,\ldots,d$, where 
$\hat{J}_{x,k}$, $\hat{J}_{y,k}$ and $\hat{J}_{z,k}$ represent the three Cartesian components of angular momentum for the 
$k$-th particle.

The set~$\{\hat{J}_{x,k},\hat{J}_{y,k},\hat{J}_{z,k}\}$, for a fixed value of $k$, forms a basis for the algebra~$\mrm{su}(2)$, 
associated with the group~$\mrm{SU}(2)$ by the exponential map~\eqref{eq:2.1}. Therefore, we can identify the dynamical group 
associated with the complete system of $d$ particles as the direct product $\mrm{SU}^{d}(2)=\mrm{SU}^{(1)}(2)\times\mrm{SU}^{(2)}(2)
\times\ldots\times\mrm{SU}^{(d)}(2)$, in which $\mrm{SU}^{(k)}(2)$ is the dynamical group for the $k$-th particle taken in isolation. 
The generators of $\mrm{SU}^{d}(2)$ are subject to the following fundamental commutation relations:\cite{Sakurai94}
\begin{equation}
 [\hat{J}_{\alpha,k},\hat{J}_{\beta,k'}]=\rmi\hbar\delta_{kk'}\suml_{\gamma}\epsilon_{\alpha\beta\gamma}\hat{J}_{\gamma,k},
 \label{eq:2.38}
\end{equation}
\noindent for $k,k'=1,2,\ldots,d$ and $\alpha,\beta,\gamma=x,y,z$. As a result of the above identity, the 
Casimir operators\cite{Gilmore74} $\hat{J}^{2}_{k}=\hat{J}_{x,k}^{2}+\hat{J}_{y,k}^{2}+\hat{J}_{z,k}^{2}$ 
and the generators~$\hat{J}_{z,k}$, for $k=1,2,\ldots,d$, constitute a complete set of commuting observables 
for our system of interest. Consequently, the simultaneous eigenvectors of $\hat{J}^{2}_{k}$ and $\hat{J}_{z,k}$ 
form an orthogonal basis for the Hilbert space~$\mds{H}$, corresponding to $d$ localized particles:
\begin{subequations}
 \label{eq:2.39}
 \begin{align}
 &\hat{J}^{2}_{k}|J_{1},J_{2},\ldots,J_{d};M_{1},M_{2},\ldots,M_{d}\ket=
 J_{k}(J_{k}+1)\hbar^2|J_{1},J_{2},\ldots,J_{d};M_{1},M_{2},\ldots,M_{d}\ket,
 \label{eq:2.39a}\\[1.2ex]
 &\hat{J}_{z,k}|J_{1},J_{2},\ldots,J_{d};M_{1},M_{2},\ldots,M_{d}\ket=
 M_{k}\hbar|J_{1},J_{2},\ldots,J_{d};M_{1},M_{2},\ldots,M_{d}\ket,
 \label{eq:2.39b}
 \end{align}
\end{subequations}
\noindent where $J_{k}=0,1/2,1,3/2,\ldots$ and $M_{k}=-J_{k},-J_{k}+1,\ldots,J_{k}-1,J_{k}$. The set of indices 
$\{J_{1},J_{2},\ldots,J_{d}\}$ determines an irreducible unitary representation of $\mrm{SU}^{d}(2)$. Therefore, 
according to the first central hypothesis of subsection~\ref{sss:2.1.1}, we must specify the values of $J_{k}$ 
in order to restrict the dynamics of the system to the support space of a single irreducible representation of 
the dynamical group. This condition corresponds to establishing the magnitude of the intrinsic angular momentum 
of each particle and, consequently, has a direct physical interpretation.

Henceforth, we shall constrain our system of interest to the Hilbert space~$\mds{H}_{J_{1},J_{2},\ldots,J_{d}}$, 
in which the values of the indices~$\{J_{1},J_{2},\ldots,J_{d}\}$ are fixed. Then, in order to simplify the 
notation, we shall omit displaying these indices in the basis states introduced by equations~\eqref{eq:2.39}. 
Note that $\mds{H}_{J_{1},J_{2},\ldots,J_{d}}$ constitutes a subspace of $\mds{H}$, such that 
$\mds{H}=\bigoplus_{J_{1},J_{2},\ldots,J_{d}}\mds{H}_{J_{1},J_{2},\ldots,J_{d}}$.

The raising and lowering operators corresponding to the $k$-th particle, respectively denoted by $\hat{J}_{+,k}$ 
and $\hat{J}_{-,k}$, are defined in terms of the $\mrm{SU}(2)$ generators by the following identity:
\begin{equation}
 \hat{J}_{\pm,k}=\hat{J}_{x,k}\pm\rmi\hat{J}_{y,k}.
 \label{eq:2.40}
\end{equation}

Next, with the aid of relations~\eqref{eq:2.38}, we determine the action of the operators~$\hat{J}_{\pm,k}$ 
on the normalized basis states:
\begin{equation}
 \hat{J}_{\pm,k}|M_{1},\ldots,M_{j},\ldots,M_{d}\ket=
 \hbar\sqrt{(J_{k}\mp M_{k})(J_{k}\pm M_{k}+1)}|M_{1},\ldots,M_{j}\pm1,\ldots,M_{d}\ket.
 \label{eq:2.41}
\end{equation}

As a result of the previous equation, we conclude that $|-J_{1},-J_{2},\ldots,-J_{d}\ket$ represents the minimum weight state of 
the space~$\mds{H}_{J_{1},J_{2},\ldots,J_{d}}$. Once the reference state is selected, we can employ the definition~\eqref{eq:2.17} 
to obtain the non-normalized $\mrm{SU}^{d}(2)$ coherent states:
\begin{equation}
 \begin{aligned}
 |z\}&=\exp\!\left(\suml_{k=1}^{d}z_{k}\hat{J}_{+,k}\right)|-J_{1},-J_{2},\ldots,-J_{d}\ket\\
 &=\bigotimes\limits_{k=1}^{d}\suml_{M_{k}=-J_{k}}^{J_{k}}
 \binom{2J_{k}}{J_{k}+M_{k}}^{\frac{1}{2}}
 z_{k}^{J_{k}+M_{k}}|M_{k}\ket.
 \end{aligned}
 \label{eq:2.42}
\end{equation}

Then, we calculate the overlap between two coherent states:
\begin{equation}
 \{z'|z\}=\prodl_{k=1}^{d}(1+{z'_{k}}\cg z_{k})^{2J_{k}}.
 \label{eq:2.43}
\end{equation}

From the above identity, it follows immediately the analytic continuation of the K\"ahler potential:
\begin{equation}
 f({z'}\cg,z)=2\suml_{k=1}^{d}J_{k}
 \ln\left(1+{z'_{k}}\cg z_{k}\right).
 \label{eq:2.44}
\end{equation}

Now, by using the formula~\eqref{eq:2.19}, we perform the normalization of the \textit{spin coherent states}\cite{Arecchi72}:
\begin{equation}
 |z\ket=\bigotimes\limits_{k=1}^{d}
 \frac{1}{(1+z\cg_{k}z_{k})^{J_{k}}}
 \suml_{M_{k}=-J_{k}}^{J_{k}}
 \binom{2J_{k}}{J_{k}+M_{k}}^{\frac{1}{2}}
 z_{k}^{J_{k}+M_{k}}|M_{k}\ket.
 \label{eq:2.45}
\end{equation}

Evaluating the cross derivatives of the result~\eqref{eq:2.44}, we obtain the analytic continuation of the metric:
\begin{equation}
 g_{kk'}({z'}\cg,z)=\frac{2J_{k}}{(1+{z'_{k}}\cg z_{k})^{2}}\delta_{kk'}.
 \label{eq:2.46}
\end{equation}

Notice that, as in the case of the canonical coherent states, the metric has diagonal form, indicating the absence of geometric 
coupling between the degrees of freedom of the system. However, unlike in the previous subsection, the matrix~\eqref{eq:2.46} 
depends explicitly on the parameters ${z'}\cg$ and $z$, evidencing the natural curvature of the quotient space. In spite of 
this complication of topological origin, the evaluation of the determinant of the metric remains trivial:
\begin{equation}
 \det[g({z'}\cg,z)]=\prod_{k=1}^{d}
 \frac{2J_{k}}{(1+{z'_{k}}\cg z_{k})^{2}}.
 \label{eq:2.47}
\end{equation}

As a direct consequence of the above equation, the integration measure acquires a factor with explicit dependence on the complex 
coordinates of the quotient space:
\begin{equation}
 \rmd\mu(z\cg,z)=\kappa(J_{1},J_{2},\ldots,J_{d})
 \left[\prodl_{k=1}^{d}\frac{2J_{k}}{(1+z\cg_{k}z_{k})^{2}}\right]
 \frac{d^{2}z}{\pi^{d}}.
 \label{eq:2.48}
\end{equation}

On account of the multiple possible choices for the irreducible unitary representation of $\mrm{SU}^{d}(2)$, the normalization 
constant becomes a function of the indices~$\{J_{1},J_{2},\ldots,J_{d}\}$:
\begin{equation}
 \kappa(J_{1},J_{2},\ldots,J_{d})=\prodl_{k=1}^{d}\frac{2J_{k}+1}{2J_{k}}.
 \label{eq:2.49}
\end{equation}

However, for large spin magnitudes, the above normalization constant approaches the result~\eqref{eq:2.35}, that is,
\begin{equation}
 \kappa(J_{1},J_{2},\ldots,J_{d})\stackrel{J_{1},J_{2},\ldots,J_{d}\gg1}{\approx}1.
 \label{eq:2.50}
\end{equation}

Finally, by substituting the expressions \eqref{eq:2.48} and \eqref{eq:2.49} into the equation~\eqref{eq:2.21}, we obtain 
the resolution of the identity in the space~$\mds{H}_{J_{1},J_{2},\ldots,J_{d}}$:
\begin{equation}
 \int\frac{d^{2}z}{\pi^{d}}
 \left[\prodl_{k=1}^{d}\frac{2J_{k}+1}{(1+z\cg_{k}z_{k})^{2}}\right]
 |z\ket\bra z|=\mds{1}.
 \label{eq:2.51}
\end{equation}

\subsubsection{\texorpdfstring{$\mrm{SU}(n)$}{SU(n)} Bosonic Coherent States}
\label{sss:2.2.3}

Consider a system composed of $N$ identical bosons in $n$ modes. In this case, unlike the example discussed 
in the subsection~\ref{sss:2.2.1}, we assume that the total number of particles remains invariant under the 
dynamics imposed by the Hamiltonian~$H$. For this reason, it is convenient to define the \textit{total 
particle number operator}:
\begin{equation}
 \hat{N}=\suml_{j=1}^{n}a\dg_{j}a_{j}.
 \label{eq:2.52}
\end{equation}

With the aid of the above definition, we can mathematically formulate the restriction on the Hamiltonian:
\begin{equation}
 [\hat{N},H]=0.
 \label{eq:2.53}
\end{equation}

Note that the elements of the orthonormal basis~$\{|m_{1},m_{2},\ldots,m_{n}\ket\}$, introduced by equations~\eqref{eq:2.27}, 
form a complete set of eigenvectors for the operator~$\hat{N}$, whose eigenvalues are described by $N=\sum_{j=1}^{n}m_{j}$. 
Given that the physical system has a fixed and well-defined total number of particles, we conclude that the dynamically 
available Hilbert space is equivalent to the eigenspace of $\hat{N}$ with eigenvalue~$N$, which we denote by $\mds{B}^{n}_{N}$.

The condition~\eqref{eq:2.53} amounts to requiring that the Hamiltonian~$H$ is a function of the \textit{bosonic bilinear 
products}~$a\dg_{j}a_{k}$, for $j,k=1,2,\ldots,n$. Therefore, by using the canonical identities~\eqref{eq:2.26}, we can 
find the fundamental commutation relations between the generating elements of the dynamics:
\begin{equation}
 [a\dg_{\alpha}a_{\beta},a\dg_{\gamma}a_{\eta}]=
 \delta_{\beta\gamma}a\dg_{\alpha}a_{\eta}-\delta_{\alpha\eta}a\dg_{\beta}a_{\gamma},
 \label{eq:2.54}
\end{equation}
\noindent in which $\alpha,\beta,\gamma,\eta=1,2,\ldots,n$. Notice that the operators~$a\dg_{j}a_{k}$, along with their 
linear combinations, constitute a vector space closed under commutation. However, taking into account the conservation 
of the total number of bosons, it becomes more convenient to select a set of generators composed only of traceless 
operators:
\begin{subequations}
 \label{eq:2.55}
 \begin{align}
 &Q_{\alpha}=a\dg_{\alpha+1}a_{\alpha+1}-a\dg_{\alpha}a_{\alpha},
 \label{eq:2.55a}\\[1.2ex]
 &X_{\beta\gamma}=a\dg_{\beta}a_{\gamma}+a\dg_{\gamma}a_{\beta},
 \label{eq:2.55b}\\[1.2ex]
 &P_{\beta\gamma}=\rmi(a\dg_{\beta}a_{\gamma}-a\dg_{\gamma}a_{\beta}),
 \label{eq:2.55c}
 \end{align}
\end{subequations}
\noindent where $\alpha=1,2,\ldots,(n-1)$, $\beta,\gamma=1,2,\ldots,n$ and $\beta>\gamma$. The relations~\eqref{eq:2.55} represent 
a homomorphism between the bosonic bilinear products and a basis for the algebra~$\mrm{su}(n)$.\cite{Lipkin65} In fact, the Hilbert 
space~$\mds{B}^{n}_{N}$ supports a \textit{fully symmetrized irreducible representation} of the group~$\mrm{SU}(n)$.\cite{Hamermesh62}
By inspecting the generators~\eqref{eq:2.55}, we conclude that a possible choice of reference state is described by the basis 
state~$|0,0,\ldots,0,N\ket$, in which only the last mode has nonzero occupation number. Then, since all necessary quantities 
are already known, we employ the definition~\eqref{eq:2.17} to obtain the non-normalized coherent states in $\mds{B}^{n}_{N}$:
\begin{equation}
 \begin{aligned}
 |z\}&=\exp\!\left(\suml_{j=1}^{n-1}z_{j}a\dg_{j}a_{n}\right)
 |0,0,\ldots,0,N\ket\\
 &=\suml_{m_{1}+m_{2}+\ldots+m_{n}=N}
 \left(\frac{N!}{m_{1}!m_{2}!\ldots m_{n}!}\right)^{\frac{1}{2}}
 \left[\prodl_{j=1}^{n-1}z_{j}^{m_{j}}\right]
 |m_{1},m_{2},\ldots,m_{n}\ket.
 \end{aligned}
 \label{eq:2.56}
\end{equation}

Observe that the quotient space is parametrized by $d=(n-1)$ complex variables, that is, the number of complex coordinates 
is smaller than the number of accessible modes in the bosonic Fock space. As expected, this difference in relation to the 
results of subsection~\ref{sss:2.1.1} is a consequence of the restriction on the total number of particles. Using the above 
expression, we calculate the overlap between two possibly distinct coherent states:
\begin{equation}
 \{z'|z\}=(1+{z'}\cg z)^{N},
 \label{eq:2.57}
\end{equation}
\noindent from which directly follows the analytic continuation of the K\"ahler potential:
\begin{equation}
 f({z'}\cg,z)=N\ln\left(1+{z'}\cg z\right).
 \label{eq:2.58}
\end{equation}

Now, by substituting this particular formulation of $f({z'}\cg,z)$ into the identity~\eqref{eq:2.19}, we obtain the normalization 
factor required for the construction of the \textit{$\mrm{SU}(n)$ bosonic coherent states}\cite{Gilmore75}:
\begin{equation}
 |z\ket=\suml_{m_{1}+m_{2}+\ldots+m_{n}=N}
 \left(\frac{N!}{m_{1}!m_{2}!\ldots m_{n}!}\right)^{\frac{1}{2}}
 \frac{\prodl_{j=1}^{n-1}z_{j}^{m_{j}}}{(1+z\cg z)^{\frac{N}{2}}}
 |m_{1},m_{2},\ldots,m_{n}\ket.
 \label{eq:2.59}
\end{equation}

Next, by performing the cross derivatives of the K\"ahler potential, we obtain the analytic continuation of the metric:\footnote{The 
symbol $\otimes$ denotes the dyadic product of two vectors. That is, the result of the multiplication $u\otimes v$ is a matrix with 
elements described by $\left[u\otimes v\right]_{jk}=u_{j}v_{k}$, for $j=1,2,\ldots,d_{1}$ and $k=1,2,\ldots,d_{2}$.}
\begin{equation}
 g({z'}\cg,z)=N\frac{(1+{z'}\cg z)\mds{1}-{z'}\cg\otimes z}{(1+{z'}\cg z)^{2}}.
 \label{eq:2.60}
\end{equation}

Unlike the previously discussed examples, the metric associated with the $\mrm{SU}(n)$ coherent states is not diagonal. 
Therefore, the degrees of freedom of the parameter space are geometrically coupled. In addition, notice that the general 
evaluation of the determinant of $g({z'}\cg,z)$ does not remain trivial:
\begin{equation}
 \det[g({z'}\cg,z)]=\frac{N^{n-1}}{(1+{z'}\cg z)^{n}}.
 \label{eq:2.61}
\end{equation}

Due to the natural curvature of the parameter space, the complex coordinates explicitly appear in the expression 
for the integration measure:
\begin{equation}
 \rmd\mu(z\cg,z)=\kappa(N)\frac{N^{n-1}}{(1+z\cg z)^{n}}
 \frac{d^{2}z}{\pi^{n-1}},
 \label{eq:2.62}
\end{equation}
\noindent whose normalization factor has the following specific form:
\begin{equation}
 \kappa(N)=\frac{(N+n-1)!}{N!N^{n-1}}.
 \label{eq:2.63}
\end{equation}

Note that, similarly to the expression~\eqref{eq:2.50}, the normalization constant approaches the unity for values 
of $N$ substantially higher than the number of modes $n$:
\begin{equation}
 \kappa(N)\stackrel{N\gg n}{\approx}1.
 \label{eq:2.64}
\end{equation}

By employing the results \eqref{eq:2.62} and \eqref{eq:2.63}, we can formulate a resolution of the identity 
in the space~$\mds{B}^{n}_{N}$:
\begin{equation}
 \int\frac{d^{2}z}{\pi^{n-1}}\frac{(N+n-1)!}{N!(1+z\cg z)^{n}}
 |z\ket\bra z|=\mds{1}.
 \label{eq:2.65}
\end{equation}

In contrast to the results \eqref{eq:2.32} and \eqref{eq:2.46}, the analytical inversion of the metric~\eqref{eq:2.60} is not 
a simple task for arbitrary values of $n$, on account of the presence of nonzero off-diagonal elements. Therefore, as a reference 
for future developments, we present the explicit form for the inverse of $g({z'}\cg,z)$:
\begin{equation}
 \xi({z'}\cg,z)=\frac{1+{z'}\cg z}{N}\left(\mds{1}+{z'}\cg\otimes z\right).
 \label{eq:2.66}
\end{equation}



\section{Semiclassical Propagator}
\label{sec:3} 

\subsection{Presentation}
\label{ssc:3.1}

The \textit{quantum propagator} in the representation of coherent states is defined as the transition amplitude between 
the initial coherent state~$|z_{i}\ket$, at time~$t_{i}$, and the final coherent state~$\bra z_{f}|$, at time~$t_{f}$, 
that is,
\begin{equation}
 K(z\cg_{f},z_{i};t_{f},t_{i})=\bra z_{f}|\hat{T}\exp\left[-\frac{\rmi}{\hbar}\intl_{t_{i}}^{t_{f}}H(t)\rmd t\right]|z_{i}\ket,
 \label{eq:3.1}
\end{equation}
\noindent where $\hat{T}$ is the time-ordering operator and $H(t)$ is the Hamiltonian of the system, which can depend 
explicitly on time.

In general, we can reformulate the quantum propagator as a \textit{path integral},\cite{Kochetov95a,Kochetov95b} 
which comprises contributions from all paths in the coherent-state phase space\footnote{From now on, in accordance 
with the discussion in subsection~\ref{sss:2.1.5}, we shall refer to the quotient space as the coherent-state phase 
space, or simply as the phase space.} connecting the points $z_{i}$ and $z_{f}$. The contribution of each path is 
simply given by a phase factor, which basically consists of an exponential of the \textit{action functional}. As 
expected, the explicit calculation of a path integral for an arbitrary physical system is an extremely complicated 
task. Exact results are known only for the simplest cases, usually identified with Hamiltonians which are linear 
in the generators of the dynamical group. However, the path integrals are of great importance as a theoretical 
tool in the development of several practical applications of quantum mechanics, particularly in the formulation 
of approximations to expression~\eqref{eq:3.1}.

The \textit{semiclassical approximation} to the quantum propagator consists of expanding the action functional 
up to the second perturbative order around appropriate \textit{classical trajectories}, thus enabling the analytical 
evaluation of the corresponding path integral. The final result of this procedure is known as the \textit{semiclassical 
propagator}:\cite{Kochetov98}
\begin{equation}
 K_{sc}(z\cg_{f},z_{i};t_{f},t_{i})=
 \suml_{\text{traj.}}\mcl{C}(z\cg_{f},z_{i};t_{f},t_{i})\,
 \rme^{\frac{\rmi}{\hbar}\left[S(z\cg_{f},z_{i};t_{f},t_{i})
 +I(z\cg_{f},z_{i};t_{f},t_{i})\right]
 +\Lambda(z\cg_{f},z_{i})}.
 \label{eq:3.2}
\end{equation}

By definition, the classical trajectories represent the solutions of the following \textit{equations of motion}: 
\begin{subequations}
 \label{eq:3.3}
 \begin{align}
 &\dot{z}=-\frac{\rmi}{\hbar}\xi\T(\b{z},z)\frac{\del\mcl{H}(\b{z},z)}{\del\b{z}},
 \label{eq:3.3a}\\[1.2ex]
 &\dot{\b{z}}=\frac{\rmi}{\hbar}\xi(\b{z},z)\frac{\del\mcl{H}(\b{z},z)}{\del z},
 \label{eq:3.3b}
 \end{align}
\end{subequations}
\noindent where $\xi(\b{z},z)=g^{-1}(\b{z},z)$. In previous identities, we introduce the \textit{effective classical 
Hamiltonian}, which is obtained from the operator~$H$ by two equivalent prescriptions:
\begin{equation}
 \mcl{H}(\b{z},z;t)=\left.\bra z|H(t)|z\ket\right|_{z\cg=\b{z}}
 =\frac{\{\b{z}\cg|H(t)|z\}}{\{\b{z}\cg|z\}}.
 \label{eq:3.4}
\end{equation}

All quantities composing the semiclassical formula~\eqref{eq:3.2} are calculated on specific classical trajectories, 
characterized by a set of \textit{two-time boundary conditions}:
\begin{subequations}
 \label{eq:3.5}
 \begin{align}
 &z(t_{i})=z_{i},
 \label{eq:3.5a}\\[1.2ex]
 &\b{z}(t_{f})=z\cg_{f}.
 \label{eq:3.5b}
 \end{align}
\end{subequations}

The differential equations~\eqref{eq:3.3} usually do not possess a unique solution under boundary conditions. 
For this reason, the expression~\eqref{eq:3.2} presents a summation symbol, which indicates the sum over 
all trajectories satisfying simultaneously the identities \eqref{eq:3.3} and \eqref{eq:3.5}.

Note that the complex vector variables $z$ and $\b{z}$ are \textit{completely independent}, that is, generally $\b{z}(t)\neq z\cg(t)$. 
This \textit{duplication of the phase space} is another direct consequence of the imposition of boundary conditions on the classical 
equations of motion. If we assume equality between $\b{z}(t)$ and $z\cg(t)$, the vector differential equations \eqref{eq:3.3a} and 
\eqref{eq:3.3b} would become equivalent and, consequently, the boundary conditions $z(t_{i})=z_{i}$ and $z\cg(t_{f})=z\cg_{f}$ would 
make the problem \textit{overdetermined}. Therefore, the \textit{duplicated phase space} represents an essential concept for solving 
the classical equations of motion in the representation of coherent states.

The action functional, from which the classical equations of motion are determined by extremization, has the following form 
in the analytic complex parametrization:
\begin{subequations}
 \label{eq:3.6}
 \begin{align}
 &\frac{\rmi}{\hbar}S(z_{f}\cg,z_{i};t_{f},t_{i})=
 \frac{\rmi}{\hbar}\intl_{t_{i}}^{t_{f}}L(\dot{\b{z}},\dot{z},\b{z},z;t)\rmd t
 +\frac{\rmi}{\hbar}\Gamma(z\cg_{f},z(t_{f}),\b{z}(t_{i}),z_{i}),
 \label{eq:3.6a}\\[1.2ex]
 &\frac{\rmi}{\hbar}L(\dot{\b{z}},\dot{z},\b{z},z;t)=
 \frac{1}{2}\left[\frac{\del f(\b{z},z)}{\del\b{z}}\dot{\b{z}}
 -\frac{\del f(\b{z},z)}{\del z}\dot{z}\right]-\frac{\rmi}{\hbar}\mcl{H}(\b{z},z;t),
 \label{eq:3.6b}\\[1.2ex]
 &\frac{\rmi}{\hbar}\Gamma(z\cg_{f},z(t_{f}),\b{z}(t_{i}),z_{i})
 =\frac{1}{2}f(z\cg_{f},z(t_{f}))+\frac{1}{2}f(\b{z}(t_{i}),z_{i}),
 \label{eq:3.6c}
 \end{align}
\end{subequations}
\noindent where we introduce the auxiliary quantities $L$, corresponding to the \textit{Lagrangian} of the system, 
and $\Gamma$, known as the \textit{boundary term}. Although the presence of the function~$\Gamma$ is unusual in most 
basic discussions on classical or quantum dynamics, this quantity turns out to be indispensable for properly obtaining 
equations~\eqref{eq:3.3} under conditions~\eqref{eq:3.5}. Notice that, in general, the classical Hamiltonian and the 
action functional take complex values on the duplicated phase space.

Another important component in the construction of the semiclassical propagator is represented by the \textit{correction
term}:\footnote{The derivative of a $d_{1}$-dimensional vector~$u$ with respect to a $d_{2}$-dimensional vector~$v$ denotes 
a matrix whose elements are described by $\left[\frac{\del u}{\del v}\right]_{jk}=\frac{\del u_{j}}{\del v_{k}}$, for 
$j=1,2,\ldots,d_{1}$ and $k=1,2,\ldots,d_{2}$.}
\begin{equation}
 \frac{\rmi}{\hbar}I(z\cg_{f},z_{i};t_{f},t_{i})
 =\frac{\rmi}{4\hbar}\intl_{t_{i}}^{t_{f}}\tr\left[
 \frac{\del}{\del\b{z}}\left(\xi\frac{\del\mcl{H}}{\del z}\right)
 +\frac{\del}{\del z}\left(\xi\T\frac{\del\mcl{H}}{\del\b{z}}\right)
 \right]\rmd t.
 \label{eq:3.7}
\end{equation}

Due to the overcompleteness of the set of coherent states, the semiclassical approximation can be properly performed by several 
different procedures, which correspond to distinct choices for the classicalization scheme. Each of these approaches, characterized 
by non-equivalent prescriptions for the classical Hamiltonian in terms of the operator~$H$, results in a completely different 
formulation for the correction term.\cite{Baranger01,Santos06}

The normalization factors of the initial and final coherent states of the quantum propagator are 
also present in the expression~\eqref{eq:3.2}. However, this information is encoded in the definition 
of the \textit{normalization term}:
\begin{equation}
 \Lambda(z_{f}\cg,z_{i})=-\frac{1}{2}f(z_{f}\cg,z_{f})-\frac{1}{2}f(z_{i}\cg,z_{i}),
 \label{eq:3.8}
\end{equation}

Finally, we present the component outside the exponential in $K_{sc}$, usually designated as the \textit{pre-factor} 
of the semiclassical propagator:
\begin{equation}
 \mcl{C}(z\cg_{f},z_{i};t_{f},t_{i})=
 \left\{\left[\frac{\det g(\b{z}(t_{i}),z(t_{i}))}{\det g(\b{z}(t_{f}),z(t_{f}))}\right]^{\frac{1}{2}}
 \det\left[\frac{\del\b{z}(t_{i})}{\del\b{z}(t_{f})}\right]\right\}^{\frac{1}{2}}.
 \label{eq:3.9}
\end{equation}

The quantity~$\del\b{z}(t_{i})/\del\b{z}(t_{f})$ is directly related to a block of the \textit{tangent matrix}~$\mbb{M}$, 
which is responsible for the dynamics of small displacements around a classical trajectory. In general, practical 
and theoretical applications of the semiclassical propagator are greatly facilitated by a thorough understanding 
of the properties of $\mbb{M}$. For this reason, in appendix~\ref{app:A}, we present a detailed discussion of some 
aspects of the linearized classical dynamics.

\subsection{Derivation}
\label{ssc:3.2}

In this subsection, we perform the derivation of all results presented at the beginning of this section. Although a meticulous 
derivation of the semiclassical propagator represents a long series of strictly formal calculations, we believe that a critical 
analysis of these theoretical principles can provide an in-depth comprehension of the correspondence between classical and quantum 
mechanics.

The expression~\eqref{eq:3.2} was first obtained by Kochetov as a direct application of his original formulation for the path 
integral in the representation of arbitrary coherent states.\cite{Kochetov95b,Kochetov98} However, in none of these mentioned 
works a detailed derivation of the generalized formula for the semiclassical propagator is presented, since the displayed 
results constitute only an immediate extension of the specific calculations for the group~$\mrm{SU}(2)$.\cite{Kochetov95a}

In certain aspects, the following subsections represent a generalization of the derivation proposed 
by Braun and Garg,\cite{Braun07a,Braun07b} which is restricted to canonical and spin coherent states. 
However, our results also incorporate some new elements introduced by a recent derivation of the 
$\mrm{SU}(n)$ semiclassical propagator.\cite{Viscondi11b}

\subsubsection{Path Integral}
\label{sss:3.2.1}

With the purpose of reformulating expression~\eqref{eq:3.1} as a path integral, we divide the propagation interval 
$(t_{f}-t_{i})$ into $M$ identical segments:
\begin{equation}
 \vep=\frac{t_{f}-t_{i}}{M}.
 \label{eq:3.10}
\end{equation}

In this way, we can rewrite the time-evolution operator as a product of $M$ exponentials:
\begin{equation}
 K(z\cg_{f},z_{i};t_{f},t_{i})=\bra z_{f}|\hat{T}\prodl_{j=1}^{M}
 \exp\left[-\frac{\rmi}{\hbar}\intl_{t_{j-1}}^{t_{j}}H(t)\rmd t\right]|z_{i}\ket,
 \label{eq:3.11}
\end{equation}
\noindent in which we define the notation for the \textit{discretized time}:
\begin{equation}
 t_{j}=t_{i}+j\vep,
 \label{eq:3.12}
\end{equation}
\noindent with $j=0,1,\ldots,M$. For a sufficiently large value of $M$, we can approximate the integral 
in equation~\eqref{eq:3.11} by $H(t_{j})\vep$. As a consequence, the propagator takes approximately the 
following value:
\begin{equation}
 K(z\cg_{f},z_{i};t_{f},t_{i})\approx\bra z_{f}|\hat{T}\prodl_{j=1}^{M}
 \exp\left[-\frac{\rmi}{\hbar}H(t_{j})\vep\right]|z_{i}\ket.
 \label{eq:3.13}
\end{equation}

In the \textit{continuous-time limit}, given by $M\rightarrow\infty$ and $\vep\rightarrow0$ 
for $M\vep=(t_{f}-t_{i})$, the above approximation must converge again to the exact result. 
Then, taking this limit, we insert a closure relation~\eqref{eq:2.21} between each pair of 
factors of the time-evolution operator in the expression~\eqref{eq:3.13}:
\begin{equation}
 K(z\cg_{f},z_{i};t_{f},t_{i})=\lim_{M\rightarrow\infty}\int\left[\prodl_{k=1}^{M-1}\rmd\mu({z^{k}}\cg,z^{k})\right]
 \prodl_{j=1}^{M}\bra z^{j}|\rme^{-\frac{\rmi}{\hbar}H(t_{j})\vep}|z^{j-1}\ket.
 \label{eq:3.14}
\end{equation}

Notice that, as a result of this last manipulation, the time-ordering operator became unnecessary and, therefore, 
was removed. In equation~\eqref{eq:3.14}, we introduce the notation for a \textit{discretized path} in phase space:
\begin{subequations}
 \label{eq:3.15}
 \begin{align}
 &z^{j}=z(t_{j}),
 \label{eq:3.15a}\\[1.2ex]
 &{z^{j}}\cg=z\cg(t_{j}),
 \label{eq:3.15b}
 \end{align}
\end{subequations}
\noindent for $j=1,2,\ldots,(M-1)$. We can also extend the above definitions to the boundary conditions 
of the propagator:
\begin{subequations}
 \label{eq:3.16}
 \begin{align}
 &z^{0}=z(t_{i})=z_{i},
 \label{eq:3.16a}\\[1.2ex]
 &{z^{M}}\cg=z\cg(t_{f})=z\cg_{f}.
 \label{eq:3.16b}
 \end{align}
\end{subequations}

Considering again the continuous-time limit, we perform the following first-order approximation with respect 
to the infinitesimal interval~$\vep$:
\begin{equation}
 \begin{aligned}
 \bra z^{j}|\rme^{-\frac{\rmi}{\hbar}H(t_{j})\vep}|z^{j-1}\ket
 &\approx\bra z^{j}|\left(\mds{1}-\frac{\rmi}{\hbar}H(t_{j})\vep\right)|z^{j-1}\ket\\
 &=\bra z^{j}|z^{j-1}\ket\left[1-\frac{\rmi}{\hbar}\mcl{H}({z^{j}}\cg,z^{j-1})\vep\right]\\
 &\approx\bra z^{j}|z^{j-1}\ket\exp\left[-\frac{\rmi}{\hbar}\mcl{H}({z^{j}}\cg,z^{j-1})\vep\right].
 \end{aligned}
 \label{eq:3.17}
\end{equation}

In the above equation, we define the \textit{discretized effective Hamiltonian}:
\begin{equation}
 \mcl{H}({z^{j}}\cg,z^{j-1})
 =\frac{\bra z^{j}|H(t_{j})|z^{j-1}\ket}{\bra z^{j}|z^{j-1}\ket}
 =\frac{\{z^{j}|H(t_{j})|z^{j-1}\}}{\{z^{j}|z^{j-1}\}},
 \label{eq:3.18}
\end{equation}
\noindent in which the relation~\eqref{eq:2.17} was employed to derive the second equality.
Now, by substituting the identity~\eqref{eq:3.17} into the propagator~\eqref{eq:3.14}, we 
obtain:
\begin{equation}
 K(z\cg_{f},z_{i};t_{f},t_{i})=\lim_{M\rightarrow\infty}
 \int\left[\prodl_{k=1}^{M-1}\rmd\mu({z^{k}}\cg,z^{k})\right]
 \exp\left[\frac{\rmi}{\hbar}\til{S}_{d}\right],
 \label{eq:3.19}
\end{equation}
\noindent where, with the purpose of simplifying the notation, the following quantity is introduced:
\begin{equation}
 \begin{aligned}
 \frac{\rmi}{\hbar}\til{S}_{d}=&
 \suml_{j=1}^{M}\left[\ln\bra z^{j}|z^{j-1}\ket-\frac{\rmi}{\hbar}
 \mcl{H}({z^{j}}\cg,z^{j-1})\vep\right]\\
 =&\suml_{j=1}^{M}\left[f({z^{j}}\cg,z^{j-1})
 -\frac{1}{2}f({z^{j}}\cg,z^{j})-\frac{1}{2}f({z^{j-1}}\cg,z^{j-1})
 -\frac{\rmi}{\hbar}\mcl{H}({z^{j}}\cg,z^{j-1})\vep\right]\\
 =&\suml_{j=1}^{M}\left[f({z^{j}}\cg,z^{j-1})
 -\frac{\rmi}{\hbar}\mcl{H}({z^{j}}\cg,z^{j-1})\vep\right]
 -\suml_{j=1}^{M-1}f({z^{j}}\cg,z^{j})\\
 &-\frac{1}{2}f(z_{f}\cg,z_{f})-\frac{1}{2}f(z_{i}\cg,z_{i}).
 \end{aligned}
 \label{eq:3.20}
\end{equation}

Although we have made several mathematical manipulations, the result~\eqref{eq:3.19} still represents an exact formulation 
for the quantum propagator. However, with the aim of performing semiclassical approximations, we now assume that only the 
\textit{continuous paths}\footnote{That is, paths in which two successive points differ only by an infinitesimal quantity.} 
in phase space are relevant in calculating the propagator.

In the quantum regime, there is no guarantee that discontinuous paths have zero or negligible contribution to the dynamics 
of a system described by a path integral. Furthermore, we expect that the discontinuous paths have an important role in 
characteristic quantum phenomena, such as entanglement and tunneling. This scenario is completely modified in classical 
and semiclassical regimes, which are characterized by large values of $|\til{S}_{d}|$ in comparison with the reduced Plank  
constant~$\hbar$. In these cases, only the paths resulting in nearly stationary values of $\til{S}_{d}$ are spared from 
cancellation by destructive phase interference.

In order to promote a privileged treatment for continuous paths, we can perform approximations on the exponent~$\til{S}_{d}$, 
under the assumption that the variations $|{z^{j}}\cg-{z^{j-1}}\cg|$ and $|z^{j}-z^{j-1}|$ have the same order of 
magnitude of $\vep$. By applying this hypothesis, we obtain the following auxiliary relations:
\begin{subequations}
 \label{eq:3.21}
 \begin{align}
 &f({z^{j}}\cg,z^{j})\approx f({z^{j+1}}\cg,z^{j})
 +\frac{\del f({z^{j+1}}\cg,z^{j})}{\del {z^{j+1}}\cg}({z^{j}}\cg-{z^{j+1}}\cg),
 \label{eq:3.21a}\\[1.2ex]
 &f({z^{j}}\cg,z^{j})\approx f({z^{j}}\cg,z^{j-1})
 +\frac{\del f({z^{j}}\cg,z^{j-1})}{\del z^{j-1}}(z^{j}-z^{j-1}),
 \label{eq:3.21b}
 \end{align}
\end{subequations}
\noindent for $j=1,2,\ldots,(M-1)$. Then, using the approximations~\eqref{eq:3.21}, 
we rewrite the second summation in the last step of equation~\eqref{eq:3.20}:
\begin{equation}
 \begin{aligned}
 \suml_{j=1}^{M-1}f({z^{j}}\cg,z^{j})\approx&\;
 \frac{1}{2}\suml_{j=1}^{M-1}\left[
 \frac{\del f({z^{j+1}}\cg,z^{j})}{\del{z^{j+1}}\cg}({z^{j}}\cg-{z^{j+1}}\cg)
 +\frac{\del f({z^{j}}\cg,z^{j-1})}{\del z^{j-1}}(z^{j}-z^{j-1})\right]
 \\&+\frac{1}{2}\suml_{j=2}^{M}f({z^{j}}\cg,z^{j-1})
 +\frac{1}{2}\suml_{j=1}^{M-1}f({z^{j}}\cg,z^{j-1}).
 \end{aligned}
 \label{eq:3.22}
\end{equation}

With the aid of the above result, we write a new expression for the phase of the propagator integrand:
\begin{equation}
 \begin{aligned}
 \frac{\rmi}{\hbar}\til{S}_{d}\approx&\;
 \frac{1}{2}f(z_{f}\cg,z^{M-1})+\frac{1}{2}f({z^{1}}\cg,z_{i})
 -\frac{1}{2}f(z_{f}\cg,z_{f})-\frac{1}{2}f(z_{i}\cg,z_{i})
 -\frac{\rmi}{\hbar}\suml_{j=1}^{M}\mcl{H}({z^{j}}\cg,z^{j-1})\vep\\
 &+\frac{1}{2}\suml_{j=1}^{M-1}\left[
 \frac{\del f({z^{j+1}}\cg,z^{j})}{\del{z^{j+1}}\cg}({z^{j+1}}\cg-{z^{j}}\cg)
 -\frac{\del f({z^{j}}\cg,z^{j-1})}{\del z^{j-1}}(z^{j}-z^{j-1})\right].
 \end{aligned}
 \label{eq:3.23}
\end{equation}

From the previous identity, we can readily obtain the continuous-time limit of the propagator phase:\footnote{Note 
that the subscript~$d$ is used to indicate functionals defined on discretized paths. Evidently, the continuous-time 
limit of these functionals is symbolized by omitting the subscript~$d$.}
\begin{equation}
 \frac{\rmi}{\hbar}\til{S}=\frac{\rmi}{\hbar}S(z_{f}\cg,z_{i};t_{f},t_{i})+\Lambda(z_{f}\cg,z_{i}),
 \label{eq:3.35}
\end{equation}
\noindent where the quantity~$\Lambda$ once again denotes the normalization term, introduced in subsection~\ref{ssc:3.1} 
by equation~\eqref{eq:3.8}. As a function of the Lagrangian~$L$ and the boundary term~$\Gamma$, the action functional~$S$ 
presented in the result~\eqref{eq:3.35} has the following form:
\begin{subequations}
 \label{eq:3.36}
 \begin{align}
 &\frac{\rmi}{\hbar}S(z_{f}\cg,z_{i};t_{f},t_{i})=
 \frac{\rmi}{\hbar}\intl_{t_{i}}^{t_{f}}L(\dot{z}\cg,\dot{z},z\cg,z;t)\rmd t
 +\frac{\rmi}{\hbar}\Gamma(z\cg_{f},z(t_{f}),z\cg(t_{i}),z_{i}),
 \label{eq:3.36a}\\[1.2ex]
 &\frac{\rmi}{\hbar}L(\dot{z}\cg,\dot{z},z\cg,z;t)=
 \frac{1}{2}\left[\frac{\del f(z\cg,z)}{\del z\cg}\dot{z}\cg
 -\frac{\del f(z\cg,z)}{\del z}\dot{z}\right]-\frac{\rmi}{\hbar}\mcl{H}(z\cg,z),
 \label{eq:3.36b}\\[1.2ex]
 &\frac{\rmi}{\hbar}\Gamma(z\cg_{f},z(t_{f}),z\cg(t_{i}),z_{i})
 =\frac{1}{2}f(z\cg_{f},z(t_{f}))+\frac{1}{2}f(z\cg(t_{i}),z_{i}).
 \label{eq:3.36c}
 \end{align}
\end{subequations}

By employing the identity~\eqref{eq:3.35}, the continuous-time limit of the propagator~\eqref{eq:3.19}
is obtained:
\begin{equation}
 \til{K}(z\cg_{f},z_{i};t_{f},t_{i})=\int\rmD\mu(z\cg,z)\exp\left[
 \frac{\rmi}{\hbar}S(z\cg_{f},z_{i};t_{f},t_{i})
 +\Lambda(z_{f}\cg,z_{i})\right],
 \label{eq:3.37}
\end{equation}
\noindent which we describe as an integral over the following \textit{path measure}:
\begin{equation}
 \rmD\mu(z\cg,z)=\lim_{M\rightarrow\infty}\prodl_{j=1}^{M-1}\rmd\mu({z^{j}}\cg,z^{j}).
 \label{eq:3.38}
\end{equation}

Note that the expression~\eqref{eq:3.37} has a distinctive tilde notation, since this result is generally 
an approximation to the quantum propagator, in which we assume that the discontinuous paths in phase space 
have irrelevant contribution to the calculation of the path integral.

However, in some situations of interest, the quantity~$\til{K}$ is in exact correspondence with 
the definition~\eqref{eq:3.1}. Particularly, in the case of the canonical coherent states, the 
equations~(\ref{eq:3.21}\-\ref{eq:3.23}) do not constitute approximate relations, so that the 
propagators \eqref{eq:3.19} and \eqref{eq:3.37} remain equivalent to each other.

\subsubsection{Classical Trajectories and Phase-Space Duplication}
\label{sss:3.2.1.add}

In subsection~\ref{sss:3.2.3}, we shall perform the semiclassical approximation of the path integral~\eqref{eq:3.37}.
As a preparatory step toward this goal, we need to identify the classical trajectories associated with the quantum 
propagator in the coherent-state representation. For this purpose, we now return to the discretized expression for 
the propagator phase, given by equation~\eqref{eq:3.23}.

By still assuming that, for a sufficiently large value of $M$, the coordinates ${z^{j}}\cg$ and $z^{j}$ form 
a continuous path for \textit{any} value of $j$, we temporarily\footnote{As discussed later in this subsection, 
the assumptions leading to equations~\eqref{eq:3.24} are incompatible with the proper determination of classical 
trajectories under the boundary conditions~\eqref{eq:3.16}. For this reason, the approximations~\eqref{eq:3.24} 
will be replaced by the identities~\eqref{eq:3.30}.} extend the approximations~\eqref{eq:3.21} to the endpoints 
of the time interval:
\begin{subequations}
 \label{eq:3.24}
 \begin{align}
 &f(z_{i}\cg,z_{i})
 \approx f({z^{1}}\cg,z_{i})
 +\frac{\del f({z^{1}}\cg,z_{i})}{\del{z^{1}}\cg}(z_{i}\cg-{z^{1}}\cg),
 \label{eq:3.24a}\\[1.2ex]
 &f(z_{f}\cg,z_{f})
 \approx f(z_{f}\cg,z^{M-1})
 +\frac{\del f(z_{f}\cg,z^{M-1})}{\del z^{M-1}}(z_{f}-z^{M-1}).
 \label{eq:3.24b}
 \end{align}
\end{subequations}

Notice that the definitions ${z^{0}}\cg=z_{i}\cg$ and $z^{M}=z_{f}$ are used in the above expressions. 
By employing the equations~\eqref{eq:3.24}, we rewrite the quantity~\eqref{eq:3.23} as a single 
summation:
\begin{equation}
 \begin{aligned}
 \frac{\rmi}{\hbar}\til{S}_{d}\approx&
 \suml_{j=1}^{M}\left\{
 \frac{1}{2}\left[\frac{\del f({z^{j}}\cg,z^{j-1})}{\del{z^{j}}\cg}({z^{j}}\cg-{z^{j-1}}\cg)
 -\frac{\del f({z^{j}}\cg,z^{j-1})}{\del z^{j-1}}(z^{j}-z^{j-1})\right]\right.\\
 &\left.-\frac{\rmi}{\hbar}\mcl{H}({z^{j}}\cg,z^{j-1})\vep\right\}.
 \end{aligned}
 \label{eq:3.25}
\end{equation}

The values of ${z^{j}}\cg$ and $z^{j}$ that extremize $\til{S}_{d}$ form a \textit{discretized classical 
trajectory}. Therefore, by calculating the first derivatives of $\til{S}_{d}$ and setting them to zero, 
we shall find a set of algebraic equations for the classical values of the complex coordinates. To this 
end, we take the derivatives of expression~\eqref{eq:3.25}:
\begin{subequations}
 \label{eq:3.26}
 \begin{align}
 &\begin{aligned}
 \frac{\rmi}{\hbar}\frac{\del\til{S}_{d}}{\del{z^{j}}\cg}\approx&\;
 \frac{1}{2}\frac{\del^{2}f({z^{j}}\cg,z^{j-1})}{{\del{z^{j}}\cg}^{2}}({z^{j}}\cg-{z^{j-1}}\cg)
 -\frac{1}{2}\frac{\del^{2}f({z^{j}}\cg,z^{j-1})}{\del{z^{j}}\cg\del z^{j-1}}(z^{j}-z^{j-1})\\
 &+\frac{1}{2}\frac{\del f({z^{j}}\cg,z^{j-1})}{\del{z^{j}}\cg}
 -\frac{1}{2}\frac{\del f({z^{j+1}}\cg,z^{j})}{\del{z^{j+1}}\cg}
 -\frac{\rmi}{\hbar}\frac{\del\mcl{H}({z^{j}}\cg,z^{j-1})}{\del{z^{j}}\cg}\vep,
 \end{aligned}
 \label{eq:3.26a}\\[1.2ex]
 &\begin{aligned}
 \frac{\rmi}{\hbar}\frac{\del\til{S}_{d}}{\del z^{j}}\approx&\;
 \frac{1}{2}\frac{\del^{2}f({z^{j+1}}\cg,z^{j})}{\del z^{j}\del{z^{j+1}}\cg}({z^{j+1}}\cg-{z^{j}}\cg)
 -\frac{1}{2}\frac{\del^{2}f({z^{j+1}}\cg,z^{j})}{{\del z^{j}}^{2}}(z^{j+1}-z^{j})\\
 &-\frac{1}{2}\frac{\del f({z^{j}}\cg,z^{j-1})}{\del z^{j-1}}
 +\frac{1}{2}\frac{\del f({z^{j+1}}\cg,z^{j})}{\del z^{j}}
 -\frac{\rmi}{\hbar}\frac{\del\mcl{H}({z^{j+1}}\cg,z^{j})}{\del z^{j}}\vep,
 \end{aligned}
 \label{eq:3.26b}
 \end{align}
\end{subequations}
\noindent for $j=1,2,\ldots,(M-1)$. By dropping the terms of quadratic order in $\vep$, 
we can significantly simplify the above equations. For this purpose, we use the following 
approximations:
\begin{subequations}
 \label{eq:3.27}
 \begin{align}
 &\begin{aligned}
 \frac{\del f({z^{j+1}}\cg,z^{j})}{\del{z^{j+1}}\cg}\approx&\;
 \frac{\del f({z^{j}}\cg,z^{j-1})}{\del{z^{j}}\cg}
 +\frac{\del^{2}f({z^{j}}\cg,z^{j-1})}{{\del{z^{j}}\cg}^{2}}({z^{j+1}}\cg-{z^{j}}\cg)\\
 &+\frac{\del^{2}f({z^{j}}\cg,z^{j-1})}{\del{z^{j}}\cg\del z^{j-1}}(z^{j}-z^{j-1}),
 \end{aligned}
 \label{eq:3.27a}\\[1.2ex]
 &\begin{aligned}
 \frac{\del f({z^{j}}\cg,z^{j-1})}{\del z^{j-1}}\approx&\;
 \frac{\del f({z^{j+1}}\cg,z^{j})}{\del z^{j}}
 +\frac{\del^{2}f({z^{j+1}}\cg,z^{j})}{{\del z^{j}}^{2}}(z^{j-1}-z^{j})\\
 &+\frac{\del^{2}f({z^{j+1}}\cg,z^{j})}{\del z^{j}\del{z^{j+1}}\cg}({z^{j}}\cg-{z^{j+1}}\cg),
 \end{aligned}
 \label{eq:3.27b}
 \end{align}
\end{subequations}
\noindent which are valid up to the first order in the infinitesimal parameter~$\vep$. 
As a consequence of equations~\eqref{eq:3.27}, the derivatives of $\til{S}_{d}$ can 
be rewritten as:
\begin{subequations}
 \label{eq:3.28}
 \begin{align}
 &\begin{aligned}
 \frac{\rmi}{\hbar}\frac{\del\til{S}_{d}}{\del{z^{j}}\cg}\approx&\;
 \frac{1}{2}\frac{\del^{2}f({z^{j}}\cg,z^{j-1})}{{\del{z^{j}}\cg}^{2}}(-{z^{j+1}}\cg+2{z^{j}}\cg-{z^{j-1}}\cg)
 -\frac{\del^{2}f({z^{j}}\cg,z^{j-1})}{\del{z^{j}}\cg\del z^{j-1}}(z^{j}-z^{j-1})\\
 &-\frac{\rmi}{\hbar}\frac{\del\mcl{H}({z^{j}}\cg,z^{j-1})}{\del{z^{j}}\cg}\vep,
 \end{aligned}
 \label{eq:3.28a}\\[1.2ex]
 &\begin{aligned}
 \frac{\rmi}{\hbar}\frac{\del\til{S}_{d}}{\del z^{j}}\approx&\;
 \frac{\del^{2}f({z^{j+1}}\cg,z^{j})}{\del z^{j}\del{z^{j+1}}\cg}({z^{j+1}}\cg-{z^{j}}\cg)
 -\frac{1}{2}\frac{\del^{2}f({z^{j+1}}\cg,z^{j})}{{\del z^{j}}^{2}}(z^{j+1}-2z^{j}+z^{j-1})\\
 &-\frac{\rmi}{\hbar}\frac{\del\mcl{H}({z^{j+1}}\cg,z^{j})}{\del z^{j}}\vep.
 \end{aligned}
 \label{eq:3.28b}
 \end{align}
\end{subequations}

Note that the terms proportional to $(z^{j+1}-2z^{j}+z^{j-1})$ and $({z^{j+1}}\cg-2{z^{j}}\cg+{z^{j-1}}\cg)$ can be readily discarded,
since they represent quantities of quadratic order in $\vep$ under the assumption of continuous paths. Therefore, considering only 
the linear terms with respect to $\vep$ in the approximations~\eqref{eq:3.28}, we finally arrive at the \textit{discretized classical 
equations of motion}:
\begin{subequations}
 \label{eq:3.29}
 \begin{align}
 &g\T({z^{j}}\cg,z^{j-1})(z^{j}-z^{j-1})
 =-\frac{\rmi}{\hbar}\frac{\del\mcl{H}({z^{j}}\cg,z^{j-1})}{\del{z^{j}}\cg}\vep,
 \label{eq:3.29a}\\[1.2ex]
 &g({z^{j+1}}\cg,z^{j})({z^{j+1}}\cg-{z^{j}}\cg)
 =\frac{\rmi}{\hbar}\frac{\del\mcl{H}({z^{j+1}}\cg,z^{j})}{\del z^{j}}\vep,
 \label{eq:3.29b}
 \end{align}
\end{subequations}
\noindent for $j=1,2,\ldots,(M-1)$. Although the coordinates~${z^{j}}\cg$ symbolize the complex conjugate values of the 
variables~$z^{j}$, we observe that, for a chosen value of the index~$j$, the expressions~\eqref{eq:3.29a} and \eqref{eq:3.29b} 
are not equivalent to each other, since these two sets of equations are partially displaced by one time step. Therefore, 
we have $2(M-1)$ complex algebraic equations for $(M-1)$ complex classical variables, given that the quantities $z^{0}=z_{i}$ 
and ${z^{M}}\cg=z_{f}\cg$ are already determined by the boundary conditions~\eqref{eq:3.16}.

The problem posed by the identities \eqref{eq:3.16} and \eqref{eq:3.29} is \textit{overdetermined} and, consequently, has 
no solution for an arbitrary Hamiltonian. To address this issue, we can assume that the values of ${z^{j}}\cg$ are completely 
independent of $z^{j}$. In general, this duplication of the number of available variables establishes a consistent set of 
algebraic equations.

The \textit{duplication of the phase space} is only made possible by the absence of the quantities $z^{M}=z_{f}$ 
and ${z^{0}}\cg=z\cg_{i}$ in the identities~\eqref{eq:3.29},\footnote{The absence of the quantities $z^{M}=z_{f}$
and ${z^{0}}\cg=z\cg_{i}$ in equations~\eqref{eq:3.29} is a consequence of the use of an analytic parametrization 
for the coherent-state set.} since these values are inextricably related to their complex conjugates by the boundary 
conditions. Also for this reason, we conclude that $|{z^{0}}\cg-{z^{1}}\cg|$ and $|z^{M}-z^{M-1}|$ do not represent 
infinitesimal quantities for an arbitrary continuous path in the duplicated phase space satisfying the 
restrictions~\eqref{eq:3.16}, in which case the classical trajectories are included.

The above conclusion invalidates the approximations~\eqref{eq:3.24} as first-order identities 
in the parameter~$\vep$ for duplicated-space continuous paths subject to conditions~\eqref{eq:3.16}. 
As a consequence, the result~\eqref{eq:3.25} does not apply to classical trajectories. Nevertheless, 
the equations~\eqref{eq:3.29} can still be properly derived from expression~\eqref{eq:3.23}. For 
this purpose, we only need to replace the identities~\eqref{eq:3.24} by the following approximations:
\begin{subequations}
 \label{eq:3.30}
 \begin{align}
 &f({z^{1}}\cg,z_{i})
 \approx f(z\cg(t_{i}),z_{i})
 +\frac{\del f({z^{1}}\cg,z_{i})}{\del {z^{1}}\cg}({z^{1}}\cg-z\cg(t_{i})),
 \label{eq:3.30a}\\[1.2ex]
 &f(z\cg_{f},z^{M-1})
 \approx f(z\cg_{f},z(t_{f}))
 -\frac{\del f(z\cg_{f},z^{M-1})}{\del z^{M-1}}(z(t_{f})-z^{M-1}),
 \label{eq:3.30b}
 \end{align}
\end{subequations}
\noindent in which $z\cg(t_{i})$ and $z(t_{f})$ represent the correct values of these dynamical variables at the endpoints 
of the time interval. Once the relevant considerations have been made, the continuous-time limit of the discretized equations 
of motion is readily obtained:
\begin{subequations}
 \label{eq:3.31}
 \begin{align}
 &g\T(z\cg,z)\dot{z}=-\frac{\rmi}{\hbar}\frac{\del\mcl{H}(z\cg,z)}{\del z\cg},
 \label{eq:3.31a}\\[1.2ex]
 &g(z\cg,z)\dot{z}\cg=\frac{\rmi}{\hbar}\frac{\del\mcl{H}(z\cg,z)}{\del z}.
 \label{eq:3.31b}
 \end{align}
\end{subequations}

In the above identities, we also introduce the continuous limit for \textit{effective classical Hamiltonian}:
\begin{equation}
 \mcl{H}(z\cg,z)=\bra z|H|z\ket.
 \label{eq:3.32}
\end{equation}

If we persist in interpreting $z\cg(t)$ as the set of complex conjugate variables in relation to $z(t)$, it is evident that 
the equations of motion~\eqref{eq:3.31} will inherit the overdetermination of their discretized versions. In this new situation, 
the identities \eqref{eq:3.31a} and \eqref{eq:3.31b} are precisely equivalent to each other, so that a single first-order vector 
differential equation is left for two boundary conditions, which are given by $z\cg(t_{f})=z_{f}\cg$ and $z(t_{i})=z_{i}$. Thus, 
with the exception of accidental cases, the dynamical system~\eqref{eq:3.31} has no solution under the constraints~\eqref{eq:3.16}.

However, similarly to the discretized case, we can assume that the vector variable~$z\cg(t)$ is completely independent of $z(t)$. 
This duplication in the number of classical degrees of freedom removes the redundancy between identities \eqref{eq:3.31a} and 
\eqref{eq:3.31b}. Consequently, we now have a well-defined problem, consisting of two first-order differential equations subject 
to two boundary conditions. Interestingly, the description of a classical trajectory in terms of the independent quantities 
$z\cg(t)$ and $z(t)$ has origin in the discretized representation, due to the partial time displacement between the expressions 
\eqref{eq:3.29a} and \eqref{eq:3.29b}.

In order to explicitly indicate the phase-space duplication in our subsequent developments, we introduce an important 
change of notation:
\begin{equation}
 z\cg(t)\rightarrow\b{z}(t).
 \label{eq:3.33}
\end{equation}

We must emphasize that, in the remainder of this paper, we shall continue to use the quantity~$z\cg(t)$, but only to 
unequivocally symbolize the complex conjugate of the vector~$z(t)$. Except for some special situations,\footnote{Many 
of the cases in which $\b{z}(t)=z\cg(t)$ will be discussed in future applications of the semiclassical propagator.} 
the notational change~\eqref{eq:3.33} generally implies that $\b{z}(t)\neq z\cg(t)$.

Aside from an evident multiplication by the matrix~$\xi(\b{z},z)\!=\!g^{-1}(\b{z},z)$, note that the equations 
of motion~\eqref{eq:3.3} represent an immediate example of application of the transformation~\eqref{eq:3.33},
since they simply follow from a change of notation in identities~\eqref{eq:3.31}.

Considering the continuous-time limit in the duplicated phase space, the boundary conditions~\eqref{eq:3.16} take the form 
presented in equations~\eqref{eq:3.5}. In this case, it should be noted that the quantities $\b{z}(t_{i})$ and $z(t_{f})$ 
are not determined by constraints on the propagator, that is, in general
\begin{subequations}
 \label{eq:3.34}
 \begin{align}
 &\b{z}(t_{i})\neq z\cg_{i},
 \label{eq:3.34a}\\[1.2ex]
 &z(t_{f})\neq z_{f}.
 \label{eq:3.34b}
 \end{align}
\end{subequations}

Notice that, unlike the definition~\eqref{eq:3.6}, which is suitable for the semiclassical propagator~\eqref{eq:3.2},
the identities~\eqref{eq:3.36} do not require the notational change~\eqref{eq:3.33}, since the phase-space duplication 
is not strictly necessary for the evaluation of the quantum propagator in the continuous-path approximation.

The path integral~\eqref{eq:3.19} consists of purely geometric paths, which are subject only to the boundary 
conditions~\eqref{eq:3.16}. On the other hand, the classical trajectories must additionally satisfy the equations 
of motion~\eqref{eq:3.31} and, consequently, they must reside in an extended phase space, as properly described by 
equations \eqref{eq:3.3} and \eqref{eq:3.5}. Therefore, the duplication of the classical degrees of freedom becomes 
indispensable for the calculation of the propagator only in the semiclassical approximation, due to the expansion 
of the action functional around its extreme values, as will be shown in subsection~\ref{sss:3.2.3}.

\subsubsection{First Variation of the Action Functional}
\label{sss:3.2.2}

In this subsection, we derive several useful identities related to the action functional. Furthermore, in view of our future 
developments, we use here the notation~\eqref{eq:3.33}, which is suitable for the duplicated phase space. Similar results for 
the \textit{simple phase space}\footnote{We use the term ``simple phase space'' in contrast to the definition of the duplicated 
phase space. In this manner, we seek to avoid possible ambiguities in references to the usual phase space, in which the duplication 
of the classical degrees of freedom is absent.} are special cases of the expressions presented below, since they can be obtained 
under the restriction~$\b{z}(t)=z\cg(t)$.

The action functional~$S$, the Lagrangian~$L$ and the boundary term~$\Gamma$ are defined, in notation appropriate to the duplicated 
phase space, by equations~\eqref{eq:3.6}. Similarly, the analytic continuation of the effective classical Hamiltonian~$\mcl{H}$ is 
described, in terms of the Hamiltonian operator~$H$, by the identity~\eqref{eq:3.4}.

In order to calculate the first variation of the functional~$S$ around one of its extreme values, including small deviations 
in the endpoints of the time interval and in the extremities of the trajectory, we introduce the following quantities:
\begin{subequations}
 \label{eq:3.39}
 \begin{align}
 &z'=z(t_{i}+\delta t_{i})\approx z(t_{i})+\dot{z}(t_{i})\delta t_{i},
 \label{eq:3.39a}\\[0.6ex]
 &\b{z}'=\b{z}(t_{i}+\delta t_{i})\approx\b{z}(t_{i})+\dot{\b{z}}(t_{i})\delta t_{i},
 \label{eq:3.39b}\\[0.6ex]
 &z''=z(t_{f}+\delta t_{f})\approx z(t_{f})+\dot{z}(t_{f})\delta t_{f},
 \label{eq:3.39c}\\[0.6ex]
 &\b{z}''=\b{z}(t_{f}+\delta t_{f})\approx\b{z}(t_{f})+\dot{\b{z}}(t_{f})\delta t_{f},
 \label{eq:3.39d}
 \end{align}
\end{subequations}
\noindent whose first-order deviations are given by
\begin{subequations}
 \label{eq:3.40}
 \begin{align}
 &\delta z'=\delta z(t_{i})+\dot{z}(t_{i})\delta t_{i},
 \label{eq:3.40a}\\[0.6ex]
 &\delta\b{z}'=\delta\b{z}(t_{i})+\dot{\b{z}}(t_{i})\delta t_{i},
 \label{eq:3.40b}\\[0.6ex]
 &\delta z''=\delta z(t_{f})+\dot{z}(t_{f})\delta t_{f},
 \label{eq:3.40c}\\[0.6ex]
 &\delta\b{z}''=\delta\b{z}(t_{f})+\dot{\b{z}}(t_{f})\delta t_{f}.
 \label{eq:3.40d}
 \end{align}
\end{subequations}

Applying the above definitions, we can readily write a general formula 
for the first variation of the action functional:
\begin{equation}
 \begin{aligned}
 \delta S=&\intl_{t_{i}}^{t_{f}}\left[
 \frac{\del L}{\del z}\delta z
 +\frac{\del L}{\del\b{z}}\delta\b{z}
 +\frac{\del L}{\del\dot{z}}\delta\dot{z}
 +\frac{\del L}{\del\dot{\b{z}}}\delta\dot{\b{z}}
 \right]\rmd t
 +\left.L\right|_{t_{f}}\delta{t_{f}}
 -\left.L\right|_{t_{i}}\delta{t_{i}}\\
 &+\frac{\del\Gamma}{\del z(t_{i})}\delta z'
 +\frac{\del\Gamma}{\del\b{z}(t_{i})}\delta\b{z}'
 +\frac{\del\Gamma}{\del z(t_{f})}\delta z''
 +\frac{\del\Gamma}{\del\b{z}(t_{f})}\delta\b{z}''\\
 =&\intl_{t_{i}}^{t_{f}}\left\{
 \left[\frac{\del L}{\del z}-\frac{\rmd}{\rmd t}\left(
 \frac{\del L}{\del\dot{z}}\right)\right]\delta z
 +\left[\frac{\del L}{\del\b{z}}-\frac{\rmd}{\rmd t}\left(
 \frac{\del L}{\del\dot{\b{z}}}\right)\right]\delta\b{z}
 \right\}\rmd t\\
 &+\left.L\right|_{t_{f}}\delta{t_{f}}
 -\left.L\right|_{t_{i}}\delta{t_{i}}
 +\left[\frac{\del L}{\del \dot{z}}\delta z\right]_{t_{i}}^{t_{f}}
 +\left[\frac{\del L}{\del \dot{\b{z}}}\delta \b{z}\right]_{t_{i}}^{t_{f}}\\
 &+\frac{\del\Gamma}{\del z(t_{i})}\delta z'
 +\frac{\del\Gamma}{\del\b{z}(t_{i})}\delta\b{z}'
 +\frac{\del\Gamma}{\del z(t_{f})}\delta z''
 +\frac{\del\Gamma}{\del\b{z}(t_{f})}\delta\b{z}''.
 \end{aligned}
 \label{eq:3.41}
\end{equation}

By calculating the derivatives appearing in expression~\eqref{eq:3.41},\footnote{The explicit 
calculation of the derivatives in equation~\eqref{eq:3.41} is given in appendix~\ref{app:B}, 
by identities~(\ref{eq:3.42}-\ref{eq:3.44}).} we rewrite the first variation of the action 
functional:
\begin{equation}
 \begin{aligned}
 \frac{\rmi}{\hbar}\delta S
 =&\intl_{t_{i}}^{t_{i}}
 \left\{\left[g(\b{z},z)\dot{\b{z}}
 -\frac{\rmi}{\hbar}\frac{\del\mcl{H}(\b{z},z)}{\del z}\right]\delta z
 +\left[-g\T(\b{z},z)\dot{z}
 -\frac{\rmi}{\hbar}\frac{\del\mcl{H}(\b{z},z)}{\del\b{z}}\right]\delta\b{z}
 \right\}\rmd t\\
 &-\frac{\rmi}{\hbar}\mcl{H}(\b{z}(t_{f}),z(t_{f}))\delta t_{f}
 +\frac{\del f(\b{z}(t_{f}),z(t_{f}))}{\del\b{z}(t_{f})}\delta\b{z}''\\
 &+\frac{\rmi}{\hbar}\mcl{H}(\b{z}(t_{i}),z(t_{i}))\delta t_{i}
 +\frac{\del f(\b{z}(t_{i}),z(t_{i}))}{\del z(t_{i})}\delta z'.
 \end{aligned}
 \label{eq:3.45}
\end{equation}

Under the restrictions $\delta\b{z}(t_{f})=\delta z(t_{i})=0$ and $\delta t_{f}=\delta t_{i}=0$, corresponding to fixed boundary 
conditions and time-interval endpoints, the identity~$\delta S=0$ again results in the equations of motion~\eqref{eq:3.31}, but 
in suitable notation for the duplicated phase space. As expected, the classical trajectories are identified as the extremizing 
values of the action functional.

The functional~$S$, when calculated on an extremizing trajectory, is designated as the \textit{classical action}. In this case, 
the integral in equation~\eqref{eq:3.45} becomes identically zero, allowing us to find the following partial derivatives:
\begin{subequations}
 \label{eq:3.46}
 \begin{align}
 &\frac{\rmi}{\hbar}\frac{\del S}{\del t_{i}}=\frac{\rmi}{\hbar}\mcl{H}(\b{z}(t_{i}),z(t_{i})),
 \label{eq:3.46a}\\[1.2ex]
 &\frac{\rmi}{\hbar}\frac{\del S}{\del t_{f}}=-\frac{\rmi}{\hbar}\mcl{H}(\b{z}(t_{f}),z(t_{f})),
 \label{eq:3.46b}\\[1.2ex]
 &\frac{\rmi}{\hbar}\frac{\del S}{\del z(t_{i})}=\frac{\del f(\b{z}(t_{i}),z(t_{i}))}{\del z(t_{i})},
 \label{eq:3.46c}\\[1.2ex]
 &\frac{\rmi}{\hbar}\frac{\del S}{\del\b{z}(t_{f})}=\frac{\del f(\b{z}(t_{f}),z(t_{f}))}{\del\b{z}(t_{f})}.
 \label{eq:3.46d}
 \end{align}
\end{subequations}

The second derivatives of the classical action, with respect to the extremities of the trajectory, follow immediately 
from the above results:
\begin{subequations}
 \label{eq:3.47}
 \begin{align}
 &\frac{\rmi}{\hbar}\frac{\del^{2}S}{\del\b{z}(t_{f})^{2}}
 =\frac{\del^{2}f(\b{z}(t_{f}),z(t_{f}))}{\del\b{z}(t_{f})^{2}}
 +\frac{\del^{2}f(\b{z}(t_{f}),z(t_{f}))}{\del\b{z}(t_{f})\del z(t_{f})}\frac{\del z(t_{f})}{\del\b{z}(t_{f})},
 \label{eq:3.47a}\\[1.2ex]
 &\frac{\rmi}{\hbar}\frac{\del^{2}S}{\del\b{z}(t_{f})\del z(t_{i})}
 =\frac{\del^{2}f(\b{z}(t_{f}),z(t_{f}))}{\del\b{z}(t_{f})\del z(t_{f})}\frac{\del z(t_{f})}{\del z(t_{i})},
 \label{eq:3.47b}\\[1.2ex]
 &\frac{\rmi}{\hbar}\frac{\del^{2}S}{\del z(t_{i})\del\b{z}(t_{f})}
 =\frac{\del^{2}f(\b{z}(t_{i}),z(t_{i}))}{\del z(t_{i})\del\b{z}(t_{i})}\frac{\del\b{z}(t_{i})}{\del\b{z}(t_{f})},
 \label{eq:3.47c}\\[1.2ex]
 &\frac{\rmi}{\hbar}\frac{\del^{2}S}{\del z(t_{i})^{2}}
 =\frac{\del^{2}f(\b{z}(t_{i}),z(t_{i}))}{\del z(t_{i})^{2}}
 +\frac{\del^{2}f(\b{z}(t_{i}),z(t_{i}))}{\del z(t_{i})\del\b{z}(t_{i})}\frac{\del\b{z}(t_{i})}{\del z(t_{i})}.
 \label{eq:3.47d}
 \end{align}
\end{subequations}

According to the previous identities, the blocks of the matrix~$\mbb{T}$, defined in appendix~\ref{app:A}, are directly 
related to the second-order derivatives of the classic action.

\subsubsection{Semiclassical Approximation}
\label{sss:3.2.3}

The semiclassical approximation of the quantum propagator basically consists in expanding the action 
functional, for fixed boundary conditions and time interval, up to the second perturbative order around 
a classical trajectory:\footnote{The subscript~$c$ is used to indicate functionals calculated on a classical 
trajectory. However, in order to simplify the notation, we shall omit the subscript~$c$ whenever its absence 
does not cause ambiguities.}
\begin{equation}
 S\approx S_{c}+\frac{1}{2}\delta^{2}S_{c},\quad\delta S_{c}=0.
 \label{eq:3.48}
\end{equation}

The above decomposition corresponds to an application of the \textit{saddle-point method}\cite{Bruijn58} to the evaluation 
of a path integral. It is expected that this asymptotic method produces satisfactory results for sufficiently large values 
of the classical action, that is, for $\left|S_{c}\right|\gg\hbar$.

As discussed in subsection~\ref{sss:3.2.1.add}, the existence of a solution to the classical equations of motion, under 
boundary conditions, is subject to the duplication of the number of degrees of freedom. Therefore, in order to include 
the classical trajectory in the path integral, we need to extend the result~\eqref{eq:3.37} to the duplicated phase 
space. So, in preparation for the expansion~\eqref{eq:3.48}, we rewrite the continuous-time limit of the quantum propagator 
according to the notation~\eqref{eq:3.33}:
\begin{equation}
 \til{K}(z\cg_{f},z_{i};t_{f},t_{i})=\int\rmD\mu(\b{z},z)\exp\left[
 \frac{\rmi}{\hbar}S(z\cg_{f},z_{i};t_{f},t_{i})+\Lambda(z_{f}\cg,z_{i})\right].
 \label{eq:3.49}
\end{equation}

Note that the most evident modification in the above equation, compared with expression~\eqref{eq:3.37}, is present 
in the path measure:
\begin{subequations}
 \label{eq:3.50}
 \begin{align}
 &\rmD\mu(\b{z},z)=\lim_{M\rightarrow\infty}
 \prodl_{j=1}^{M-1}\rmd\mu(\b{z}^{j},z^{j}),
 \label{eq:3.50a}\\[0.0ex]
 &\rmd\mu(\b{z}^{j},z^{j})=\kappa(l)\det\left[g(\b{z}^{j},z^{j})\right]
 \prodl_{k=1}^{d}\frac{\rmd\b{z}^{j}_{k}\rmd z^{j}_{k}}{2\pi\rmi}.
 \label{eq:3.50b}
 \end{align}
\end{subequations}

The insertion of the notation~\eqref{eq:3.33} into the identities~\eqref{eq:3.50} implicitly indicates that, for each value 
of $j$ and $k$, the integration domain is transferred from the complex plane to a two-dimensional surface in the duplicated 
space, which contains a single point of the classical trajectory. This deformation of the integration domain is allowed only 
if the integrand of the propagator constitutes an analytic function of $\b{z}^{j}_{k}$ and $z^{j}_{k}$ in the region between 
the two considered surfaces.

With the purpose of calculating the second variation~$\delta^{2}S_{c}$, we introduce new variables for the path integral:
\begin{subequations}
 \label{eq:3.51}
 \begin{align}
 &\eta(t)=z(t)-z_{c}(t),
 \label{eq:3.51a}\\[1.2ex]
 &\b{\eta}(t)=\b{z}(t)-\b{z}_{c}(t).
 \label{eq:3.51b}
 \end{align}
\end{subequations}

The functions $\b{z}_{c}(t)$ and $z_{c}(t)$ denote a solution of the equations of motion~\eqref{eq:3.3} under the boundary 
conditions~\eqref{eq:3.5}. Consequently, the quantities $\b{\eta}(t)$ and $\eta(t)$ correspond to deviations from the classical 
trajectory. Under the above transformation, the boundary conditions acquire the following form:
\begin{subequations}
 \label{eq:3.52}
 \begin{align}
 &\eta(t_{i})=0,
 \label{eq:3.52a}\\[1.2ex]
 &\b{\eta}(t_{f})=0.
 \label{eq:3.52b}
 \end{align}
\end{subequations}

Substituting the decomposition~\eqref{eq:3.48} into the propagator~\eqref{eq:3.49} and performing 
the transformation~\eqref{eq:3.51} on the integration variables, we obtain the initial expression 
for the \textit{semiclassical propagator}:
\begin{equation}
 K_{sc}(z\cg_{f},z_{i};t_{f},t_{i})=\suml_{\text{traj.}}
 \exp\left[\frac{\rmi}{\hbar}S_{c}(z\cg_{f},z_{i};t_{f},t_{i})
 +\Lambda(z_{f}\cg,z_{i})\right]
 \int\rmD\mu(\b{\eta},\eta)\;\rme^{\frac{\rmi}{2\hbar}\delta^{2}S_{c}}.
 \label{eq:3.53}
\end{equation}

As discussed earlier, the equations of motion~\eqref{eq:3.3}, when subjected to boundary conditions, can result in multiple 
solutions. For this reason, according to the usual prescription of the saddle-point method, it is necessary to insert a summation 
sign in the semiclassical propagator, which indicates the sum of the resulting contributions from every classical trajectory 
specified by identities~\eqref{eq:3.5}.

By applying the transformation~\eqref{eq:3.51} to the equations~\eqref{eq:3.50}, we 
explicitly write the integration element corresponding to the propagator~\eqref{eq:3.53}:
\begin{equation}
 \rmD\mu(\b{\eta},\eta)\approx
 \lim_{M\rightarrow\infty}\prodl_{j=1}^{M-1}\left\{
 \kappa(l)\det\left[g(\b{z}_{c}^{j},z_{c}^{j})\right]
 \prodl_{k=1}^{d}\frac{\rmd\b{\eta}^{j}_{k}\rmd\eta^{j}_{k}}{2\pi\rmi}\right\}.
 \label{eq:3.54}
\end{equation}

Notice that the determinant of the metric is calculated directly on the classical trajectory. This approximation is also 
prescribed by the saddle-point method, under the assumption that the function~$\det\left[g(\b{z}^{j},z^{j})\right]$ varies 
slowly in the region of the integration domain with relevant contribution to the evaluation of the integral.

As a consequence of the result~\eqref{eq:3.54}, the derivation of the semiclassical propagator 
amounts to evaluating a Gaussian path integral, which we designate as the \textit{reduced propagator}:
\begin{equation}
 K_{red}=\int\rmD\mu(\b{\eta},\eta)\exp\left[\frac{\rmi}{2\hbar}\delta^{2}S\right].
 \label{eq:3.55}
\end{equation}

By using the definition~\eqref{eq:3.6a}, we obtain a general expression 
for the second variation of the action around a classical trajectory:
\begin{equation}
 \begin{aligned}
 \delta^{2}S=&\intl_{t_{i}}^{t_{f}}\rmd t
 \left[\eta\frac{\del^{2}L}{\del z^{2}}\eta
 +2\eta\frac{\del^{2}L}{\del z\del\b{z}}\b{\eta}
 +\b{\eta}\frac{\del^{2}L}{\del\b{z}^{2}}\b{\eta}
 +2\eta\frac{\del^{2}L}{\del z\del\dot{z}}\dot{\eta}
 +2\b{\eta}\frac{\del^{2}L}{\del\b{z}\del\dot{\b{z}}}\dot{\b{\eta}}
 \right.\\
 &\left.
 +2\eta\frac{\del^{2}L}{\del z\del\dot{\b{z}}}\dot{\b{\eta}}
 +2\b{\eta}\frac{\del^{2}L}{\del\b{z}\del\dot{z}}\dot{\eta}\right]
 +\eta(t_{f})\frac{\del^{2}\Gamma}{\del z(t_{f})^{2}}\eta(t_{f})
 +\b{\eta}(t_{i})\frac{\del^{2}\Gamma}{\del\b{z}(t_{i})^{2}}\b{\eta}(t_{i})\\
 =&\intl_{t_{i}}^{t_{f}}\rmd t\left\{\eta\left[\frac{\del^{2}L}{\del z^{2}}
 -\frac{\rmd}{\rmd t}\left(\frac{\del^{2}L}{\del z\del\dot{z}}\right)\right]\eta
 +\b{\eta}\left[\frac{\del^{2}L}{\del\b{z}^{2}}
 -\frac{\rmd}{\rmd t}\left(\frac{\del^{2}L}{\del\b{z}\del\dot{\b{z}}}\right)\right]\b{\eta}\right.\\
 &\left.+2\eta\frac{\del^{2}L}{\del z\del\b{z}}\b{\eta}
 +2\eta\frac{\del^{2}L}{\del z\del\dot{\b{z}}}\dot{\b{\eta}}
 +2\b{\eta}\frac{\del^{2}L}{\del\b{z}\del\dot{z}}\dot{\eta}\right\}
 +\left[\eta\frac{\del^{2}L}{\del z\del\dot{z}}\eta\right]_{t_{i}}^{t_{f}}\\
 &+\left[\b{\eta}\frac{\del^{2}L}{\del\b{z}\del\dot{\b{z}}}\b{\eta}\right]_{t_{i}}^{t_{f}}
 +\eta(t_{f})\frac{\del^{2}\Gamma}{\del z(t_{f})^{2}}\eta(t_{f})
 +\b{\eta}(t_{i})\frac{\del^{2}\Gamma}{\del\b{z}(t_{i})^{2}}\b{\eta}(t_{i}),
 \end{aligned}
 \label{eq:3.56}
\end{equation}
\noindent in which we have omitted the second-order derivatives of $L$ and $\Gamma$ with identically zero 
values. Moreover, in formulating the second equality of the above equation, we assume that the Lagrangian 
has the following properties:
\begin{subequations}
 \label{eq:3.57}
 \begin{align}
 &\frac{\del^{2}L}{\del z\del\dot{z}}
 =\left[\frac{\del^{2}L}{\del z\del\dot{z}}\right]\T
 =\frac{\del^{2}L}{\del\dot{z}\del z},
 \label{eq:3.57a}\\[1.2ex]
 &\frac{\del^{2}L}{\del\b{z}\del\dot{\b{z}}}
 =\left[\frac{\del^{2}L}{\del\b{z}\del\dot{\b{z}}}\right]\T
 =\frac{\del^{2}L}{\del\dot{\b{z}}\del\b{z}}.
 \label{eq:3.57b}
 \end{align}
\end{subequations}

Observe that, according to the identities \eqref{eq:3.58e} and \eqref{eq:3.58f} 
in appendix~\ref{app:B}, the assumptions~\eqref{eq:3.57} are absolutely valid. 

Then, by calculating the second-order derivatives in expression~\eqref{eq:3.56},\footnote{The explicit 
calculation of the second derivatives in equation~\eqref{eq:3.56} is given in appendix~\ref{app:B}, 
by identities~(\ref{eq:3.58}\-\ref{eq:3.60}).} we concisely rewrite the second variation of the action 
functional:
\begin{equation}
 \frac{\rmi}{\hbar}\delta^{2}S=\intl_{t_{i}}^{t_{f}}\rmd t
 \left[\eta g(\b{z},z)\dot{\b{\eta}}-\b{\eta}g\T(\b{z},z)\dot{\eta}
 +\eta A\eta+2\eta B\b{\eta}+\b{\eta}C\b{\eta}\right],
 \label{eq:3.61}
\end{equation}
\noindent in which the matrices $A$, $B$ and $C$ are given by
\begin{subequations}
 \label{eq:3.62}
 \begin{align}
 &\begin{aligned}
 A=&\,\frac{\rmi}{\hbar}\left[\frac{\del^{2}L}{\del z^{2}}
 -\frac{\rmd}{\rmd t}\left(\frac{\del^{2}L}{\del z\del\dot{z}}\right)\right]\\
 =&\,\frac{\del}{\del z}\left[g(\b{z},z)\dot{\b{z}}\right]
 -\frac{\rmi}{\hbar}\frac{\del^{2}\mcl{H}(\b{z},z)}{\del z^{2}}\\
 =&-\frac{\rmi}{\hbar}g(\b{z},z)\frac{\del}{\del z}
 \left[\xi(\b{z},z)\frac{\del\mcl{H}(\b{z},z)}{\del z}\right],
 \end{aligned}
 \label{eq:3.62a}\\[1.2ex]
 &\begin{aligned}
 B=&\,\frac{\rmi}{\hbar}\frac{\del^{2}L}{\del z\del\b{z}}\\
 =&\,\frac{1}{2}\left(\dot{\b{z}}\frac{\del}{\del\b{z}}
 -\dot{z}\frac{\del}{\del z}\right)g(\b{z},z)
 -\frac{\rmi}{\hbar}\frac{\del^{2}\mcl{H}(\b{z},z)}{\del z\del\b{z}}\\
 =&-\frac{\rmi}{2\hbar}g(\b{z},z)\frac{\del}{\del\b{z}}
 \left[\xi(\b{z},z)\frac{\del\mcl{H}(\b{z},z)}{\del z}\right]\\
 &-\frac{\rmi}{2\hbar}\left\{\frac{\del}{\del z}\left[
 \xi\T(\b{z},z)\frac{\del\mcl{H}(\b{z},z)}{\del\b{z}}\right]
 \right\}\T g(\b{z},z),
 \end{aligned}
 \label{eq:3.62b}\\[1.2ex]
 &\begin{aligned}
 C=&\,\frac{\rmi}{\hbar}\left[\frac{\del^{2}L}{\del\b{z}^{2}}-\frac{\rmd}{\rmd t}\left(
 \frac{\del^{2}L}{\del\b{z}\del\dot{\b{z}}}\right)\right]\\
 =&-\frac{\del}{\del\b{z}}\left[g\T(\b{z},z)\dot{z}\right]
 -\frac{\rmi}{\hbar}\frac{\del^{2}\mcl{H}(\b{z},z)}{\del\b{z}^{2}}\\
 =&-\frac{\rmi}{\hbar}g\T(\b{z},z)\frac{\del}{\del\b{z}}
 \left[\xi\T(\b{z},z)\frac{\del\mcl{H}(\b{z},z)}{\del\b{z}}\right].
 \end{aligned}
 \label{eq:3.62c}
 \end{align}
\end{subequations}

Note that each of the matrices $A$, $B$ and $C$ receives three different, but completely equivalent, formulations. The first form 
of each matrix is its fundamental definition, which follows directly from identity~\eqref{eq:3.56}. The second representation will 
be employed in the next subsection, during the evaluation of the path integral corresponding to the reduced propagator. The last 
form presented for each of the above matrices proves to be quite useful in explicit calculations of these quantities, considering 
a specific metric and Hamiltonian. The third prescription for $B$ will also be used in subsection~\ref{sss:3.2.6}, in the 
development of the final expression for the semiclassical propagator. Finally, in reference only to the second line of equations 
\eqref{eq:3.62a} and \eqref{eq:3.62c}, we observe that the velocities $\dot{\b{z}}$ and $\dot{z}$ are independent of the variables 
$\b{z}$ and $z$ with respect to the evaluation of derivatives, in accordance with the functional character of the Lagrangian~$L$.

Employing the result~\eqref{eq:3.61}, we reformulate the original expression for the reduced propagator:
\begin{equation}
 K_{red}=\int\rmD\mu(\b{\eta},\eta)\exp\left\{
 \frac{1}{2}\intl_{t_{i}}^{t_{f}}\rmd t\left[
 \eta g(\b{z},z)\dot{\b{\eta}}-\b{\eta}g\T(\b{z},z)\dot{\eta}
 +\eta A\eta+2\eta B\b{\eta}+\b{\eta}C\b{\eta}
 \right]\right\}.
 \label{eq:3.63}
\end{equation}

At this point, it is convenient to introduce a second change of the integration variables, which is described by 
the following linear transformation:
\begin{subequations}
 \label{eq:3.64}
 \begin{align}
 &\nu(t)=\Theta\T(\b{z}_{c}(t),z_{c}(t))\eta(t),
 \label{eq:3.64a}\\[1.2ex]
 &\b{\nu}(t)=\Theta(\b{z}_{c}(t),z_{c}(t))\b{\eta}(t),
 \label{eq:3.64b}
 \end{align}
\end{subequations}
\noindent where the matrix~$\Theta$ is formally related to the analytic continuation of the metric:
\begin{equation}
 \Theta(\b{z},z)=g^{\frac{1}{2}}(\b{z},z).
 \label{eq:3.65}
\end{equation}

The boundary conditions for the new variables follow immediately from equations~\eqref{eq:3.52}:
\begin{subequations}
 \label{eq:3.66}
 \begin{align}
 &\nu(t_{i})=0,
 \label{eq:3.66a}\\[1.2ex]
 &\b{\nu}(t_{f})=0.
 \label{eq:3.66b}
 \end{align}
\end{subequations}

By applying the inverse of the transformation~\eqref{eq:3.64} to the first two terms in the exponent 
of identity~\eqref{eq:3.63}, we obtain the following expression:
\begin{equation}
 \eta g\dot{\b{\eta}}-\b{\eta}g\T\dot{\eta}=
 \nu\dot{\b{\nu}}-\b{\nu}\dot{\nu}
 +\nu(\Theta^{-1}\dot{\Theta}-\dot{\Theta}\Theta^{-1})\b{\nu}.
 \label{eq:3.67}
\end{equation}

Notice that, in the above equation, we use the matrix relation~$\rmd\Theta^{-1}/\rmd t=-\Theta^{-1}\dot{\Theta}\Theta^{-1}$. 
Also note that the matrices $g(\b{z},z)$ and $\Theta(\b{z},z)$ are considered as functions exclusively dependent on time, 
since they are calculated on a classical trajectory. Evidently, this same reasoning applies to all other quantities involved 
in the calculation of the second variation of the action. 

Now, with the aid of identity~\eqref{eq:3.67}, we redefine the other 
terms of $\delta^{2}S$ in terms of the new integration variables:
\begin{subequations}
 \label{eq:3.68}
 \begin{align}
 &\eta A\eta=\nu\Theta^{-1}A(\Theta^{-1})\T\nu=\nu\til{A}\nu,
 \label{eq:3.68a}\\[1.2ex]
 &2\eta B\b{\eta}+\nu(\Theta^{-1}\dot{\Theta}-\dot{\Theta}\Theta^{-1})\b{\nu}=
 2\nu\left[\Theta^{-1}B\Theta^{-1}
 +\frac{1}{2}(\Theta^{-1}\dot{\Theta}-\dot{\Theta}\Theta^{-1})\right]\b{\nu}=2\nu\til{B}\b{\nu},
 \label{eq:3.68b}\\[1.2ex]
 &\b{\eta}C\b{\eta}=\b{\nu}(\Theta^{-1})\T C\Theta^{-1}\b{\nu}=\b{\nu}\til{C}\b{\nu}.
 \label{eq:3.68c}
 \end{align}
\end{subequations}

Then, by inserting the results \eqref{eq:3.67} and \eqref{eq:3.68} into the equation~\eqref{eq:3.61}, 
we rewrite the second variation of the classical action:
\begin{equation}
 \frac{\rmi}{\hbar}\delta^{2}S=
 \intl_{t_{i}}^{t_{f}}\rmd t\left[
 \nu\dot{\b{\nu}}-\b{\nu}\dot{\nu}
 +\nu\til{A}\nu+2\nu\til{B}\b{\nu}+\b{\nu}\til{C}\b{\nu}\right].
 \label{eq:3.69}
\end{equation}

As a consequence of the transformation~\eqref{eq:3.64}, the integration element~\eqref{eq:3.54} takes the following form:
\begin{equation}
 \begin{aligned}
 \rmD\mu(\b{\nu},\nu)&=\lim_{M\rightarrow\infty}\prodl_{j=1}^{M-1}\kappa(l)
 \frac{\rmd\b{\nu}^{j}\rmd\nu^{j}}{(2\pi\rmi)^{d}}\\
 &\approx\lim_{M\rightarrow\infty}\prodl_{j=1}^{M-1}
 \frac{\rmd\b{\nu}^{j}\rmd\nu^{j}}{(2\pi\rmi)^{d}}.
 \end{aligned}
 \label{eq:3.70}
\end{equation}

Note that, in the second line of the above equation, we perform the approximation
\begin{equation}
 \kappa(l)\stackrel{l\gg1}{\approx}1,
 \label{eq:3.71}
\end{equation}
\noindent in order to avoid divergence ($\kappa(l)>1$) or cancellation ($\kappa(l)<1$) of the reduced propagator due to the 
infinite product of normalization factors found in the path measure~\eqref{eq:3.70}. As discussed in subsection~\ref{sss:2.1.4}, 
the constant~$\kappa(l)$ is specified by the same set of indices~$l$ that determines the Hilbert space accessible to the 
system of interest. In general, this collection of indices consists of fundamental quantum numbers, such as angular momentum, 
number of particles and energy. Therefore, the conjecture~$l\gg1$, incorporated in identity~\eqref{eq:3.71}, is equivalent 
to the hypothesis of \textit{large quantum numbers}. This assumption is in agreement with the regime of applicability of 
the semiclassical approximation, since the condition~$l\gg1$ generally implies that $\left|S_{c}\right|\gg\hbar$.

In the case of the canonical coherent states, presented in subsection~\ref{sss:2.2.1}, the factor~$\kappa$ does not depend 
on possible quantum numbers associated with the space~$\mds{B}^{d}$. However, according to identity~\eqref{eq:2.35}, the 
conjecture~\eqref{eq:3.71} is exactly satisfied. Since the set of values~$l$ does not exist, the semiclassical approximation 
corresponding to the coherent states~\eqref{eq:2.31} is not clearly grounded in the hypothesis of large quantum numbers. 
For this reason, we are often compelled to solely rely on the assumption~$|S_{c}|\gg\hbar$, which leads only to superficial 
conclusions.

As indicated by equation~\eqref{eq:2.50}, the spin coherent states satisfy the assumption~\eqref{eq:3.71} for large magnitudes 
of angular momentum. So, in this particular case, it is expected that the semiclassical approximation will be effectively valid 
in the \textit{regime of large spins}. Similarly, the $\mrm{SU}(n)$ bosonic coherent states are expected to exhibit characteristics 
appropriate for the semiclassical dynamics only in the \textit{regime of many particles}, as suggested by identity~\eqref{eq:2.64}. 
Notice that, in the last two examples, the assumption of large quantum numbers has a precise meaning, since we can systematically 
analyze the fundamental properties of the system of interest as a function of the parameters $J_{k}$ and $N$.

The path measure~\eqref{eq:3.70}, after performing the approximation~\eqref{eq:3.71}, assumes the same form of an integration element 
in a flat space, that is, the same result would be readily obtained from equation \eqref{eq:3.54} for $g(\b{z},z)=\mds{1}$. However, 
the curvature of the phase space is still implicitly contained in the classical trajectory and, consequently, in the second variation 
of the action. Therefore, the approximations made in the propagator do not completely remove the geometric properties introduced by 
the coherent states, although the derivation presented in this paper is based on reducing the semiclassical propagator to its specific
algebraic form for a flat space.

Substituting the results~\eqref{eq:3.69} and \eqref{eq:3.70} into the equation~\eqref{eq:3.63}, we 
obtain the reduced propagator, in the continuous-time limit, for the third set of integration variables:
\begin{equation}
 K_{red}=\int\rmD\mu(\b{\nu},\nu)\exp\left\{
 \frac{1}{2}\intl_{t_{i}}^{t_{f}}\rmd t\left[
 \nu\dot{\b{\nu}}-\b{\nu}\dot{\nu}
 +\nu\til{A}\nu+2\nu\til{B}\b{\nu}+\b{\nu}\til{C}\b{\nu}\right]\right\}.
 \label{eq:3.72}
\end{equation}

The above expression for $K_{red}$ is perfectly suitable for solving the Gaussian path integral in terms of well-defined classical 
quantities. However, the next step of the derivation corresponds to a necessary digression, in which we resort to the time-discretized 
formulation of the semiclassical propagator.

\subsubsection{Evaluation of the Reduced Propagator}
\label{sss:3.2.4}

In the present subsection, with the purpose of evaluating the path integral~\eqref{eq:3.72}, we shall resort 
to a time-discretized form of expression~\eqref{eq:3.55}. In this case, it is worth remembering that the 
semiclassical approximation of the propagator~\eqref{eq:3.1} relies on the hypothesis of continuous paths, 
that is, only the paths whose consecutive points differ by an infinitesimal amount are relevant in the 
semiclassical regime.

With the aid of the continuous-path hypothesis, we shall reformulate the time-discretized reduced 
propagator as a sequentially-solvable set of multidimensional Gaussian integrals. As a result of 
the integration procedure, the value of the reduced propagator will be determined by the solution 
of a recurrence relation, whose continuous-time limit will be related to the Jacobi equation in 
subsection~\ref{sss:3.2.5}.

As a first step in solving the path integral~\eqref{eq:3.72}, we rewrite the reduced 
propagator~\eqref{eq:3.55} in a time-discretized form:
\begin{equation}
 K_{red}=\lim_{M\rightarrow\infty}
 \int\left[\prod_{j=1}^{M-1}\rmd\mu(\b{\eta}^{j},\eta^{j})\right]
 \exp\left[\frac{\rmi}{2\hbar}\delta^{2}S_{d}\right].
 \label{eq:3.73}
\end{equation}

At each instant of time, according to the formula~\eqref{eq:3.54}, the integration measure 
is represented by the following expression:
\begin{equation}
 \rmd\mu(\b{\eta}^{j},\eta^{j})\approx
 \kappa(l)\det\!\left[g(\b{z}^{j},z^{j})\right]
 \prodl_{k=1}^{d}\frac{\rmd\b{\eta}^{j}_{k}\rmd\eta^{j}_{k}}{2\pi\rmi}.
 \label{eq:3.74}
\end{equation}

Notice that the derivatives of the exponent~$\til{S}_{d}$, described by equation~\eqref{eq:3.23}, are identical to the derivatives 
of $S_{d}$, since these two quantities differ only by the normalization constant~$\Lambda(z_{f}\cg,z_{i})$, as established by the 
definition~\eqref{eq:3.35}. Therefore, considering only the terms up to the first order in the parameter~$\vep$, we obtain, directly 
from the identities~\eqref{eq:3.28}, the first derivatives of the discretized action functional:
\begin{subequations}
 \label{eq:3.75}
 \begin{align}
 &\frac{\rmi}{\hbar}\frac{\del S_{d}}{\del z^{j}}\approx
 g(\b{z}^{j+1},z^{j})(\b{z}^{j+1}-\b{z}^{j})
 -\frac{\rmi}{\hbar}\frac{\del\mcl{H}(\b{z}^{j+1},z^{j})}{\del z^{j}}\vep,
 \label{eq:3.75a}\\[1.2ex]
 &\frac{\rmi}{\hbar}\frac{\del S_{d}}{\del\b{z}^{j}}\approx
 -g\T(\b{z}^{j},z^{j-1})(z^{j}-z^{j-1})
 -\frac{\rmi}{\hbar}\frac{\del\mcl{H}(\b{z}^{j},z^{j-1})}{\del\b{z}^{j}}\vep,
 \label{eq:3.75b}
 \end{align}
\end{subequations}
\noindent in which we use the proper notation for the duplicated phase space. Then, in order to evaluate the 
second variation~$\delta^{2}S_{d}$, we take the derivatives of the expressions~\eqref{eq:3.75}:
\begin{subequations}
 \label{eq:3.76}
 \begin{align}
 &\frac{\rmi}{\hbar}
 \frac{\del^{2}S_{d}}{{\del z^{j}}^{2}}\approx
 \frac{\del}{\del z^{j}}\left[g(\b{z}^{j+1},z^{j})(\b{z}^{j+1}-\b{z}^{j})\right]
 -\frac{\rmi}{\hbar}\frac{\del^{2}\mcl{H}(\b{z}^{j+1},z^{j})}{{\del z^{j}}^{2}}\vep
 =-D_{j,j},
 \label{eq:3.76a}\\[1.2ex]
 &\frac{\rmi}{\hbar}
 \frac{\del^{2}S_{d}}{{\del\b{z}^{j}}^{2}}\approx
 -\frac{\del}{\del\b{z}^{j}}\left[g\T(\b{z}^{j},z^{j-1})(z^{j}-z^{j-1})\right]
 -\frac{\rmi}{\hbar}\frac{\del^{2}\mcl{H}(\b{z}^{j},z^{j-1})}{{\del\b{z}^{j}}^{2}}\vep
 =-D_{\wb{j},\wb{j}},
 \label{eq:3.76b}\\[1.2ex]
 &\frac{\rmi}{\hbar}\frac{\del^{2}S_{d}}{\del z^{j}\del\b{z}^{j}}\approx
 -g(\b{z}^{j+1},z^{j})
 =-D_{j,\wb{j}},
 \label{eq:3.76c}\\[1.2ex]
 &\frac{\rmi}{\hbar}\frac{\del^{2}S_{d}}{\del\b{z}^{j}\del z^{j}}\approx
 -g\T(\b{z}^{j},z^{j-1})
 =-D_{\wb{j},j}.
 \label{eq:3.76d}
 \end{align}
\end{subequations}

The previous equations are valid only for $j=1,2,\ldots,(M-1)$. In a slightly different way, the remaining second derivatives 
of the discretized action functional are restricted to $j=1,2,\ldots,(M-2)$:
\begin{subequations}
 \label{eq:3.77}
 \begin{align}
 &\frac{\rmi}{\hbar}
 \frac{\del^{2}S_{d}}{\del z^{j}\del\b{z}^{j+1}}\approx
 \frac{\del}{\del\b{z}^{j+1}}\left[
 g(\b{z}^{j+1},z^{j})(\b{z}^{j+1}-\b{z}^{j})\right]
 -\frac{\rmi}{\hbar}
 \frac{\del^{2}\mcl{H}(\b{z}^{j+1},z^{j})}{\del z^{j}\del\b{z}^{j+1}}\vep
 =-D_{j,\wb{j+1}},
 \label{eq:3.77a}\\[1.2ex]
 &\frac{\rmi}{\hbar}
 \frac{\del^{2}S_{d}}{\del\b{z}^{j+1}\del z^{j}}\approx
 -\frac{\del}{\del z^{j}}\left[
 g\T(\b{z}^{j+1},z^{j})(z^{j+1}-z^{j})\right]
 -\frac{\rmi}{\hbar}
 \frac{\del^{2}\mcl{H}(\b{z}^{j+1},z^{j})}{\del\b{z}^{j+1}\del z^{j}}\vep
 =-D_{\wb{j+1},j}.
 \label{eq:3.77b}
 \end{align}
\end{subequations}

By employing the equations \eqref{eq:3.76} and \eqref{eq:3.77}, we write the 
second variation of the discretized action functional around a classical trajectory:
\begin{equation}
 \begin{aligned}
 \frac{\rmi}{\hbar}\delta^{2}S_{d}
 =&\,\frac{\rmi}{\hbar}\suml_{j=1}^{M-1}\left[
 \eta^{j}\frac{\del^{2}S_{d}}{{\del z^{j}}^{2}}\eta^{j}
 +\b{\eta}^{j}\frac{\del^{2}S_{d}}{{\del\b{z}^{j}}^{2}}\b{\eta}^{j}
 +\eta^{j}\frac{\del^{2}S_{d}}{\del z^{j}\del\b{z}^{j}}\b{\eta}^{j}
 +\b{\eta}^{j}\frac{\del^{2}S_{d}}{\del\b{z}^{j}\del z^{j}}\eta^{j}\right]\\
 &+\frac{\rmi}{\hbar}\suml_{j=1}^{M-2}\left[
 \eta^{j}\frac{\del^{2}S_{d}}{\del z^{j}\del\b{z}^{j+1}}\b{\eta}^{j+1}
 +\b{\eta}^{j+1}\frac{\del^{2}S_{d}}{\del\b{z}^{j+1}\del z^{j}}\eta^{j}\right]\\
 \approx&-\suml_{j=1}^{M-1}\left[
 \eta^{j}D_{j,j}\eta^{j}
 +\b{\eta}^{j}D_{\wb{j},\wb{j}}\b{\eta}^{j}
 +\eta^{j}D_{j,\wb{j}}\b{\eta}^{j}
 +\b{\eta}^{j}D_{\wb{j},j}\eta^{j}\right]\\
 &-\suml_{j=1}^{M-2}\left[
 \eta^{j}D_{j,\wb{j+1}}\b{\eta}^{j+1}
 +\b{\eta}^{j+1}D_{\wb{j+1},j}\eta^{j}\right],
 \end{aligned}
 \label{eq:3.78}
\end{equation}
\noindent in which we introduce the time-discretized version of the transformation~\eqref{eq:3.51}:
\begin{subequations}
 \label{eq:3.79}
 \begin{align}
 &\eta^{j}=z^{j}-z^{j}_{c},
 \label{eq:3.79a}\\[1.2ex]
 &\b{\eta}^{j}=\b{z}^{j}-\b{z}^{j}_{c}.
 \label{eq:3.79b}
 \end{align}
\end{subequations}

In order to simplify our future calculations, we can eliminate the terms in the expression~\eqref{eq:3.78}
of order higher than linear in the parameter~$\vep$. For this purpose, we present the following identities:
\begin{subequations}
 \label{eq:3.80}
 \begin{align}
 &D_{j,\wb{j}}\approx g(\b{z}^{j},z^{j})
 +\frac{\del g(\b{z}^{j+1},z^{j})}{\del\b{z}^{j+1}}
 (\b{z}^{j+1}-\b{z}^{j}),
 \label{eq:3.80a}\\[1.2ex]
 &D_{\wb{j},j}\approx g\T(\b{z}^{j},z^{j})
 +\frac{\del g\T(\b{z}^{j},z^{j-1})}{\del z^{j-1}}
 (z^{j-1}-z^{j}),
 \label{eq:3.80b}
 \end{align}
\end{subequations}
\noindent from which we formulate two further approximations:
\begin{subequations}
 \label{eq:3.81}
 \begin{align}
 &\eta^{j}D_{j,\wb{j}}\b{\eta}^{j}+\eta^{j}D_{j,\wb{j+1}}\b{\eta}^{j+1}
 \approx\eta^{j}g(\b{z}^{j},z^{j})\b{\eta}^{j}
 +\eta^{j}\left[\frac{\rmi}{\hbar}
 \frac{\del^{2}\mcl{H}(\b{z}^{j+1},z^{j})}{\del z^{j}\del\b{z}^{j+1}}\vep
 -g(\b{z}^{j+1},z^{j})
 \right]\b{\eta}^{j+1},
 \label{eq:3.81a}\\[1.2ex]
 &\b{\eta}^{j}D_{\wb{j},j}\eta^{j}+\b{\eta}^{j}D_{\wb{j},j-1}\eta^{j-1}
 \approx\b{\eta}^{j}g\T(\b{z}^{j},z^{j})\eta^{j}
 +\b{\eta}^{j}\left[\frac{\rmi}{\hbar}
 \frac{\del^{2}\mcl{H}(\b{z}^{j},z^{j-1})}{\del\b{z}^{j}\del z^{j-1}}\vep
 -g\T(\b{z}^{j},z^{j-1})
 \right]\eta^{j-1},
 \label{eq:3.81b}
 \end{align}
\end{subequations}
\noindent which are also valid only up to the first order in $\vep$. In the above equations, we assume that the values 
of $|\b{\eta}^{j+1}-\b{\eta}^{j}|$ and $|\eta^{j}-\eta^{j-1}|$ are linearly proportional to $\vep$ in the continuous-time limit, 
when considering only the relevant paths for the semiclassical calculation of the integral~\eqref{eq:3.73}. In other 
words, we use again the hypothesis of continuous paths, discussed in subsection~\ref{sss:3.2.1}.

Note that the approximations~\eqref{eq:3.81} are well defined at the time instants $j=1$ and $j=(M-1)$, in which the 
quantities $D_{\wb{j},j-1}$ and $D_{j,\wb{j+1}}$, respectively, are not established by equations~\eqref{eq:3.77}. 
In these cases, in accordance with the boundary conditions~\eqref{eq:3.52}, we can make use of the identities 
$\eta^{0}=0$ and $\b{\eta}^{M}=0$, so that the occurrence of indefinite quantities corresponds only to the 
addition of null terms. 

With the purpose of simplifying the implementation of the above results, it becomes convenient to reformulate 
the notation:
\begin{subequations}
 \label{eq:3.82}
 \begin{align}
 &D'_{j,\wb{j}}=g(\b{z}^{j},z^{j}),
 \label{eq:3.82a}\\[1.2ex]
 &D'_{j,\wb{j+1}}=\frac{\rmi}{\hbar}
 \frac{\del^{2}\mcl{H}(\b{z}^{j+1},z^{j})}{\del z^{j}\del\b{z}^{j+1}}\vep
 -g(\b{z}^{j+1},z^{j}).
 \label{eq:3.82b}
 \end{align}
\end{subequations}

In terms of these new definitions, we rewrite the second variation of the action in its time-discretized form:
\begin{equation}
 \frac{\rmi}{\hbar}\delta^{2}S_{d}\approx
 -\suml_{j=1}^{M-1}\left[
 \eta^{j}D_{j,j}\eta^{j}
 +\b{\eta}^{j}D_{\wb{j},\wb{j}}\b{\eta}^{j}
 +2\eta^{j}D'_{j,\wb{j}}\b{\eta}^{j}\right]
 -2\suml_{j=1}^{M-2}
 \eta^{j}D'_{j,\wb{j+1}}\b{\eta}^{j+1}.
 \label{eq:3.83}
\end{equation}

Now, in analogy to equation~\eqref{eq:3.64}, we perform an appropriate transformation of the integration variables:
\begin{subequations}
 \label{eq:3.84}
 \begin{align}
 &\nu^{j}=\Theta\T(\b{z}^{j}_{c},z^{j}_{c})\eta^{j}=\Theta\T_{j}\eta^{j},
 \label{eq:3.84a}\\[1.2ex]
 &\b{\nu}^{j}=\Theta(\b{z}^{j}_{c},z^{j}_{c})\b{\eta}^{j}=\Theta_{j}\b{\eta}^{j},
 \label{eq:3.84b}
 \end{align}
\end{subequations}
\noindent in which we reintroduce the matrix~$\Theta(\b{z}^{j},z^{j})=g^{\frac{1}{2}}(\b{z}^{j},z^{j})$, 
originally defined by identity~\eqref{eq:3.65}. Similarly to the result~\eqref{eq:3.70}, the integration 
measure~\eqref{eq:3.74} takes the following form:
\begin{equation}
 \rmd\mu(\b{\nu}^{j},\nu^{j})\approx\prodl_{k=1}^{d}\frac{\rmd\b{\nu}^{j}_{k}\rmd\nu^{j}_{k}}{2\pi\rmi}.
 \label{eq:3.85}
\end{equation}

Notice that, in the above identity, we again employ the conjecture~\eqref{eq:3.71}, which is inherently associated 
with the semiclassical approximation. Then, by applying the transformation~\eqref{eq:3.84} to equation~\eqref{eq:3.83}, 
we obtain the final expression for the second variation of the discretized action:
\begin{equation}
 \frac{\rmi}{\hbar}\delta^{2}S_{d}
 \approx-\suml_{j=1}^{M-1}
 \left[\nu^{j}\til{D}_{j,j}\nu^{j}
 +\b{\nu}^{j}\til{D}_{\wb{j},\wb{j}}\b{\nu}^{j}
 +2\nu^{j}\b{\nu}^{j}\right]
 -2\suml_{j=1}^{M-2}
 \nu^{j}\til{D}_{j,\wb{j+1}}\b{\nu}^{j+1},
 \label{eq:3.86}
\end{equation}
\noindent in which we define the following matrix quantities:
\begin{subequations}
 \label{eq:3.87}
 \begin{align}
 &\til{D}_{j,j}=\Theta_{j}^{-1}D_{j,j}(\Theta\T_{j})^{-1},
 \label{eq:3.87a}\\[1.2ex]
 &\til{D}_{\wb{j},\wb{j}}=(\Theta\T_{j})^{-1}D_{\wb{j},\wb{j}}\Theta_{j}^{-1},
 \label{eq:3.87b}\\[1.2ex]
 &\til{D}_{j,\wb{j+1}}=\Theta_{j}^{-1}D'_{j,\wb{j+1}}\Theta_{j+1}^{-1}.
 \label{eq:3.87c}
 \end{align}
\end{subequations}

Next, by inserting the identities \eqref{eq:3.85} and \eqref{eq:3.86} into the reduced propagator~\eqref{eq:3.73}, 
we write a general expression for the Gaussian integral corresponding to each instant of time:
\begin{equation}
 \mcl{I}_{j}=\int\frac{\rmd\b{\nu}^{j}\rmd\nu^{j}}{(2\pi\rmi)^{d}}
 \exp\left[\frac{1}{2}\left(\begin{array}{c c}\b{\nu}^{j}&\nu^{j}\end{array}\right)
 G^{j}\left(\begin{array}{c}\b{\nu}^{j}\\ \nu^{j}\end{array}\right)
 +\left(\begin{array}{c c}\b{\nu}^{j}&\nu^{j}\end{array}\right)V^{j}\right].
 \label{eq:3.88}
\end{equation}

The matrix~$G^{j}$ cannot be immediately determined, since its elements explicitly depend on the integration at the time~$(j-1)$. 
That is, the dynamical character of the reduced propagator is encoded in $G^{j}$, 
whose specific expression remains undefined, except for its initial value:
\begin{equation}
 G^{1}=-\left(\begin{array}{c c}
 \til{D}_{\wb{1},\wb{1}}&\mds{1}\\
 \mds{1}&\til{D}_{1,1}
 \end{array}\right).
 \label{eq:3.89}
\end{equation}

Unlike the matrix~$G^{j}$, the vector~$V^{j}$ is independent of the preceding instants of time:
\begin{equation}
 V^{j}=-\left(
 \begin{array}{c}
 0\\ \til{D}_{j,\wb{j+1}}\b{\nu}^{j+1}
 \end{array}\right).
 \label{eq:3.90}
\end{equation}

The integral~\eqref{eq:3.88} can be solved under the following transformation of variables:
\begin{equation}
 \beta^{j}-\gamma^{j}=
 -\rmi\left(\begin{array}{c}
 \b{\nu}^{j}\\ \nu^{j}
 \end{array}\right),
 \label{eq:3.91}
\end{equation}
\noindent where $\gamma^{j}$ represents a constant vector. By employing the above equation, we promptly rewrite 
the identity~\eqref{eq:3.88} as
\begin{equation}
 \mcl{I}_{j}=\int\frac{\rmi^{d}\rmd\beta^{j}}{(2\pi)^{d}}
 \exp\!\left[-\frac{1}{2}\beta^{j}G^{j}\beta^{j}
 +\frac{1}{2}\gamma^{j}G^{j}\beta^{j}
 +\frac{1}{2}\beta^{j}G^{j}\gamma^{j}
 -\frac{1}{2}\gamma^{j}G^{j}\gamma^{j}
 +\rmi\beta^{j}V^{j}-\rmi\gamma^{j}V^{j}\right].
 \label{eq:3.92}
\end{equation}

Then, in order to simplify the exponent in the integrand of expression~\eqref{eq:3.92}, 
we require the terms linear in $\beta^{j}$ to cancel each other:
\begin{equation}
 \frac{1}{2}\gamma^{j}G^{j}\beta^{j}
 +\frac{1}{2}\beta^{j}G^{j}\gamma^{j}
 +\rmi\beta^{j}V^{j}=0.
 \label{eq:3.93}
\end{equation}

In this way, noting that $G^{j}$ constitutes a symmetric matrix by construction, 
we determine the value of the constant~$\gamma^{j}$ at each instant of time:
\begin{equation}
 G^{j}\gamma^{j}=-\rmi V^{j}
 =\rmi\left(\begin{array}{c}
 0\\ \til{D}_{j,\wb{j+1}}\b{\nu}^{j+1}
 \end{array}\right).
 \label{eq:3.94}
\end{equation}

With the aid of the above result, we obtain a new formulation for the integral~\eqref{eq:3.92}:
\begin{equation}
 \mcl{I}_{j}=\int\frac{\rmi^{d}\rmd\beta^{j}}{(2\pi)^{d}}
 \exp\left[-\frac{1}{2}\beta^{j}G^{j}\beta^{j}
 +\frac{1}{2}\gamma^{j}G^{j}\gamma^{j}\right].
 \label{eq:3.95}
\end{equation}

Using again equation~\eqref{eq:3.94}, we find the following identity:
\begin{equation}
 \begin{aligned}
 \gamma^{j}G^{j}\gamma^{j}
 =&-V^{j}(G^{j})^{-1}V^{j}\\
 =&-\left(\begin{array}{c c}
 0&\b{\nu}^{j+1}\til{D}\T_{j,\wb{j+1}}\end{array}\right)
 (G^{j})^{-1}\left(\begin{array}{c}
 0\\ \til{D}_{j,\wb{j+1}}\b{\nu}^{j+1}\end{array}\right)\\
 =&-\b{\nu}^{j+1}\til{D}\T_{j,\wb{j+1}}[(G^{j})^{-1}]_{22}
 \til{D}_{j,\wb{j+1}}\b{\nu}^{j+1},
 \end{aligned}
 \label{eq:3.96}
\end{equation}
\noindent in which we assume that the matrix~$G^{j}$ is invertible. By employing the above equation along with 
the third fundamental hypothesis described in subsection~\ref{sss:2.1.5}, we calculate the Gaussian integral 
corresponding to a single instant of time in the reduced propagator:
\begin{equation}
 \begin{aligned}
 \mcl{I}_{j}=&\int\frac{\rmi^{d}\rmd\beta^{j}}{(2\pi)^{d}}
 \exp\left\{-\frac{1}{2}\beta^{j}G^{j}\beta^{j}
 -\frac{1}{2}\b{\nu}^{j+1}\til{D}\T_{j,\wb{j+1}}[(G^{j})^{-1}]_{22}
 \til{D}_{j,\wb{j+1}}\b{\nu}^{j+1}\right\}\\
 =&\;\rmi^{d}\left\{\det G^{j}\right\}^{-\frac{1}{2}}
 \exp\left\{-\frac{1}{2}\b{\nu}^{j+1}\til{D}\T_{j,\wb{j+1}}[(G^{j})^{-1}]_{22}
 \til{D}_{j,\wb{j+1}}\b{\nu}^{j+1}\right\}.
 \end{aligned}
 \label{eq:3.97}
\end{equation}

Then, by performing a comparison between the identities \eqref{eq:3.86}, \eqref{eq:3.88} and \eqref{eq:3.97}, 
we determine a recurrence relation for the matrices~$G^{j}$:
\begin{equation}
 G^{j+1}=-\left(\begin{array}{c c}
 \til{D}_{\wb{j+1},\wb{j+1}}
 +\til{D}\T_{j,\wb{j+1}}[(G^{j})^{-1}]_{22}\til{D}_{j,\wb{j+1}}
 &\mds{1}\\
 \mds{1}&\til{D}_{j+1,j+1}
 \end{array}\right),
 \label{eq:3.98}
\end{equation}
\noindent which is valid for $j=1,2,\ldots,(M-2)$. Notice that only the block~$G^{j}_{11}$ is explicitly modified by the preceding 
time instants. Therefore, we can establish a general form for $G^{j}$:
\begin{equation}
 G^{j}=\left(
 \begin{array}{c c}
 G^{j}_{11}&G^{j}_{12}\\
 G^{j}_{21}&G^{j}_{22}
 \end{array}\right)
 =\left(\begin{array}{c c}
 G^{j}_{11}&-\mds{1}\\
 -\mds{1}&-\til{D}_{j,j}
 \end{array}\right),
 \label{eq:3.99}
\end{equation}
\noindent which is defined for $j=1,2,\ldots,(M-1)$. Also note that a single block of the inverse matrix~$(G^{j})^{-1}$ is relevant 
in evaluating the integral~\eqref{eq:3.97}. Thus, it becomes opportune to introduce the following formal representation:
\begin{equation}
 \begin{aligned}
 (G^{j})^{-1}&=\left(
 \begin{array}{c c}
 {[(G^{j})^{-1}]}_{11}&{[(G^{j})^{-1}]}_{12}\\[0.2ex]
 {[(G^{j})^{-1}]}_{21}&{[(G^{j})^{-1}]}_{22}
 \end{array}\right)\\
 &=-\left(\begin{array}{c c}
 (\mds{1}-G^{j}_{22}G^{j}_{11})^{-1}G^{j}_{22}&
 (\mds{1}-G^{j}_{22}G^{j}_{11})^{-1}\\[0.5ex]
 (\mds{1}-G^{j}_{11}G^{j}_{22})^{-1}&
 (\mds{1}-G^{j}_{11}G^{j}_{22})^{-1}G^{j}_{11}
 \end{array}\right).
 \end{aligned}
 \label{eq:3.100}
\end{equation}

By using the identities \eqref{eq:3.99} and \eqref{eq:3.100}, we can extract all the dynamical information 
contained in the recurrence relation~\eqref{eq:3.98}:
\begin{equation}
 G^{j+1}_{11}=-\til{D}_{\wb{j+1},\wb{j+1}}+\til{D}\T_{j,\wb{j+1}}
 (\mds{1}+G^{j}_{11}\til{D}_{j,j})^{-1}G^{j}_{11}\til{D}_{j,\wb{j+1}}.
 \label{eq:3.101}
\end{equation}

The partitioning of the matrix~$G^{j}$ into blocks also enables a simplified description of its determinant:
\begin{equation}
 \begin{aligned}
 \det G^{j}
 =&\,\det\left(\begin{array}{c c}
 -G^{j}_{21}&-G^{j}_{22}\\
 G^{j}_{11}&G^{j}_{12}
 \end{array}\right)\\
 =&\;(-1)^{d}\det\left(\begin{array}{c c}
 \mds{1}&\til{D}_{jj}\\
 -G^{j}_{11}&\mds{1}
 \end{array}\right)\\
 =&\;(-1)^{d}\det\left(\begin{array}{c c}
 \mds{1}+\til{D}_{jj}G^{j}_{11}&0\\
 -G^{j}_{11}&\mds{1}
 \end{array}\right)\\
 =&\;(-1)^{d}\det(
 \mds{1}+\til{D}_{jj}G^{j}_{11}).
 \end{aligned}
 \label{eq:3.102}
\end{equation}

Consequently, we can rewrite the result~\eqref{eq:3.97} solely as a function of the block~$G^{j}_{11}$:
\begin{equation}
 \mcl{I}_{j}=\left[\det(\mds{1}+\til{D}_{jj}G^{j}_{11})\right]^{-\frac{1}{2}}
 \exp\left[\frac{1}{2}\b{\nu}^{j+1}\til{D}_{\wb{j+1}j}
 (\mds{1}+G^{j}_{11}\til{D}_{jj})^{-1}G^{j}_{11}
 \til{D}_{j\wb{j+1}}\b{\nu}^{j+1}\right].
 \label{eq:3.103}
\end{equation}

According to the above identity, the propagator~\eqref{eq:3.73} is completely determined 
by solving the equation~\eqref{eq:3.101}:
\begin{equation}
 K_{red}=\lim_{M\rightarrow\infty}\prodl_{j=1}^{M-1}
 \left[\det(\mds{1}+\til{D}_{jj}G^{j}_{11})
 \right]^{-\frac{1}{2}}.
 \label{eq:3.104}
\end{equation}

Notice that, during the formulation of the previous expression, we made use of the boundary 
condition~$\b{\nu}^{M}=0$, corresponding to the discretized version of equation~\eqref{eq:3.66b}.

As the next step in the derivation of the semiclassical propagator, we need to solve the recurrence 
relation for the block~$G^{j}_{11}$ in terms of familiar classical quantities. For this purpose, 
we must perform the continuous-time limit of all constituents of identity~\eqref{eq:3.101}. First, 
by employing the equations \eqref{eq:3.62a} and \eqref{eq:3.62c}, we calculate the continuous forms 
of the definitions \eqref{eq:3.87a} and \eqref{eq:3.87b}:
\begin{subequations}
 \label{eq:3.105}
 \begin{align}
 &\begin{aligned}
 \lim_{\vep\rightarrow0}\frac{\til{D}_{j,j}}{\vep}
 =&\,\lim_{\vep\rightarrow0}\,\Theta^{-1}_{j}\left\{
 -\frac{\del}{\del z^{j}}\left[g(\b{z}^{j+1},z^{j})\frac{(\b{z}^{j+1}-\b{z}^{j})}{\vep}\right]
 +\frac{\rmi}{\hbar}\frac{\del^{2}\mcl{H}(\b{z}^{j+1},z^{j})}{{\del z^{j}}^{2}}
 \right\}(\Theta\T_{j})^{-1}\\
 =&\;\Theta^{-1}\left\{-\frac{\del}{\del z}\left[
 g(\b{z},z)\dot{\b{z}}\right]+\frac{\rmi}{\hbar}
 \frac{\del^{2}\mcl{H}(\b{z},z)}{{\del z}^{2}}
 \right\}(\Theta\T)^{-1}\\
 =&-\Theta^{-1}A(\Theta\T)^{-1}=-\til{A},
 \end{aligned}
 \label{eq:3.105a}\\[1.2ex]
 &\begin{aligned}
 \lim_{\vep\rightarrow0}\frac{\til{D}_{\wb{j},\wb{j}}}{\vep}
 =&\,\lim_{\vep\rightarrow0}\,(\Theta\T_{j})^{-1}\left\{\frac{\del}{\del\b{z}^{j}}
 \left[g\T(\b{z}^{j},z^{j-1})\frac{(z^{j}-z^{j-1})}{\vep}\right]
 +\frac{\rmi}{\hbar}\frac{\del^{2}\mcl{H}(\b{z}^{j},z^{j-1})}{{\del\b{z}^{j}}^{2}}
 \right\}\Theta^{-1}_{j}\\
 =&\;(\Theta\T)^{-1}\left\{
 \frac{\del}{\del\b{z}}\left[g\T(\b{z},z)\dot{z}\right]
 +\frac{\rmi}{\hbar}\frac{\del^{2}\mcl{H}(\b{z},z)}{{\del\b{z}}^{2}}
 \right\}\Theta^{-1}\\
 =&-(\Theta\T)^{-1}C\Theta^{-1}=-\til{C}.
 \end{aligned}
 \label{eq:3.105b}
 \end{align}
\end{subequations}

Due to the relative complexity in taking the continuous limit of identity~\eqref{eq:3.87c}, we rewrite 
this expression in a less concise manner:
\begin{equation}
 \til{D}_{j,\wb{j+1}}=\Theta^{-1}_{j}\left[-g(\b{z}^{j+1},z^{j})
 +\frac{\rmi}{\hbar}\frac{\del^{2}\mcl{H}(\b{z}^{j+1},z^{j})}{\del z^{j}\del\b{z}^{j+1}}
 \vep\right]\Theta^{-1}_{j+1}.
 \label{eq:3.106}
\end{equation}

Then, retaining only the terms up to the first order in $\vep$, we establish the following matrix relation:
\begin{equation}
 \begin{aligned}
 \Omega_{j,\wb{j+1}}=&\;\Theta^{-1}(\b{z}^{j},z^{j})g(\b{z}^{j+1},z^{j})\Theta^{-1}(\b{z}^{j+1},z^{j+1})\\
 \approx&\;\mds{1}+\left\{\left[(\b{z}^{j}-\b{z}^{j+1})\frac{\del}{\del\b{z}^{j+1}}\right]
 \Theta^{-1}(\b{z}^{j+1},z^{j})\right\}g(\b{z}^{j+1},z^{j})\Theta^{-1}(\b{z}^{j+1},z^{j})\\
 &+\Theta^{-1}(\b{z}^{j+1},z^{j})g(\b{z}^{j+1},z^{j})\left\{\left[
 (z^{j+1}-z^{j})\frac{\del}{\del z^{j}}\right]\Theta^{-1}(\b{z}^{j+1},z^{j})\right\},
 \end{aligned}
 \label{eq:3.107}
\end{equation}
\noindent from which we obtain the continuous-time limit associated with the auxiliary quantity~$\Omega_{j,\wb{j+1}}$:
\begin{equation}
 \begin{aligned}
 \lim_{\vep\rightarrow0}\frac{(\Omega_{j,\wb{j+1}}-\mds{1})}{\vep}=&
 -\left[\left(\dot{\b{z}}\frac{\del}{\del\b{z}}\right)\Theta^{-1}(\b{z},z)\right]g(\b{z},z)\Theta^{-1}(\b{z},z)\\
 &+\Theta^{-1}(\b{z},z)g(\b{z},z)\left[\left(\dot{z}\frac{\del}{\del z}\right)\Theta^{-1}(\b{z},z)\right]\\
 =&\;\frac{1}{2}\Theta^{-1}(\b{z},z)\left[\left(\dot{\b{z}}\frac{\del}{\del\b{z}}
 -\dot{z}\frac{\del}{\del z}\right)g(\b{z},z)\right]\Theta^{-1}(\b{z},z)\\
 &+\frac{1}{2}\left[\Theta^{-1}(\b{z},z)\dot{\Theta}(\b{z},z)-\dot{\Theta}(\b{z},z)\Theta^{-1}(\b{z},z)\right].
 \end{aligned}
 \label{eq:3.108}
\end{equation}

In the transition to the second equality of the above result, we performed several algebraic manipulations on the derivatives 
of the matrices $\Theta(\b{z},z)$ and $g(\b{z},z)$. Now, using the identity~\eqref{eq:3.108} in comparison with the definitions 
\eqref{eq:3.62b} and \eqref{eq:3.68b}, we find the continuous form for the expression~\eqref{eq:3.106}:
\begin{equation}
 \lim_{\vep\rightarrow0}\frac{(\til{D}_{j,\wb{j+1}}+\mds{1})}{\vep}=
 \Theta^{-1}(\b{z},z)\frac{\rmi}{\hbar}\frac{\del^{2}\mcl{H}(\b{z},z)}{\del z\del\b{z}}\Theta^{-1}(\b{z},z)
 -\lim_{\vep\rightarrow0}\frac{(\Omega_{j,\wb{j+1}}-\mds{1})}{\vep}
 =-\til{B}.
 \label{eq:3.109}
\end{equation}

The linear behavior of the recurrence relation~\eqref{eq:3.101} with respect to the parameter~$\vep$ follows directly 
from equations \eqref{eq:3.105} and \eqref{eq:3.109}:
\begin{equation}
 \begin{aligned}
 G_{11}^{j+1}\approx&\;
 G_{11}(t_{j})+\dot{G}_{11}(t_{j})\vep\\
 \approx&\;\til{C}(t_{j})\vep
 +[-\mds{1}-\til{B}\T(t_{j})\vep]
 [\mds{1}-G_{11}(t_{j})\til{A}(t_{j})\vep]^{-1}
 G_{11}(t_{j})
 [-\mds{1}-\til{B}(t_{j})\vep]\\
 \approx&\;\til{C}(t_{j})\vep
 +[\mds{1}+\til{B}\T(t_{j})\vep]
 [\mds{1}+G_{11}(t_{j})\til{A}(t_{j})\vep]
 G_{11}(t_{j})
 [\mds{1}+\til{B}(t_{j})\vep]\\
 \approx&\;G_{11}(t_{j})
 +\left[\til{C}(t_{j})
 +\til{B}\T(t_{j})G_{11}(t_{j})
 +G_{11}(t_{j})\til{A}(t_{j})G_{11}(t_{j})
 +G_{11}(t_{j})\til{B}(t_{j})\right]\vep.
 \end{aligned}
 \label{eq:3.110}
\end{equation}

Therefore, by taking the continuous-time limit of the above identity, the block~$G_{11}(t)$ becomes the solution 
of a first-order differential equation:
\begin{equation}
 \dot{G}_{11}(t)=\til{C}(t)+\til{B}\T(t)G_{11}(t)
 +G_{11}(t)\til{A}(t)G_{11}(t)+G_{11}(t)\til{B}(t),
 \label{eq:3.111}
\end{equation}
\noindent whose initial condition results from expression~\eqref{eq:3.89} calculated at $\vep=0$:
\begin{equation}
 G_{11}(t_{i})=0.
 \label{eq:3.112}
\end{equation}

Employing again the equation~\eqref{eq:3.105a}, we rewrite the reduced propagator~\eqref{eq:3.104} in its continuous form:
\begin{equation}
 \begin{aligned}
 \ln K_{red}=&
 \lim_{M\rightarrow\infty}-\frac{1}{2}\suml_{j=1}^{M-1}\ln\left[\det(\mds{1}+\til{D}_{jj}G^{j}_{11})\right]
 =\lim_{M\rightarrow\infty}-\frac{1}{2}\suml_{j=1}^{M-1}\tr\left[\ln(\mds{1}+\til{D}_{jj}G^{j}_{11})\right]\\
 =&\lim_{M\rightarrow\infty}-\frac{1}{2}\suml_{j=1}^{M-1}\tr[\til{D}_{jj}G^{j}_{11}]
 =\frac{1}{2}\intl_{t_{i}}^{t_{f}}\tr[\til{A}(t)G_{11}(t)]\rmd t.
 \end{aligned}
 \label{eq:3.113}
\end{equation}

With the purpose of establishing a connection between the above results and the apparent digression 
presented in the next subsection, we introduce three successive transformations to be applied on 
equation~\eqref{eq:3.111}:
\begin{subequations}
 \label{eq:3.114}
 \begin{align}
 &F(t)=-\til{A}(t)G_{11}(t),
 \label{eq:3.114a}\\[1.2ex]
 &\dot{X}(t)X^{-1}(t)=\mcl{U}^{-1}(t)F(t)\mcl{U}(t),
 \label{eq:3.114b}\\[1.2ex]
 &Y(t)=\mcl{U}(t)X(t).
 \label{eq:3.114c}
 \end{align}
\end{subequations}

The matrices $F(t)$, $X(t)$ and $Y(t)$ represent the new dynamical variables, 
while $\mcl{U}(t)$ constitutes a time-ordered exponential of a predetermined 
classical function:
\begin{equation}
 \mcl{U}(t)=\hat{T}\exp\!\left[-\intl_{t_{i}}^{t}\til{B}(t')\rmd t'\right].
 \label{eq:3.115}
\end{equation}

As a consequence of the above definition, the quantity~$\mcl{U}(t)$ satisfies the following identity:
\begin{equation}
 \dot{\mcl{U}}(t)=-\til{B}(t)\mcl{U}(t).
 \label{eq:3.116}
\end{equation}

First, by using the transformation~\eqref{eq:3.114a}, we reformulate the differential equation~\eqref{eq:3.111} 
in terms of the matrix~$F(t)$:
\begin{equation}
 \begin{aligned}
 \dot{F}=&-\dot{\til{A}}G_{11}-\til{A}\dot{G}_{11}\\
 =&-F^{2}+(\dot{\til{A}}\til{A}^{-1}+\til{A}\til{B}\T\til{A}^{-1})F
 +F\til{B}-\til{A}\til{C}.
 \end{aligned}
 \label{eq:3.117}
\end{equation}

Then, with the aid of identities~\eqref{eq:3.114b} and \eqref{eq:3.116}, we perform a second substitution 
of the dynamical variables:
\begin{equation}
 \mcl{U}\ddot{X}=
 (\dot{\til{A}}\til{A}^{-1}+\til{A}\til{B}\T\til{A}^{-1}
 +\til{B})\mcl{U}\dot{X}
 -\til{A}\til{C}\mcl{U}X.
 \label{eq:3.118}
\end{equation}

Finally, by employing the expression~\eqref{eq:3.114c}, we rewrite the above equation as a function of $Y(t)$:
\begin{equation}
 \ddot{Y}=(\til{A}\til{B}\T\til{A}^{-1}+\dot{\til{A}}\til{A}^{-1}-\til{B})\dot{Y}
 +(\til{A}\til{B}\T\til{A}^{-1}\til{B}-\til{A}\til{C}+\dot{\til{A}}\til{A}^{-1}
 \til{B}-\dot{\til{B}})Y.
 \label{eq:3.119}
\end{equation}

In agreement with the result~\eqref{eq:3.112} and the transformations~\eqref{eq:3.114}, we obtain the initial conditions 
corresponding to the new sets of variables:
\begin{subequations}
 \label{eq:3.120}
 \begin{align}
 &F(t_{i})=0,
 \label{eq:3.120a}\\[1.2ex]
 &\dot{X}(t_{i})=0,
 \; X(t_{i})=\mds{1},
 \label{eq:3.120b}\\[1.2ex]
 &\dot{Y}(t_{i})=-\til{B}(t_{i}),
 \; Y(t_{i})=\mds{1}.
 \label{eq:3.120c}
 \end{align}
\end{subequations}

Note that, due to the introduction of the matrix variable~$X(t)$, the original differential equation acquires second-order 
time derivatives. At first, this change causes an indetermination, since the number of initial conditions required for an 
unambiguous specification of a solution is doubled. This complication is appropriately eliminated by choosing the initial 
value $X(t_{i})=\mds{1}$, which enables a direct comparison between the identity~\eqref{eq:3.120c} and the subsequent results 
of the present section.

As a preparation for the next stages of this derivation, we perform the transformations~\eqref{eq:3.114} 
on the expression~\eqref{eq:3.113}:
\begin{equation}
 \begin{aligned}
 \ln K_{red}=&
 -\frac{1}{2}\intl_{t_{i}}^{t_{f}}\rmd t\,\tr F(t)
 =-\frac{1}{2}\intl_{t_{i}}^{t_{f}}\rmd t\,\tr\!\left[\dot{X}(t)X^{-1}(t)\right]\\
 =&-\frac{1}{2}\tr\left[\ln X(t_{f})\right]
 =-\frac{1}{2}\ln\left[\det X(t_{f})\right]\\
 =&-\frac{1}{2}\ln\left\{
 \det\!\left[\hat{T}\exp\left(\intl_{t_{i}}^{t_{f}}\rmd t\,\til{B}(t)\right)\right]
 \det Y(t_{f})\right\}\\
 =&\ln\left\{\exp\!\left[-\frac{1}{2}\intl_{t_{i}}^{t_{f}}\rmd t\,\tr\til{B}(t)\right]
 \left[\det Y(t_{f})\right]^{-\frac{1}{2}}\right\}.
 \end{aligned}
 \label{eq:3.121}
\end{equation}

\subsubsection{Jacobi Equation}
\label{sss:3.2.5}

In the previous subsection, we have explicitly evaluated the path integral corresponding to the reduced propagator 
in the continuous-time limit. However, the complete determination of $K_{red}$ still depends on the resolution of 
a second-order differential equation, whose dynamical variables do not yet have an evident connection with usual 
classical quantities. So, in order to establish a physical interpretation for the matrix~$Y(t)$, we now make a 
small digression on the Jacobi equation.

The second variation of the action, in the form presented by equation~\eqref{eq:3.69}, constitutes a functional 
of the path described by the variables $\b{\nu}(t)$ and $\nu(t)$. With the purpose of introducing a suitable 
notation for the manipulation of $\delta^{2}S$, we define the following Lagrangian function:
\begin{equation}
 \mcl{L}(\dot{\b{\nu}},\dot{\nu},\b{\nu},\nu;t)=
 \dot{\b{\nu}}\nu-\b{\nu}\dot{\nu}
 +\nu\til{A}\nu+2\nu\til{B}\b{\nu}+\b{\nu}\til{C}\b{\nu}.
 \label{eq:3.122}
\end{equation}

Similarly to the variables $\b{\eta}(t)$ and $\eta(t)$, we can interpret $\b{\nu}(t)$ and $\nu(t)$ as deviations 
from the classical trajectory. Although the values of these quantities are completely arbitrary in the evaluation 
of the path integral~\eqref{eq:3.72}, there must be a set of particular solutions that extremize the second variation 
of the action. In order to identify the extremizing trajectories, we calculate the first variation of $\delta^{2}S$:
\begin{equation}
 \begin{aligned}
 \frac{\rmi}{\hbar}\delta[\delta^{2}S]=&
 \intl_{t_{i}}^{t_{f}}\left[
 \frac{\del\mcl{L}}{\del\nu}\delta\nu
 +\frac{\del\mcl{L}}{\del\dot{\nu}}\delta\dot{\nu}
 +\frac{\del\mcl{L}}{\del\b{\nu}}\delta\b{\nu}
 +\frac{\del\mcl{L}}{\del\dot{\b{\nu}}}\delta\dot{\b{\nu}}
 \right]\rmd t\\
 =&\intl_{t_{i}}^{t_{f}}\left\{
 \left[\frac{\del\mcl{L}}{\del\nu}
 -\frac{\rmd}{\rmd t}\left(
 \frac{\del\mcl{L}}{\del\dot{\nu}}
 \right)\right]\delta\nu
 +\left[\frac{\del\mcl{L}}{\del\b{\nu}}
 -\frac{\rmd}{\rmd t}\left(
 \frac{\del\mcl{L}}{\del\dot{\b{\nu}}}
 \right)\right]\delta\b{\nu}
 \right\}\rmd t\\
 &+\left[\frac{\del\mcl{L}}{\del\dot{\nu}}
 \delta\nu\right]_{t_{i}}^{t_{f}}
 +\left[\frac{\del\mcl{L}}{\del\dot{\b{\nu}}}
 \delta\b{\nu}\right]_{t_{i}}^{t_{f}}.
 \end{aligned}
 \label{eq:3.123}
\end{equation}

As expected, by equating the above expression to zero, we obtain the Euler-Lagrange equations:
\begin{subequations}
 \label{eq:3.124}
 \begin{align}
 &\frac{\del\mcl{L}}{\del\b{\nu}}
 -\frac{\rmd}{\rmd t}\left(
 \frac{\del\mcl{L}}{\del\dot{\b{\nu}}}
 \right)=0,
 \label{eq:3.124a}\\[1.2ex]
 &\frac{\del\mcl{L}}{\del\nu}
 -\frac{\rmd}{\rmd t}\left(
 \frac{\del\mcl{L}}{\del\dot{\nu}}
 \right)=0,
 \label{eq:3.124b}
 \end{align}
\end{subequations}
\noindent which are true only if we assume the following restrictions on the Lagrangian:
\begin{subequations}
 \label{eq:3.125}
 \begin{align}
 &\frac{\del\mcl{L}}{\del\dot{\nu}}(t_{f})=0,
 \label{eq:3.125a}\\[1.2ex]
 &\frac{\del\mcl{L}}{\del\dot{\b{\nu}}}(t_{i})=0.
 \label{eq:3.125b}
 \end{align}
\end{subequations}

Notice that, according to the boundary conditions~\eqref{eq:3.66} and the definition~\eqref{eq:3.122}, 
the function~$\mcl{L}$ strictly satisfies the equations~\eqref{eq:3.125}.

Analogously to the results of subsection~\ref{sss:3.2.2}, the validity of identities~\eqref{eq:3.124} 
is also subject to the imposition of two-time boundary conditions on the extremizing trajectory:
\begin{subequations}
 \label{eq:3.126}
 \begin{align}
 &\delta\nu(t_{i})=0,
 \label{eq:3.126a}\\[1.2ex]
 &\delta\b{\nu}(t_{f})=0.
 \label{eq:3.126b}
 \end{align}
\end{subequations}

Substituting the expression~\eqref{eq:3.122} into the Euler-Lagrange equations, we obtain the equations of motion corresponding 
to the dynamical variables $\b{\nu}(t)$ and $\nu(t)$:
\begin{subequations}
 \label{eq:3.127}
 \begin{align}
 &\dot{\nu}=\til{B}\T\nu+\til{C}\b{\nu},
 \label{eq:3.127a}\\[1.2ex]
 &\dot{\b{\nu}}=-\til{A}\nu-\til{B}\b{\nu}.
 \label{eq:3.127b}
 \end{align}
\end{subequations}

Note that, in the above identities, we do not introduce any specific notation for the extremizing trajectories 
of the functional~$\delta^{2}S$. However, especially in the calculations performed in subsections \ref{sss:3.2.3} 
and \ref{sss:3.2.4}, we must remember that the vectors $\b{\nu}(t)$ and $\nu(t)$ represent integration variables, 
whose values are arbitrary in their respective domains. Therefore, a solution of the equations~\eqref{eq:3.127}, 
which is defined only in the context of the present digression, constitutes a particular value for the deviation 
around a classical trajectory as a function of time.

During the derivation of equations~\eqref{eq:3.127}, the following properties were employed:
\begin{subequations}
 \label{eq:3.128}
 \begin{align}
 &\til{A}\T=\til{A},
 \label{eq:3.128a}\\[1.2ex]
 &\til{C}\T=\til{C},
 \label{eq:3.128b}
 \end{align}
\end{subequations}
\noindent which follow immediately from the definitions \eqref{eq:3.68a} and \eqref{eq:3.68c}, since the relations 
$A\T=A$ and $C\T=C$ are also valid, according to the identities \eqref{eq:3.57}, \eqref{eq:3.62a} and \eqref{eq:3.62c}.

The extremization of the functional $\delta^{2}S$ is in exact correspondence with with the linearization of the equations 
of motion~\eqref{eq:3.3}, since both procedures provide entirely equivalent solutions, which we designate as \textit{classical 
deviations}. Consequently, the identities~\eqref{eq:3.127} are in complete analogy with the initial definitions of appendix~\ref{app:A}, 
except for an obvious transformation of variables. In order to emphasize the similarities between these apparently different results, 
we rewrite the differential equations for $\b{\nu}(t)$ and $\nu(t)$ in matrix notation:
\begin{equation}
 \left(\begin{array}{c}
 \dot{\nu}(t)\\ \dot{\b{\nu}}(t)
 \end{array}\right)=
 \left(\begin{array}{c c}
 \til{B}\T(t) & \til{C}(t)\\
 -\til{A}(t)  & -\til{B}(t)
 \end{array}\right)
 \left(\begin{array}{c}
 \nu(t)\\ \b{\nu}(t)
 \end{array}\right).
 \label{eq:3.129}
\end{equation}

Then, similarly to the identity~\eqref{eq:A.2}, we can establish a connection between the classical deviations at the endpoints 
of the propagation interval:
\begin{equation}
 \left(\begin{array}{c}
 \nu(t_{f})\\ \b{\nu}(t_{f})
 \end{array}\right)
 =\left(\begin{array}{c c}
 \til{M}_{11}(t_{f},t_{i})&\til{M}_{12}(t_{f},t_{i})\\
 \til{M}_{21}(t_{f},t_{i})&\til{M}_{22}(t_{f},t_{i})
 \end{array}\right)
 \left(\begin{array}{c}
 \nu(t_{i})\\ \b{\nu}(t_{i})
 \end{array}\right)
 =\til{\mbb{M}}(t_{f},t_{i})
 \left(\begin{array}{c}
 \nu(t_{i})\\ \b{\nu}(t_{i})
 \end{array}\right).
 \label{eq:3.130}
\end{equation}

Now, considering $t=t_{f}$, we insert the above expression into both sides of equation~\eqref{eq:3.129}:
\begin{equation}
 \left(\begin{array}{c c}
 \dot{\til{M}}_{11}(t,t_{i}) & \dot{\til{M}}_{12}(t,t_{i})\\
 \dot{\til{M}}_{21}(t,t_{i}) & \dot{\til{M}}_{22}(t,t_{i})
 \end{array}\right)=
 \left(\begin{array}{c c}
 \til{B}\T(t) & \til{C}(t)\\
 -\til{A}(t)  & -\til{B}(t)
 \end{array}\right)
 \left(\begin{array}{c c}
 \til{M}_{11}(t,t_{i}) & \til{M}_{12}(t,t_{i})\\
 \til{M}_{21}(t,t_{i}) & \til{M}_{22}(t,t_{i})
 \end{array}\right).
 \label{eq:3.131}
\end{equation}

The initial condition for the matrix~$\til{\mbb{M}}$ follows directly from definition~\eqref{eq:3.130}:
\begin{equation}
 \til{\mbb{M}}(t_{i},t_{i})=\mds{1}.
 \label{eq:3.132}
\end{equation}

Notice that the system of differential equations~\eqref{eq:3.131} couples the blocks of $\til{\mbb{M}}$ only in pairs. 
So, with the purpose of highlighting the relevant quantities in the evaluation of the semiclassical propagator, we 
rewrite relations between $\til{M}_{12}$ and $\til{M}_{22}$:
\begin{subequations}
 \label{eq:3.133}
 \begin{align}
 &\dot{\til{M}}_{12}=\til{B}\T\til{M}_{12}+\til{C}\til{M}_{22},
 \label{eq:3.133a}\\[0.0ex]
 &\dot{\til{M}}_{22}=-\til{A}\til{M}_{12}-\til{B}\til{M}_{22}.
 \label{eq:3.133b}
 \end{align}
\end{subequations}

By manipulating the above expressions, we can isolate the time evolution of $\til{M}_{22}$ into a single 
second-order differential equation:
\begin{equation}
 \ddot{\til{M}}_{22}
 =\left(\til{A}\til{B}\T\til{A}^{-1}+\dot{\til{A}}\til{A}^{-1}-\til{B}\right)\dot{\til{M}}_{22}
 +\left(\til{A}\til{B}\T\til{A}^{-1}\til{B}-\til{A}\til{C}+\dot{\til{A}}\til{A}^{-1}\til{B}-\dot{\til{B}}\right)\til{M}_{22},
 \label{eq:3.134}
\end{equation}
\noindent which is usually designated as the \textit{Jacobi equation}. With the aid of identities \eqref{eq:3.132} 
and \eqref{eq:3.133}, we can readily find the initial conditions corresponding to the result~\eqref{eq:3.134}:
\begin{subequations}
 \label{eq:3.135}
 \begin{align}
 &\til{M}_{22}(t_{i},t_{i})=\mds{1},
 \label{eq:3.135a}\\[0.0ex]
 &\dot{\til{M}}_{22}(t_{i},t_{i})=-\til{B}(t_{i}).
 \label{eq:3.135b}
 \end{align}
\end{subequations}

Note that the equation of motion and the initial conditions for the block~$\til{M}_{22}$ are completely identical 
to the expressions \eqref{eq:3.119} and \eqref{eq:3.120c}, which determine the time evolution of the matrix~$Y(t)$. 
Since these two sets of dynamical variables have the same dimension, we immediately conclude that:
\begin{equation}
 Y(t)=\til{M}_{22}(t,t_{i}).
 \label{eq:3.136}
\end{equation}

Therefore, in accordance with the original goal of the present subsection, we have established a connection between 
the explicit solution of the reduced propagator and a classical function of precise physical meaning.

\subsubsection{Result}
\label{sss:3.2.6}

According to the equations~\eqref{eq:3.64}, the quantities $\b{\nu}(t)$ and $\nu(t)$ are related to the deviations $\b{\eta}(t)$ 
and $\eta(t)$ by a simple linear transformation. Consequently, in comparison with the identities~\eqref{eq:3.51}, we can define 
new dynamical variables $\b{v}(t)$ and $v(t)$ with the following properties:
\begin{subequations}
 \label{eq:3.137}
 \begin{align}
 &\nu(t)=v(t)-v_{c}(t),
 \label{eq:3.137a}\\
 &\b{\nu}(t)=\b{v}(t)-\b{v}_{c}(t).
 \label{eq:3.137b}
 \end{align}
\end{subequations}

Notice that the vectors $\b{v}_{c}(t)$ and $v_{c}(t)$, which are used as a reference for the 
deviations $\b{\nu}(t)$ and $\nu(t)$, represent an alternative description for the classical 
trajectory.

Since the classical equations of motion for $\b{v}(t)$ and $v(t)$ remain unknown, we need to rewrite these quantities in terms 
of the original variables $\b{z}(t)$ and $z(t)$. So, by substituting the expressions \eqref{eq:3.51} and \eqref{eq:3.137} into 
the equations~\eqref{eq:3.64}, we obtain the relation between these two parametrizations of the duplicated phase space:
\begin{subequations}
 \label{eq:3.138}
 \begin{align}
 &v(t)=\Theta\T(t)z(t),
 \label{eq:3.138a}\\
 &\b{v}(t)=\Theta(t)\b{z}(t).
 \label{eq:3.138b}
 \end{align}
\end{subequations}

Similarly to the identity~\eqref{eq:A.2}, in order to apply the above transformation to the result~\eqref{eq:3.136}, 
we represent the blocks of the matrix~$\til{\mbb{M}}$ in the form of derivatives between the endpoints of the classical 
trajectory:
\begin{equation}
\renewcommand\arraystretch{2.2}
 \til{\mbb{M}}(t_{f},t_{i})=
 \left(\begin{array}{c c}
 \ds{\frac{\del v(t_{f})}{\del v(t_{i})}}&
 \ds{\frac{\del v(t_{f})}{\del\b{v}(t_{i})}}\\
 \ds{\frac{\del\b{v}(t_{f})}{\del v(t_{i})}}&
 \ds{\frac{\del\b{v}(t_{f})}{\del\b{v}(t_{i})}}
 \end{array}\right).
 \label{eq:3.139}
\renewcommand\arraystretch{1}
\end{equation}

Note that, in this last equation, we have omitted again a possible additional notation to indicate the functions calculated 
on a classical trajectory. As suggested by the expressions \eqref{eq:3.138} and \eqref{eq:3.139}, we can identify the initial 
values $\b{z}(t_{i})$ and $z(t_{i})$ as the independent variables associated with the elements of the matrix~$\til{\mbb{M}}$. 
However, as a consequence of the boundary conditions~\eqref{eq:3.5}, all constituent quantities of the semiclassical 
propagator must have explicit functional dependence on $\b{z}(t_{f})$ and $z(t_{i})$. For this reason, analogously 
to equation~\eqref{eq:A.5}, it becomes appropriate to define a second coupling matrix between classical deviations:
\begin{equation}
\renewcommand\arraystretch{2.2}
 \til{\mbb{T}}(t_{f},t_{i})
 =\left(\begin{array}{c c}
 \ds{\frac{\del v(t_{f})}{\del\b{v}(t_{f})}}&
 \ds{\frac{\del v(t_{f})}{\del v(t_{i})}}\\
 \ds{\frac{\del\b{v}(t_{i})}{\del\b{v}(t_{f})}}&
 \ds{\frac{\del\b{v}(t_{i})}{\del v(t_{i})}}
 \end{array}\right).
 \label{eq:3.140}
\renewcommand\arraystretch{1}
\end{equation}

Except for an evident change of notation, the relations between the matrices $\til{\mbb{M}}(t_{f},t_{i})$ 
and $\til{\mbb{T}}(t_{f},t_{i})$ are equivalently described by the equations~\eqref{eq:A.8}. Consequently, 
we can establish the following identity:
\begin{equation}
 \til{M}_{22}(t_{f},t_{i})=\til{T}_{21}^{-1}(t_{f},t_{i})
 =\left[\frac{\del\b{v}(t_{i})}{\del\b{v}(t_{f})}\right]^{-1}.
 \label{eq:3.141}
\end{equation}

By successively inserting the results \eqref{eq:3.136} and \eqref{eq:3.141} into the expression~\eqref{eq:3.121}, 
we obtain the formal solution for the path integral~\eqref{eq:3.72}:
\begin{equation}
 K_{red}=\left\{\det\left[\frac{\del\b{v}(t_{i})}{\del\b{v}(t_{f})}\right]\right\}^{\frac{1}{2}}
 \exp\left[-\frac{1}{2}\intl_{t_{i}}^{t_{f}}\rmd t\;\tr\til{B}(t)\right].
 \label{eq:3.142}
\end{equation}

At this point, we can perform the last significant simplifications on the factors of $K_{red}$. First, by using 
the transformations~\eqref{eq:3.138}, we redefine the block~$\til{\mbb{T}}_{21}(t_{f},t_{i})$ as a function of 
the original set of dynamical variables:
\begin{equation}
 \frac{\del\b{v}(t_{i})}{\del\b{v}(t_{f})}=
 \frac{\del\b{v}(t_{i})}{\del\b{z}(t_{i})}
 \frac{\del\b{z}(t_{i})}{\del\b{z}(t_{f})}
 \frac{\del\b{z}(t_{f})}{\del\b{v}(t_{f})}=
 \Theta(t_{i})\frac{\del\b{z}(t_{i})}{\del\b{z}(t_{f})}
 \Theta^{-1}(t_{f}).
 \label{eq:3.143}
\end{equation}

Next, by substituting the definition~\eqref{eq:3.65} into the above identity, we rewrite the determinant corresponding 
to the first factor of the reduced propagator:
\begin{equation}
 \det\left[\frac{\del\b{v}(t_{i})}{\del\b{v}(t_{f})}\right]=
 \left[\frac{\det g(t_{i})}{\det g(t_{f})}\right]^{\frac{1}{2}}
 \det\left[\frac{\del\b{z}(t_{i})}{\del\b{z}(t_{f})}\right].
 \label{eq:3.144}
\end{equation}

Now, considering the equation~\eqref{eq:3.68b}, we perform some simple manipulations on the trace 
of the quantity~$\til{B}(t)$:
\begin{equation}
 \begin{aligned}
 \tr[\til{B}(t)]\!=&
 \tr\left[\Theta^{-1}(t)B(t)\Theta^{-1}(t)\right]\\
 =&\tr\left[\xi(t)B(t)\right].
 \end{aligned}
 \label{eq:3.145}
\end{equation}

Then, by using the expression~\eqref{eq:3.62b}, we obtain a new description for the integrand 
in the second factor of the propagator~\eqref{eq:3.142}:
\begin{equation}
 \begin{aligned}
 \tr\left[\xi(\b{z},z)B\right]=&
 -\frac{\rmi}{2\hbar}\tr\left\{
 \frac{\del}{\del\b{z}}\left[\xi(\b{z},z)\frac{\del\mcl{H}}{\del z}\right]
 +\frac{\del}{\del z}\left[\xi\T(\b{z},z)\frac{\del\mcl{H}}{\del\b{z}}\right]\right\}\\
 =&\;\frac{1}{2}\tr\left[\frac{\del\dot{z}}{\del z}
 -\frac{\del\dot{\b{z}}}{\del\b{z}}\right]
 =\frac{1}{2}\tr\left[R_{11}(t)-R_{22}(t)\right].
 \end{aligned}
 \label{eq:3.146}
\end{equation}

In the second line of this last equation, we introduce the blocks of the matrix~$\mbb{R}(t)$, which 
are defined in appendix~\ref{app:A}. By applying the simplifications~(\ref{eq:3.144}\-\ref{eq:3.146}) 
to the identity~\eqref{eq:3.142}, we find the final expression for the reduced propagator:
\begin{equation}
 K_{red}=\left\{\left[\frac{\det g(t_{i})}{\det g(t_{f})}\right]^{\frac{1}{2}}
 \det\left[\frac{\del\b{z}(t_{i})}{\del\b{z}(t_{f})}\right]\right\}^{\frac{1}{2}}
 \exp\left\{\frac{1}{4}\intl_{t_{i}}^{t_{f}}
 \tr\left[R_{22}(t)-R_{11}(t)\right]\rmd t\right\}.
 \label{eq:3.147}
\end{equation}

At last, according to the definition~\eqref{eq:3.55}, the straightforward substitution 
of the above result into equation~\eqref{eq:3.53} precisely provides the semiclassical 
propagator in the form presented by identity~\eqref{eq:3.2}. In this way, we conclude 
the derivation of the formulae shown in subsection~\ref{ssc:3.1}.



\section{Conclusion}
\label{sec:4} 

In section~\ref{sec:2}, a brief review of the generalized concept of coherent states 
was presented. In this context, the main assumptions determining the allowed coherent-state 
sets for the proper derivation of the semiclassical propagator were carefully introduced and 
discussed. Particularly, we observe that the use of an \textit{analytic complex parametrization} 
is essential in performing the phase-space duplication, described in subsection~\ref{sss:3.2.1.add}. 
Moreover, the \textit{unrestricted domain} of the variables~$z$ in the complex plane is an indispensable 
requirement for the evaluation of the Gaussian integrals constituting the reduced propagator, as shown 
in subsection~\ref{sss:3.2.4}.

With the purpose of enabling immediate applications of the semiclassical propagator 
in the theoretical investigation of a wide range of quantum-mechanical systems, three 
examples of coherent-state sets were described in subsection~\ref{ssc:2.2}, namely, the 
canonical, spin, and $\mrm{SU}(n)$ bosonic coherent states.\footnote{Other examples of 
coherent-state sets, including appropriate formulations for the semiclassical treatment 
of fermionic systems, are readily found in the literature.\cite{Zhang90a,VanVoorhis04}} 
The applicability of these examples can be further extended by noting that the tensor 
product of two or more coherent states is also a coherent state, whose dynamical group 
is given by the direct product of the dynamical groups of the original coherent states. 
For instance, the product of canonical and spin coherent states can be used in the study 
of the interaction between quantized electromagnetic fields and multilevel systems, as 
portrayed by Dicke\cite{Dicke54} and Jaynes-Cummings\cite{Jaynes63} models.

In section~\ref{sec:3}, we presented the main result of this paper, a comprehensive derivation 
of the semiclassical propagator in the generalized coherent-state representation. En route we 
obtained several intermediate results with high relevance in developing semiclassical approximations
and in studying classical-quantum correspondence, such as the generalized expression for the 
classical action, as well as its first and second derivatives, shown in subsection~\ref{sss:3.2.2}.

During the derivation of the coherent-state path integral and its semiclassical approximation, 
some important concepts were examined in detail, three of which are worth mentioning. First, 
in subsection~\ref{sss:3.2.1}, we discussed the \textit{continuous-path hypothesis}, which 
was necessary in formulating a continuous-time expression for the path integral. Second, 
in subsection~\ref{sss:3.2.1.add}, for the purpose of properly determining the solutions 
of the classical equations of motion under two-time boundary conditions, we introduced 
the \textit{phase-space duplication}. Third, in subsection~\ref{sss:3.2.3}, we employed 
the \textit{hypothesis of large quantum numbers} in order to reformulate the reduced 
propagator as a well-behaved product of Gaussian integrals. Furthermore, we observed 
that the conjecture~\eqref{eq:3.71} can be used in a systematic analysis of the validity 
regime of the semiclassical approximation.

The semiclassical propagator, as presented in equation~\eqref{eq:3.2}, is not easily applicable to the explicit calculation 
of the approximate time evolution of quantum systems, due to the occurrence of three technical difficulties. First, obtaining 
analytical or numerical solutions to the classical equations of motion under boundary conditions represents a fairly involved 
and computationally expensive process, especially in systems with multiple degrees of freedom.\footnote{This practical problem 
is aggravated in the coherent-state representation by the phase-space duplication.} Second, there are classical trajectories 
that, although properly satisfying the equations of motion and the boundary conditions, do not provide correct physical 
contributions to the evaluation of the semiclassical propagator.\cite{Adachi89,Rubin95,Shudo95,Shudo96,Ribeiro04a,Aguiar05} 
For this reason, these \textit{spurious trajectories} must be identified and removed from the semiclassical calculations. 
Third, the quantity~$\det\!\left[\del\b{z}(t_{i})/\del\b{z}(t_{f})\right]$, found in the pre-factor~\eqref{eq:3.9}, 
diverges at the so-called \textit{focal points} in the variables $\b{z}$.

Although a direct implementation of the semiclassical propagator presents some significant obstacles,
the expression~\eqref{eq:3.2} constitutes a very useful tool in developing semiclassical formulas 
endowed with immediate practical applicability. For examples of effective use of the semiclassical 
propagator, in which all the issues mentioned in the previous paragraph are appropriately addressed,
we refer the reader to previous works of the present authors.\cite{Aguiar10,Viscondi11c}



\section*{Acknowledgements}

This work was supported by FAPESP under grant numbers 2008/09491-9, 
2011/20065-4, 2012/20452-0, and 2014/04036-2. In addition, MAMA also 
received support from CNPq.


\appendix


\section{Matrix Definitions}
\label{app:A}

In this appendix, we present some definitions relevant to the dynamics of small deviations around a classical trajectory. 
In general, the matrices introduced in this context represent important elements in the construction of semiclassical 
formulas.

First, by performing the linearization of identities~\eqref{eq:3.3}, we find the equations of motion 
for the \textit{classical deviations}:
\renewcommand\arraystretch{1.5}
\begin{equation}
 \begin{aligned}
 \left(\begin{array}{c}
 \delta\dot{z}(t)\\ \delta\dot{\b{z}}(t)
 \end{array}\right)=&
 \left(\begin{array}{c c}
 \frac{\del\dot{z}}{\del z}&
 \frac{\del\dot{z}}{\del\b{z}}\\
 \frac{\del\dot{\b{z}}}{\del z}&
 \frac{\del\dot{\b{z}}}{\del\b{z}}
 \end{array}\right)
 \left(\begin{array}{c}
 \delta z(t)\\ \delta\b{z}(t)
 \end{array}\right)\\
 =&\left(\begin{array}{c c}
 -\frac{\rmi}{\hbar}\frac{\del}{\del z}\left[\xi\T(\b{z},z)
 \frac{\del\mcl{H}(\b{z},z)}{\del\b{z}}\right]&
 -\frac{\rmi}{\hbar}\frac{\del}{\del\b{z}}\left[\xi\T(\b{z},z)
 \frac{\del\mcl{H}(\b{z},z)}{\del\b{z}}\right]\\
 \frac{\rmi}{\hbar}\frac{\del}{\del z}\left[\xi(\b{z},z)
 \frac{\del\mcl{H}(\b{z},z)}{\del z}\right]&
 \frac{\rmi}{\hbar}\frac{\del}{\del\b{z}}\left[\xi(\b{z},z)
 \frac{\del\mcl{H}(\b{z},z)}{\del z}\right]
 \end{array}\right)
 \left(\begin{array}{c}
 \delta z(t)\\ \delta\b{z}(t)
 \end{array}\right)\\
 =&\left(\begin{array}{c c}
 R_{11}(t)&R_{12}(t)\\
 R_{21}(t)&R_{22}(t)
 \end{array}\right)
 \left(\begin{array}{c}
 \delta z(t)\\ \delta\b{z}(t)
 \end{array}\right)
 =\mbb{R}(t)
 \left(\begin{array}{c}
 \delta z(t)\\ \delta\b{z}(t)
 \end{array}\right).
 \end{aligned}
 \label{eq:A.1}
\end{equation}
\renewcommand\arraystretch{1}

Equivalently, we could have obtained the above result by extremization of the functional~$\delta^{2}S$, 
considering specifically the expression described by identity~\eqref{eq:3.61}. Therefore, the variables 
$\delta\b{z}(t)$ and $\delta z(t)$ represent the extremizing trajectories associated with the arbitrary 
deviations $\b{\eta}(t)$ and $\eta(t)$.

Next, we define the \textit{tangent matrix}~$\mbb{M}(t_{f},t_{i})$ as the linear transformation between 
the classical deviations evaluated at the initial and final times:
\renewcommand\arraystretch{1.5}
\begin{equation}
 \begin{aligned}
 \left(\begin{array}{c}
 \delta z(t_{f})\\
 \delta\b{z}(t_{f})
 \end{array}\right)
 =&\left(\begin{array}{c c}
 M_{11}(t_{f},t_{i})&M_{12}(t_{f},t_{i})\\
 M_{21}(t_{f},t_{i})&M_{22}(t_{f},t_{i})
 \end{array}\right)
 \left(\begin{array}{c}
 \delta z(t_{i})\\
 \delta\b{z}(t_{i})
 \end{array}\right)
 =\mbb{M}(t_{f},t_{i})
 \left(\begin{array}{c}
 \delta z(t_{i})\\
 \delta\b{z}(t_{i})
 \end{array}\right)\\
 =&\left(\begin{array}{c c}
 \frac{\del z(t_{f})}{\del z(t_{i})}&
 \frac{\del z(t_{f})}{\del\b{z}(t_{i})}\\
 \frac{\del\b{z}(t_{f})}{\del z(t_{i})}&
 \frac{\del\b{z}(t_{f})}{\del\b{z}(t_{i})}
 \end{array}\right)
 \left(\begin{array}{c}
 \delta z(t_{i})\\
 \delta\b{z}(t_{i})
 \end{array}\right).
 \end{aligned}
 \label{eq:A.2}
\end{equation}
\renewcommand\arraystretch{1}

Alternatively, we can calculate the tangent matrix as the solution of a system of differential equations 
subject to initial conditions. For this purpose, we insert the expression~\eqref{eq:A.2} into both sides 
of identity~\eqref{eq:A.1} for $t=t_{f}$:
\begin{equation}
 \frac{\rmd\mbb{M}(t_{f},t_{i})}{\rmd t_{f}}=\mbb{R}(t_{f})\mbb{M}(t_{f},t_{i}).
 \label{eq:A.3}
\end{equation}

Note that the vector of initial classical deviations was removed during the formulation of this last equation. 
Now, by rewriting the definition~\eqref{eq:A.2} for the particular case in which $t_{f}=t_{i}$, we find the 
initial conditions required for the determination of $\mbb{M}$:
\begin{equation}
 \mbb{M}(t_{i},t_{i})=\mds{1}.
 \label{eq:A.4}
\end{equation}

However, we must remember that the matrix~$\mbb{R}$ is evaluated on a classical trajectory, which generally 
results from the imposition of initial or boundary conditions on the identities~\eqref{eq:3.3}. Therefore, 
although the equation of motion for the tangent matrix is explicitly determined by an initial value, the 
corresponding calculation of the quantity~$\mbb{R}$ can depend implicitly on a two-time boundary condition.

In general, the results based on classical trajectories subject to boundary conditions, such as the semiclassical 
propagator, are composed of quantities with functional dependence on the variables $\b{z}(t_{f})$ and $z(t_{i})$. 
In this case, it is convenient to define a third matrix, which relates the deviations in the independent endpoints 
of the trajectory to the remaining extremities:
\renewcommand\arraystretch{1.5}
\begin{equation}
 \begin{aligned}
 \left(\begin{array}{c}
 \delta z(t_{f})\\
 \delta\b{z}(t_{i})
 \end{array}\right)
 =&\left(\begin{array}{c c}
 T_{11}(t_{f},t_{i})&T_{12}(t_{f},t_{i})\\
 T_{21}(t_{f},t_{i})&T_{22}(t_{f},t_{i})
 \end{array}\right)
 \left(\begin{array}{c}
 \delta\b{z}(t_{f})\\
 \delta z(t_{i})
 \end{array}\right)
 =\mbb{T}(t_{f},t_{i})
 \left(\begin{array}{c}
 \delta\b{z}(t_{f})\\
 \delta z(t_{i})
 \end{array}\right)\\
 =&\left(\begin{array}{c c}
 \frac{\del z(t_{f})}{\del\b{z}(t_{f})} &
 \frac{\del z(t_{f})}{\del z(t_{i})}\\
 \frac{\del\b{z}(t_{i})}{\del\b{z}(t_{f})}&
 \frac{\del\b{z}(t_{i})}{\del z(t_{i})}
 \end{array}\right)
 \left(\begin{array}{c}
 \delta\b{z}(t_{f})\\
 \delta z(t_{i})
 \end{array}\right).
 \end{aligned}
 \label{eq:A.5}
\end{equation}
\renewcommand\arraystretch{1}

On the other hand, in formulations determined by initial values, the relevant independent variables are $\b{z}(t_{i})$ 
and $z(t_{i})$. Consequently, in many situations, which are usually associated with the transformation of the functional 
dependence of a classical quantity, we must reformulate the constituent derivatives of $\mbb{T}$ in terms of the blocks 
of the matrix $\mbb{M}$. For this purpose, we rewrite the expression~\eqref{eq:A.2} as follows:
\begin{subequations}
 \label{eq:A.6}
 \begin{align}
 &\delta z(t_{f})=M_{11}\delta z(t_{i})+M_{12}\delta\b{z}(t_{i}),
 \label{eq:A.6a}\\[1.2ex]
 &\delta\b{z}(t_{f})=M_{21}\delta z(t_{i})+M_{22}\delta\b{z}(t_{i}).
 \label{eq:A.6b}
 \end{align}
\end{subequations}

Then, by assuming that the block~$M_{22}$ is invertible, we reorganize 
the classical deviations in the form indicated by identity~\eqref{eq:A.5}:
\begin{subequations}
 \label{eq:A.7}
 \begin{align}
 &\delta z(t_{f})=M_{12}M_{22}^{-1}\delta\b{z}(t_{f})+\left(M_{11}-M_{12}M_{22}^{-1}M_{21}\right)\delta z(t_{i}),
 \label{eq:A.7a}\\[1.2ex]
 &\delta\b{z}(t_{i})=M_{22}^{-1}\delta\b{z}(t_{f})-M_{22}^{-1}M_{21}\delta z(t_{i}).
 \label{eq:A.7b}
 \end{align}
\end{subequations}

In this way, the relations between the matrices $\mbb{T}$ and $\mbb{M}$ become evident:
\begin{subequations}
 \label{eq:A.8}
 \begin{align}
 &T_{11}=M_{12}M_{22}^{-1},
 \label{eq:A.8a}\\[1.2ex]
 &T_{12}=M_{11}-M_{12}M_{22}^{-1}M_{21},
 \label{eq:A.8b}\\[1.2ex]
 &T_{21}=M_{22}^{-1},
 \label{eq:A.8c}\\[1.2ex]
 &T_{22}=-M_{22}^{-1}M_{21}.
 \label{eq:A.8d}
 \end{align}
\end{subequations}

Finally, notice that the definitions presented in this appendix for the matrices $\mbb{R}$, $\mbb{M}$ and $\mbb{T}$ 
are described in appropriate notation for the duplicated phase space. The analogous expressions for the simple phase 
space are given by the immediate particularization of equations \eqref{eq:A.1}, \eqref{eq:A.2} and \eqref{eq:A.5} 
under the restriction~$\b{z}(t)=z\cg(t)$.



\section{Auxiliary Results}
\label{app:B}

For the sake of completeness, we present here the first and second derivatives of the Lagrangian~\eqref{eq:3.6b}
and the boundary term~\eqref{eq:3.6c}. These results are employed in subsections \ref{sss:3.2.2} and \ref{sss:3.2.3},
during the discussion of the first and second variations of the action functional in the duplicated phase space.

First, by using the expression~\eqref{eq:3.6b}, we calculate the first derivatives of the Lagrangian:
\begin{subequations}
 \begin{align}
 &\frac{\rmi}{\hbar}\frac{\del L}{\del z}=
 \frac{1}{2}\frac{\del^{2}f(\b{z},z)}{\del z\del\b{z}}\dot{\b{z}}
 -\frac{1}{2}\frac{\del^{2}f(\b{z},z)}{\del z^{2}}\dot{z}
 -\frac{\rmi}{\hbar}\frac{\del\mcl{H}(\b{z},z)}{\del z},
 \label{eq:3.42a}\\[1.2ex]
 &\frac{\rmi}{\hbar}\frac{\del L}{\del\b{z}}=
 \frac{1}{2}\frac{\del^{2}f(\b{z},z)}{\del\b{z}^{2}}\dot{\b{z}}
 -\frac{1}{2}\frac{\del^{2}f(\b{z},z)}{\del\b{z}\del z}\dot{z}
 -\frac{\rmi}{\hbar}\frac{\del\mcl{H}(\b{z},z)}{\del\b{z}},
 \label{eq:3.42b}\\[1.2ex]
 &\frac{\rmi}{\hbar}\frac{\del L}{\del\dot{z}}=
 -\frac{1}{2}\frac{\del f(\b{z},z)}{\del z},
 \label{eq:3.42c}\\[1.2ex]
 &\frac{\rmi}{\hbar}\frac{\del L}{\del\dot{\b{z}}}=
 \frac{1}{2}\frac{\del f(\b{z},z)}{\del\b{z}}.
 \label{eq:3.42d}
 \end{align}
 \label{eq:3.42} 
\end{subequations}

Then, by employing the above results, we determine the time derivatives 
required for the construction of the first variation of the action:
\begin{subequations}
 \label{eq:3.43}
 \begin{align}
 &\frac{\rmi}{\hbar}\frac{\rmd}{\rmd t}\left(\frac{\del L}{\del\dot{z}}\right)=
 -\frac{1}{2}\left[\frac{\del^{2}f(\b{z},z)}{\del z\del\b{z}}\dot{\b{z}}
 +\frac{\del^{2}f(\b{z},z)}{\del z^{2}}\dot{z}\right],
 \label{eq:3.43a}\\[1.2ex]
 &\frac{\rmi}{\hbar}\frac{\rmd}{\rmd t}\left(\frac{\del L}{\del\dot{\b{z}}}\right)=
 \frac{1}{2}\left[\frac{\del^{2}f(\b{z},z)}{\del\b{z}\del z}\dot{z}
 +\frac{\del^{2}f(\b{z},z)}{\del\b{z}^{2}}\dot{\b{z}}\right].
 \label{eq:3.43b}
 \end{align}
\end{subequations}

Next, considering the definition~\eqref{eq:3.6c} for $z_{i}=z(t_{i})$ and $z\cg_{f}=\b{z}(t_{f})$, 
we obtain the first derivatives of the boundary term:
\begin{subequations}
 \label{eq:3.44}
 \begin{align}
 &\frac{\rmi}{\hbar}\frac{\del\Gamma}{\del z(t_{i})}=
 \frac{1}{2}\frac{\del f(\b{z}(t_{i}),z(t_{i}))}{\del z(t_{i})},
 \label{eq:3.44a}\\[1.2ex]
 &\frac{\rmi}{\hbar}\frac{\del\Gamma}{\del\b{z}(t_{i})}=
 \frac{1}{2}\frac{\del f(\b{z}(t_{i}),z(t_{i}))}{\del\b{z}(t_{i})},
 \label{eq:3.44b}\\[1.2ex]
 &\frac{\rmi}{\hbar}\frac{\del\Gamma}{\del z(t_{f})}=
 \frac{1}{2}\frac{\del f(\b{z}(t_{f}),z(t_{f}))}{\del z(t_{f})},
 \label{eq:3.44c}\\[1.2ex]
 &\frac{\rmi}{\hbar}\frac{\del\Gamma}{\del\b{z}(t_{f})}=
 \frac{1}{2}\frac{\del f(\b{z}(t_{f}),z(t_{f}))}{\del\b{z}(t_{f})}.
 \label{eq:3.44d}
 \end{align}
\end{subequations}

The identities~(\ref{eq:3.42}\-\ref{eq:3.44}) are used in subsection~\ref{sss:3.2.2}
for the derivation of equation~\eqref{eq:3.45}. 

With the purpose of explicitly calculating the second variation of the action functional, 
given initially by expression~\eqref{eq:3.56}, we use the equations~\eqref{eq:3.42} to 
obtain the nonzero second-order derivatives of the Lagrangian:
\begin{subequations}
 \label{eq:3.58}
 \begin{align}
 &\frac{\rmi}{\hbar}\frac{\del^{2}L}{\del z^{2}}=
 \frac{1}{2}\frac{\del}{\del z}
 \left[\frac{\del^{2}f(\b{z},z)}{\del z\del\b{z}}\dot{\b{z}}
 -\frac{\del^{2}f(\b{z},z)}{\del z^{2}}\dot{z}\right]
 -\frac{\rmi}{\hbar}\frac{\del^{2}\mcl{H}(\b{z},z)}{\del z^{2}},
 \label{eq:3.58a}\\[1.2ex]
 &\frac{\rmi}{\hbar}\frac{\del^{2}L}{\del\b{z}^{2}}=
 \frac{1}{2}\frac{\del}{\del\b{z}}\left[
 \frac{\del^{2}f(\b{z},z)}{\del\b{z}^{2}}\dot{\b{z}}
 -\frac{\del^{2}f(\b{z},z)}{\del\b{z}\del z}\dot{z}\right]
 -\frac{\rmi}{\hbar}\frac{\del^{2}\mcl{H}}{\del\b{z}^{2}},
 \label{eq:3.58b}\\[1.2ex]
 &\frac{\rmi}{\hbar}\frac{\del^{2}L}{\del z\del\b{z}}=
 \frac{1}{2}\frac{\del}{\del\b{z}}\left[
 \frac{\del^{2}f(\b{z},z)}{\del z\del\b{z}}\dot{\b{z}}
 -\frac{\del^{2}f(\b{z},z)}{\del z^{2}}\dot{z}\right]
 -\frac{\rmi}{\hbar}\frac{\del^{2}\mcl{H}(\b{z},z)}{\del z\del\b{z}},
 \label{eq:3.58c}\\[1.2ex]
 &\frac{\rmi}{\hbar}\frac{\del^{2}L}{\del\b{z}\del z}=
 \frac{1}{2}\frac{\del}{\del z}\left[
 \frac{\del^{2}f(\b{z},z)}{\del\b{z}^{2}}\dot{\b{z}}
 -\frac{\del^{2}f(\b{z},z)}{\del\b{z}\del z}\dot{z}\right]
 -\frac{\rmi}{\hbar}\frac{\del^{2}\mcl{H}(\b{z},z)}{\del\b{z}\del z},
 \label{eq:3.58d}\\[1.2ex]
 &\frac{\rmi}{\hbar}\frac{\del^{2}L}{\del z\del\dot{z}}=
 -\frac{1}{2}\frac{\del^{2}f(\b{z},z)}{\del z^{2}},
 \label{eq:3.58e}\\[1.2ex]
 &\frac{\rmi}{\hbar}\frac{\del^{2}L}{\del\b{z}\del\dot{\b{z}}}=
 \frac{1}{2}\frac{\del^{2}f(\b{z},z)}{\del\b{z}^{2}},
 \label{eq:3.58f}\\[1.2ex]
 &\frac{\rmi}{\hbar}\frac{\del^{2}L}{\del z\del\dot{\b{z}}}
 =\frac{1}{2}\frac{\del^{2}f(\b{z},z)}{\del z\del\b{z}}
 =\frac{1}{2}g(\b{z},z),
 \label{eq:3.58g}\\[1.2ex]
 &\frac{\rmi}{\hbar}\frac{\del^{2}L}{\del\b{z}\del\dot{z}}
 =-\frac{1}{2}\frac{\del^{2}f(\b{z},z)}{\del\b{z}\del z}
 =-\frac{1}{2}g\T(\b{z},z).
 \label{eq:3.58h}
 \end{align}
\end{subequations}

By employing the results \eqref{eq:3.58e} and \eqref{eq:3.58f}, we 
calculate the time derivatives required for the evaluation of $\delta^{2}S$:
\begin{subequations}
 \label{eq:3.59}
 \begin{align}
 &\frac{\rmi}{\hbar}\frac{\rmd}{\rmd t}\left(\frac{\del^{2}L}{\del z\del\dot{z}}\right)=
 -\frac{1}{2}\left(\dot{z}\frac{\del}{\del z}\right)\left[\frac{\del^{2}f(\b{z},z)}{\del z^{2}}\right]
 -\frac{1}{2}\left(\dot{\b{z}}\frac{\del}{\del\b{z}}\right)\left[\frac{\del^{2}f(\b{z},z)}{\del z^{2}}\right],
 \label{eq:3.59a}\\[1.2ex]
 &\frac{\rmi}{\hbar}\frac{\rmd}{\rmd t}\left(\frac{\del^{2}L}{\del\b{z}\del\dot{\b{z}}}\right)=
 \frac{1}{2}\left(\dot{z}\frac{\del}{\del z}\right)\left[\frac{\del^{2}f(\b{z},z)}{\del\b{z}^{2}}\right]
 +\frac{1}{2}\left(\dot{\b{z}}\frac{\del}{\del\b{z}}\right)\left[\frac{\del^{2}f(\b{z},z)}{\del\b{z}^{2}}\right].
 \label{eq:3.59b}
 \end{align}
\end{subequations}

Then, by using the definition~\eqref{eq:3.6c} for fixed boundary conditions, 
we determine the nonzero second derivatives of the boundary term:
\begin{subequations}
 \label{eq:3.60}
 \begin{align}
 &\frac{\rmi}{\hbar}\frac{\del^{2}\Gamma}{\del z(t_{f})^{2}}=
 \frac{1}{2}\frac{\del^{2}f(\b{z}(t_{f}),z(t_{f}))}{\del z(t_{f})^{2}},
 \label{eq:3.60a}\\[1.2ex]
 &\frac{\rmi}{\hbar}\frac{\del^{2}\Gamma}{\del\b{z}(t_{i})^{2}}=
 \frac{1}{2}\frac{\del^{2}f(\b{z}(t_{i}),z(t_{i}))}{\del\b{z}(t_{i})^{2}}.
 \label{eq:3.60b}
 \end{align}
\end{subequations}

As shown in subsection~\ref{sss:3.2.3}, the equations~(\ref{eq:3.58}\-\ref{eq:3.60})
are employed in the formulation of identities \eqref{eq:3.61} and \eqref{eq:3.62}.



\bibliographystyle{unsrt}
\renewcommand{\bibname}{References}


\end{document}